\pdfoutput=1
\documentclass{PoS}

\newcommand{\ee}{e^{+}e^{-}}
\newcommand{\leplep}{\ell^{+}\ell^{-}}
\newcommand{\jp}{J/\psi}

\newcommand{\psip}{\psi^{\prime}}
\newcommand{\etap}{\eta^{\prime}}

\newcommand{\pipi}{\pi^{+}\pi^{-}}

\newcommand{\pim}{\pi^{-}}
\newcommand{\pip}{\pi^{+}}
\newcommand{\piz}{\pi^{0}}
\newcommand{\ppbar}{p\bar{p}}
\newcommand{\pbar}{\bar{p}}

\newcommand{\ubar}{\bar{u}}

\newcommand{\dbar}{\bar{d}}
\newcommand{\DDbar}{D\bar{D}}

\newcommand{\ccbar}{c\bar{c}}
\newcommand{\cbar}{\bar{c}}
\newcommand{\bbbar}{b\bar{b}}

\newcommand{\qqbar}{q\bar{q}}
\newcommand{\qbar}{\bar{q}}

\newcommand{\rt}{\rightarrow}

\newcommand{\etal}{\em et al.}
\newcommand{\jpsi}{J/\psi}

\title{XYZ Meson Spectroscopy}

\ShortTitle{XYZ Mesons}

\author{\speaker{Stephen Lars Olsen}%
         \thanks{This work supported under Project Code IBS-R001-D1.}\\
        Center for Underground Physics, Institute for Basic Science\\
        Daejeon 305-811, Republic of Korea\\ 
        E-mail: \email{solsensnu@gmail.com}}

\abstract{
Many candidate multiquark mesons, {\it i.e.}, mesons with substructures that are more complex
than the quark-antiquark prescription that is in the textbooks, have recently been observed.
Many of the most recently observed candidate states are electrically charged and have the
same spin and parity, namely $J^{P}=1^{+}$. In this talk I give an overview of the current
experimental situation and identify patterns among the recently discovered $J^{P}=1^{+}$ 
states, compare these patterns with expectations from proposed theoretical models, and the 
existence of additional, related states that might be accessible at current and future
experiments.  In addition, models that attribute the observed states to kinematically
induced cusps are discussed.
}

\FullConference{53rd International Winter Meeting on Nuclear Physics\\
                 26-30 January 2015\\
                 Bormio, Italy}

\begin{document}

\section{Introduction}
\noindent
The strongly interacting particles of the Standard Model are colored quarks and gluons. In 
ontrast, the strongly interacting particles in nature are color-singlet ({\it i.e.}, white)
mesons and baryons.  In the theory, quarks and gluons are related to mesons and baryons by
the long-distance regime of Quantum Chromodynamics (QCD), which remains the least understood
aspect of the theory.  Since first-principle lattice-QCD (LQCD) calculations are still not
practical for most long-distance phenomena,\footnote{This might not the case for very much
longer. Recent progress in this area has been impressive~\cite{sasa}.} a number of models
motivated by the color structure of QCD have been proposed.  However, so far at least,
predictions of these QCD-motivated models that pertain to the spectrum of hadrons have not
had great success.

\begin{figure}[htb]
\begin{center}
  \includegraphics[height=0.9\textwidth,width=0.7\textwidth]{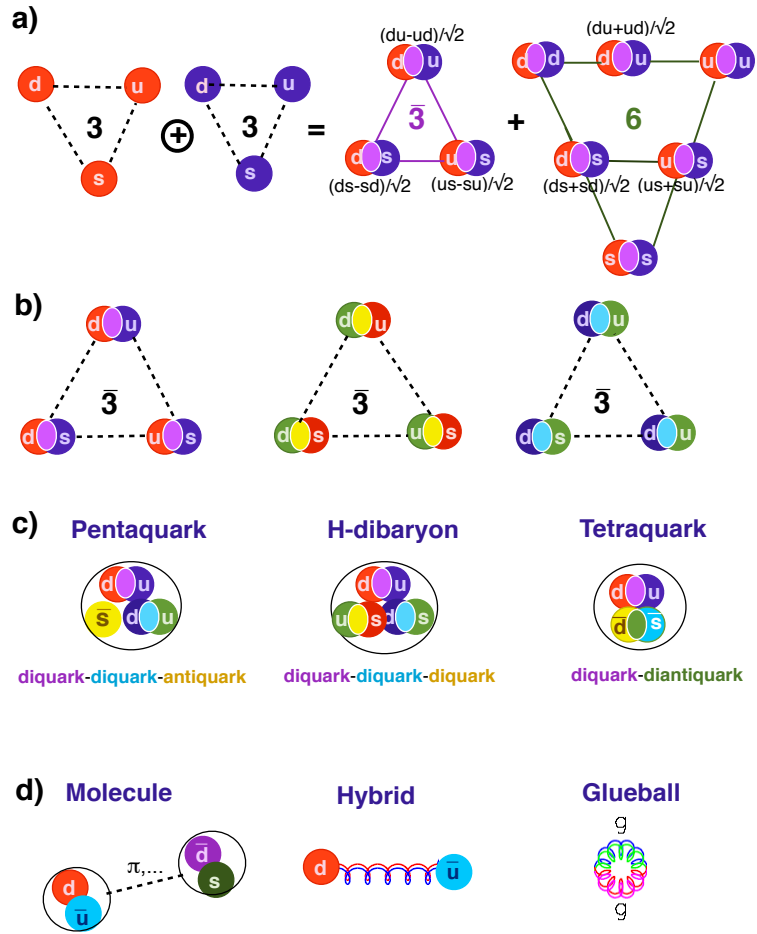}
\caption{\footnotesize {\bf a)} Combining a red and blue quark triplet produces a
magenta (antigreen) antitriplet and sextet. 
{\bf b)}  The three anticolored diquark antitriplets.
{\bf c)}  Some of the multiquark, color-singlet states that can be formed from quarks,
antiquarks, diquarks and diantiquarks.
{\bf d)} Other possible multiquark/gluon systems.}
\label{fig:diquarks}
\end{center}
\end{figure}

For example, it is well known that combining a $q=u,d,s$ light-quark triplet with a
$\bar{q}=\bar{u},\bar{d},\bar{s}$ light-antiquark antitriplet gives the familiar meson octet of flavor-$SU(3)$.
Using similar considerations based on QCD,  two quark triplets can be combined to form a ``diquark''
antitriplet of antisymmetric $qq$ states and a sextet of symmetric $qq$ states as illustrated in
Fig.~\ref{fig:diquarks}a. In QCD, these diquarks have color: combining a red triplet with a blue triplet
-- as shown in the figure -- produces a magenta (anti-green) diquark and, for the antisymmetric triplet
configuration, the color force between the two quarks is expected to be attractive. 
Likewise, green-red and blue-green diquarks form yellow (anti-blue) and cyan (anti-red)
antitriplets as shown in Fig.~\ref{fig:diquarks}b.

Since these diquarks are not color-singlets, they cannot exist as free particles but, on the
other hand, the  anticolored diquark antitriplets should be able to combine with other colored
objects in a manner similar to antiquark antitriplets, thereby forming multiquark color-singlet
states with a more complex substructure than the $\qqbar$ mesons and $qqq$ baryons of the original
quark model~\cite{jaffe_diquark}.   These so-called ``exotic'' states  include 
pentaquark baryons, six-quark
$H$-dibaryons, and tetraquark mesons, as illustrated in Fig.~\ref{fig:diquarks}c.
Other proposed exotic states are: {\it glueballs}, which are mesons made only from gluons;
{\it hybrids} formed from a $q$, $\bar{q}$ and a gluon; and {\it molecules}, which are deuteron-like
bound states of color-singlet ``normal'' hadrons~\cite{molecule_pre1994}. These
are illustrated in Fig.~\ref{fig:diquarks}d.   Glueball and hybrid mesons are motivated by QCD;
molecules are a generalization of classical nuclear physics to systems of subatomic particles.

\section{Searches for Exotic Hadrons in Light-Quark Systems}
\noindent
Of the proposed ``multiquark'' states,  pentaquarks have probably attracted the most 
theoretical and experimental attention~\cite{hicks_pentaquark}.  However, in
spite of some dramatic false alarms~\cite{nakano}, there is no strong evidence for the existence
of pentaquarks in nature~\cite{penta, trilling}.  The absence of pentaquarks led Wilczek to remark ``{\it The
story of the pentaquark shows how poorly we understand QCD}''~\cite{wilczek_quote}.   As for
the six-quark $H$-dibaryon state, its strong theoretical motivation~\cite{jaffe_H} has inspired numerous
experimental searches~\cite{belle_H}, However, to date, no evidence for it has been seen.
The lack of any sign of the $H$-dibaryon (among other things) led Jaffe to observe that ``{\it The absence
of exotics is one of the most obvious features of QCD}''~\cite{jaffe_quote}.  

The experimental case for baryonium, a nucleon-antinucleon bound state~\cite{fermi-yang}, is more promising.
While searches for narrow gamma-ray lines produced inclusively by at-rest $\ppbar$ annihilations
found no signals~\cite{LEAR}, a strong threshold enhancement in the $\ppbar$ mass spectrum for
radiative $\jpsi\rt\gamma\ppbar$ decays, reported by the BESII collaboration in 2003~\cite{bes2_gppb},
may be the tail of an $S$-wave $\ppbar$ bound state.   Similar structures are not seen in the
$\ppbar$ systems produced in $\jpsi\rt\omega\ppbar$~\cite{bes2_wppb} or
$\Upsilon(1S)\rt\gamma\ppbar$~\cite{cleo_u1s_gppb} decays, which suggests that the observed
structure cannot be entirely attributable to final-state-interactions between the $p$ and $\pbar$.
Different theoretical attempts to understand this threshold structure give contradictory results.
An analysis based on an $N\bar{N}$ interaction
derived in the framework of chiral effective field theory~\cite{eft} finds an isovector $N\bar{N}$
bound state 37~MeV below the $2m_p$ mass threshold~\cite{kang}. 
In contrast, an analysis based on the Paris $N\bar{N}$ potential~\cite{paris}, attributes the threshold
enhancement to the tail of  an isospin singlet $^1S_0$ $N\bar{N}$ state with a mass 
that is 4.8~MeV below threshold~\cite{dedonder}.\footnote{In this report the convention $c=1$ is used.}
This latter analysis also predicts a nearby triplet $P$-wave state.

Two candidates for exotic light-quark mesons, the $J^{PC}=1^{-+}$ $\pi_1(1600)$~\cite{pdg}, and a
new result from COMPASS, the $1^{++}$ $a_(1420)$~\cite{compass_a1}, were discussed at this meeting
by Ketzer~\cite{ketzer}. The $\pi_1(1600)$ has explicity exotic\footnote{Quantum numbers that
cannot be accessed by a $\qqbar$ system, {\it e.g.} $0^{--}$, $0^{+-}$, $1^{-+}$, $2^{+-}$ etc.,
are called ``exotic.''} quantum numbers.  However, because strong rescattering effects are
expected to provide significant backgrounds with $1^{-+}$ quantum numbers, questions have been
raised about the interpretation of the $\pi_1(1600)$ signals as a true resonance~\cite{dzierba}.
COMPASS sees strong $\pi_1(1600)\rt\etap\pi$ signals in $\pim p\rt \pim\etap p$ reactions with
large four-momentum transfer to the proton, where rescattering effects
are expected to be small. However, a full analysis of these new data is not yet available.
The $a_1(1420)$ shows up as a peak in the $1^{++}$ $f_0(980)\pi$ $P$-wave produced in
$\pim p\rt f_0(980)\pim p$ reactions.  Although it does not have exotic quantum numbers,
it is unusual in that its mass (1412$\sim$1422~MeV) is too low and width (130$\sim$150~MeV) too
narrow to be considered as a reasonable candidate for a radial excitation of the well established
$1^{++}$ ground state meson, the $a_1(1260)$.  Ketzer cautioned thet the $a_1(1420)$ mass peak is
just above the $K^*(890)\bar{K}$ mass threshold ($m_{K^*(890)}+m_{K}\simeq 1390$~MeV), and the rescattering
process $a_1(1260)\rt K^*(890)\bar{K}\rt f0(980)\pi$ can produce a cusp-like peak in the $f_0(980)\pi$
invariant mass distribution that is unrelated to any resonance and dangerously close to 1420~MeV. 
Moreover, the finite width and Breit-Wigner phase of the $K^*(890)$ resonance can produce a phase
motion that might mimic that of a real resonance.  I discuss rescattering induced, near-threshold
cusp effects below, albeit in a somewhat different context.

\section{Charmonium and the $\mathbf{XYZ}$ Mesons}
\begin{figure}[htb]
\begin{center}
  \includegraphics[height=0.8\textwidth,width=0.7\textwidth]{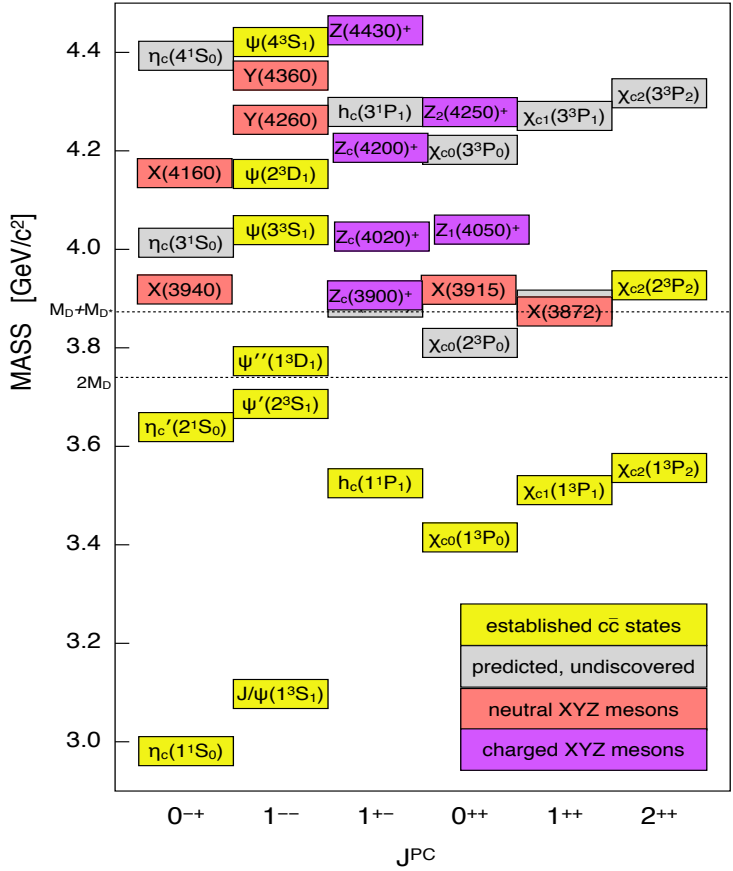}
\caption{\footnotesize The spectrum of charmonium and charmoniumlike mesons.
}
\label{fig:charmonium}
\end{center}
\end{figure}

\noindent
Charmonium mesons are states that can be formed from a charmed ($c$) and anticharmed
($\bar{c}$) quark pair.   Since the charmed quark is relatively massive ($m_c\simeq 1.3$~GeV), the
constituent velocities in charmonium mesons are relatively small ($v^2/c^2\simeq 0.2$) and relativisitic effects
can be treated as perturbative corrections to ordinary Quantum Mechanical calculations~\cite{gi,barnes}.
The mass spectrum of experimentally established charmonium states is indicated by the yellow rectangles
in Fig.~\ref{fig:charmonium}; all of the expected states with mass below the $2m_D$ open-charmed threshold
have been assigned.   The gray rectangles indicate remaining unassigned levels that are below
4.5~GeV. 

The large $c$-quark mass is expected to suppress the production of $\ccbar$ pairs via
quark$\rt$hadron fragmentation processes at $E_{\rm cm}\le\sim 10$~GeV to an insignificant level~\cite{seymour}.
Thus, if a newly observed meson state decays into final states that contain a $c$- and 
$\cbar$-quark pair, those quarks must be present in the initial-state particle.  If the
initial-state particle's constituents are only the $c$- and the $\cbar$-quarks, then the particle
is necessarily a charmonium meson and must occupy one of the unassigned levels in the 
charmonium spectrum, ({\it i.e.}, one of the gray rectangles in Fig.~\ref{fig:charmonium}).
Similar remarks apply to meson states that are seen to decay to final states containing
a $b$ and $\bar{b}$ quark pair and the bottomonium ($\bbbar$) meson spectrum.

\subsection{The $\mathbf{XYZ}$ mesons}
\noindent
The $XYZ$ mesons are an assortment of recently discovered resonance-like structures in hadronic final
states that contain either a $c$ and $\bar{c}$, or a $b$ and $\bar{b}$ quark pair, with properties
that do not match to expectations for any of the currently unassigned $\ccbar$ charmonium or $\bbbar$
bottomonium states.  In Fig.~\ref{fig:charmonium}, the charmoniumlike $XYZ$ mesons are indicated as red
and purple rectangles aligned according to my best guess at their $J^{PC}$ quantum numbers.  A reasonably
up-to-date list of the $XYZ$ candidate states, together with some of their essential properties, is provided
in Table~\ref{tab:Q-Qbar} and some recent reviews can be found in
Refs.~\cite{polosa, olsen_fop,nora_qwg,bodwin}.\footnote{In Table~\ref{tab:Q-Qbar} and the rest of this
report, the inclusion of charge conjugate states is always implied.}
The designation of these states as $X$, $Y$, or $Z$
was initially haphazard, but now has settled into a pattern in which researchers
engaged in this field (but not the Particle Data Group (PDG)~\cite{pdg}) designate $J^{PC}=1^{--}$
neutral states as $Y$, those with isospin=1 as $Z$, and all of the rest as $X$.  However, a few exceptions
to this pattern persist.

\subsection{A whirlwind tour}
\noindent
Moving from left to right in Fig.~\ref{fig:charmonium}, I review reasons that the $XYZ$ states
are poor matches for any of the unassigned charmonium states. (Experimental references are given
in Table~\ref{tab:Q-Qbar}.)

\vspace{0.2in}
\begin{table}[htb]
\caption{\footnotesize The $XYZ$ mesons. 
Masses $M$ and widths $\Gamma$ are
weighted averages with uncertainties added in quadrature.
Only $ \pipi\jpsi $ decays are used in the $X(3872)$ mass average.
Ellipses (...) indicate inclusive reactions.
Question marks indicate best guesses or no $J^{PC}$ information.
For charged states, $C$ refers to the neutral isospin partner.  
 } 
\setlength{\tabcolsep}{0.21pc}
\label{tab:Q-Qbar}
\begin{tabular}{lcccll}
\hline\hline
\rule[10pt]{-1mm}{0mm}
 State & $M$~(MeV) & $\Gamma$~(MeV) & $J^{PC}$ & Process~(decay mode) & 
     References \\ 
\hline
\rule[10pt]{-1mm}{0mm}
${X(3872)}$& 3871.68$\pm$0.17 & $<1.2$ &
    $1^{++}$
    & $B \to K + (J/\psi\, \pi^+\pi^-)$ &
    \cite{belle_x3872,belle_x3872-pipi,babar_x3872,lhcb_x3872-jpc}  \\ 
& & & & $p\bar p \to (J/\psi\, \pi^+\pi^-)+ ...$ &
    \cite{cdf_x3872,cdf_x3872-pipi,cdf_x3872-jpc,cdf_x3872-mass,d0_x3872}  \\ 
& & &   & $B \to K + (J/\psi\, \pi^+\pi^-\pi^0)$ &
    \cite{belle_x3872-wjpsi,babar_y3940}  \\ 
& & & & $B \to K + (D^0 \bar D^0 \pi^0)$ &
    \cite{belle_x3872-ddstr,babar_x3872-ddstr}  \\ 
& & & & $B \to K + (J/\psi\, \gamma)$ &
    \cite{babar_x3872-gjpsi,belle_x3872-gjpsi,lhcb_x3872-gjpsi} \\ 
& & & & $B \to K + (\psip \, \gamma)$ &
    \cite{babar_x3872-gjpsi,belle_x3872-gjpsi,lhcb_x3872-gjpsi} \\ 
& & & &  $pp \to (J/\psi\, \pi^+\pi^-)+ ...$   & 
    \cite{lhcb_x3872,cms_x3872}  \\
${X(3915)}$ & $3917.4\pm2.7$ & 28$^{+10}_{-~9}$ & $0^{++}$ &
    $B \to K + (J/\psi\, \omega)$ &
    \cite{belle_y3940,babar_y3940} \\ 
     & & & & $e^+e^- \to e^+e^- + (J/\psi\, \omega)$ &
    \cite{belle_2g_x3915,babar_jpc_x3915} \\ 
$X(3940)$ & $3942^{+9}_{-8}$ & $37^{+27}_{-17}$ & $0(?)^{-(?)+}$ &
     $e^+e^- \to J/\psi + (D^* \bar D)$ &
     \cite{belle_x4160} \\ 
&&&& $e^+e^- \to J/\psi + (...)$ &
     \cite{belle_x3940} \\ 
${G(3900)}$ & $3943\pm21$ & 52$\pm$11 & $1^{--}$ &
     $e^+e^- \to \gamma + (D \bar D)$ &
     \cite{babar_g3900,belle_g3900}
      \\ 
$Y(4008)$ & $4008^{+121}_{-\ 49}$ & 226$\pm$97 & $1^{--}$ &
     $e^+e^- \to \gamma + (J/\psi\, \pi^+\pi^-)$ &
     \cite{belle_y4260}
      \\
$Y(4140)$ & $4144\pm3$  & $17 \pm 9$ & $?^{?+}$ &
     $B \to K + (J/\psi\, \phi)$ &
     \cite{cdf_y4140-1,cdf_y4140-2,cms_y4140} \\ 
$X(4160)$ & $4156^{+29}_{-25} $ & $139^{+113}_{-65}$ & $0(?)^{-(?)+}$ &
     $e^+e^- \to J/\psi + (D^* \bar D)$ &
     \cite{belle_x4160} \\ 
${Y(4260)}$ & $4263^{+8}_{-9}$ & 95$\pm$14 & $1^{--}$ &
     $e^+e^- \to \gamma + (J/\psi\, \pi^+\pi^-)$ &
     \cite{babar_y4260,babar_y4260-1,cleo_y4260,belle_y4260}  \\ 
& & & & $e^+e^-\to (J/\psi\, \pi^+\pi^-)$ & \cite{cleoc_y4260}\\ 
& & & & $e^+e^-\to (J/\psi\, \pi^0\pi^0)$ & \cite{cleoc_y4260}\\ 
\rule[10pt]{-1mm}{0mm}
${Y(4360)}$ & $4361\pm13$ & 74$\pm$18 & $1^{--}$ &
     $e^+e^-\to\gamma + (\psip \, \pi^+\pi^-)$ &
     \cite{babar_y4360,belle_y4360} \\ 
$X(4630)$ & $4634^{+\ 9}_{-11}$ & $92^{+41}_{-32}$ & $1^{--}$ &
     $e^+e^-\to\gamma\, (\Lambda_c^+ \Lambda_c^-)$ &
     \cite{belle_x4630} \\ 
$Y(4660)$ & 4664$\pm$12 & 48$\pm$15 & $1^{--}$ &
     $e^+e^-\to\gamma + (\psip \, \pi^+\pi^-)$ &
     \cite{belle_y4360} \\ 
\hline
${Z_c^+(3900)}$ & $3890\pm 3$ & $33\pm 10$ & $1^{+-}$ &
     $Y(4260) \to \pi^- + (J/\psi\, \pi^+)$ &
     \cite{bes_z3900,belle_z3900}  \\ 
 &  &  &  &
     $Y(4260)\to \pi^- + (D\bar{D}^{*})^+$ &
     \cite{bes_z3885}  \\ 
${Z_c^+(4020)}$ & $4024\pm 2$ & $10\pm 3$ & $1(?)^{+(?)-}$ &
     $Y(4260) \to \pi^- + (h_c\, \pi^+)$ &
     \cite{bes_z4020}  \\ 
 &  &  &  &
     $Y(4260) \to \pi^- + (D^*\bar{D}^{*})^+$ &
     \cite{bes_z4025}  \\ 
${Z_c^0(4020)}$ & $4024\pm 4$ & $10\pm 3$ & $1(?)^{+(?)-}$ &
     $Y(4260) \to \pi^0 + (h_c\, \pi^0)$ &
     \cite{bes_z4020-0}  \\ 
$Z_1^+(4050)$ & $4051^{+24}_{-43}$ & $82^{+51}_{-55}$ & $?^{?+}$&
     $ B \to K + (\chi_{c1}\, \pi^+)$ &
     \cite{belle_z1z2,babar_z1z2} \\ 
$Z^+(4200)$ & $4196^{+35}_{-32}$ & $370^{+99}_{-149}$ & $1^{+-}$&
     $ B \to K + (\jpsi \, \pi^+)$ &
     \cite{belle_z4200} \\ 
$Z_2^+(4250)$ & $4248^{+185}_{-\ 45}$ &
     177$^{+321}_{-\ 72}$ &  $?^{?+}$ &
     $ B \to K + (\chi_{c1}\, \pi^+)$ &
     \cite{belle_z1z2,babar_z1z2} \\ 
$Z^+(4430)$ & $4477\pm 20$ & $181 \pm 31$ & $1^{+-}$&
     $B \to K + (\psip \, \pi^+)$ &
     \cite{belle_z4430,belle_z_dalitz1,belle_z_dalitz2,LHCb_z4430} \\ 
 &  &  & &
     $B \to K + (J\psi\, \pi^+)$ &
     \cite{belle_z4200}  \\ 
\hline\hline
$Y_b(10890)$ & 10888.4$\pm$3.0 & 30.7$^{+8.9}_{-7.7}$ & $1^{--}$ &
      $e^+e^- \to (\Upsilon(nS)\, \pi^+\pi^-)$ &
      \cite{belle_kfchen1} \\ 
\hline
$Z_{b}^+(10610)$ & 10607.2$\pm$2.0 & 18.4$\pm$2.4 & $1^{+-}$ &
       $\Upsilon(5S) \to \pi^- + (\Upsilon(1,2,3S)\,\pi^+)$&
      \cite{belle_zb,belle_zb-jpc}  \\ 
 &  &  &  &
      $\Upsilon(5S) \to \pi^- + (h_b(1,2P)\,\pi^+)$ & 
       \cite{belle_zb}  \\ 
 &  &  &  &
      $\Upsilon(5S) \to \pi^- + (B\bar{B}^*)^+$ & 
       \cite{belle_bbstr}  \\ 
$Z_{b}^0(10610)$ & 10609$\pm$ 6 &  & $1^{+-}$ &
       $\Upsilon(5S) \to \pi^0 + (\Upsilon(1,2,3S)\,\pi^0)$~~ &
      \cite{belle_zb-0}  \\ 
$Z_{b}^+(10650)$ & 10652.2$\pm$1.5 & 11.5$\pm$2.2 & $1^{+-}$ &
       $\Upsilon(5S) \to\pi^- + (\Upsilon(1,2,3S)\,\pi^+)$~ &
    \cite{belle_zb}  \\ 
 &  &  &  &
       $\Upsilon(5S) \to \pi^- + (h_b(1,2P)\,\pi^+)$ & 
       \cite{belle_zb} \\ 
 &  &  &  &
      $\Upsilon(5S) \to \pi^- + (B^*\bar{B}^*)^+$ & 
       \cite{belle_bbstr}  \\ 

\hline\hline
\end{tabular}
\end{table}

\clearpage

\begin{description}
\item[The $\mathbf{X(3940)}$ and $\mathbf{X(4160)}$] are seen in the invariant
mass distributions of the $D\bar{D}^*$ and $D^*\bar{D}^*$ systems that recoil from the $\jpsi$ in
$\ee\rt\jpsi D^{(*)}\bar{D}^*$ reactions at $E_{\rm cm}\simeq 10.6$~GeV, respectively.  The
only known charmonium states that are seen recoiling from a $\jpsi$ in these processes, the $\eta_c$,
$\chi_{c0}$ and $\eta_{c}(2S)$, all have spin=0~\cite{belle_ee2jpsiccbar,babar_ee2jpsiccbar}.  This,
plus the fact that neither the $X(3940)$ nor the $X(4160)$ is seen to decay to
$D\bar{D}$~\cite{belle_x4160}, provides circumstantial evidence for $J^{PC}=0^{-+}$. The unassigned
$0^{-+}$ charmonium states are the $\eta_c(3S)$ and $\eta_c(4S)$, which are expected to have masses around
$4010$ and $4390$~MeV, respectively~\cite{barnes}.  The $X(3940)=\eta_c(3S)$ and $X(4160)=\eta_c(4S)$
assignments would imply anomalously large $\psi (nS)$-$\eta_c (nS)$ mass splittings for $n=3$ of
$\sim 120$~MeV, and $n=4$ of $\sim 260$~MeV, which are both huge compared to theoretical expectations
of $\sim 30$ and $\sim 25$~MeV, respectively~\cite{barnes,li-chao_2009}.

\item[The $\mathbf{Y(4260)}$ and $\mathbf{Y(4360)}$] were first seen in the $\pipi\jp$ and $\pipi\psip$ 
mass distributions, respectively, in the initital-state-radiation (isr) processes
$\ee\rt\gamma_{\rm isr}\pipi\jpsi (\psip)$.  This production mechanism ensures that $J^{PC}$
quantum numbers for these state are $1^{--}$.  All of the $1^{--}$ charmonium states with masses below
4.5~GeV have already been established~\cite{huamin_R}; there are no available slots for either
the $Y(4260)$ or the $Y(4360)$.

\item[The $\mathbf{Z_c(3900)}$, $\mathbf{Z_c(4020)}$, $\mathbf{Z_c(4200)}$, $\mathbf{Z(4430)}$, 
$\mathbf{Z_1(4050)}$ and $\mathbf{Z_2(4250)}$] are seen in $\pip \jpsi$, $\pip h_c$, $\pip\psip$
or $\pip\chi_{c1}$ invariant mass specta and, thus, have a non-zero electric charge. Charmonium
states, by definition, are $\ccbar$ states with zero charge (and isospin). The first four have
$J^{PC}=1^{++}$ quantum numbers,\footnote{This includes an ``informed guess'' for the $J^P$ of the
$Z_c(4020)$ that is discussed below in Section~\ref{sect:zc4020}.}  where here, and in the rest
of this report, $C$ refers to the $C$-parity of the neutral member of the isospin triplet.
Although the $Z_1$ and $Z_2$ necessarily have even $C$-parity, they could have any $J^P$ value
other than $0^{-}$, which is forbidden by parity.   The BESIII and Belle experimental signals for
the charged $Z$ states were discussed at this meeting by Gradl~\cite{gradl} and Lange~\cite{lange}.

\noindent
\item[The $\mathbf{X(3915)}$] is seen as $\omega\jpsi$ invariant mass peaks in $B\rt K\omega\jpsi$
decays and in $\gamma\gamma\rt \omega\jpsi$ two-photon fusion reactions. A BaBar study of the latter
process concluded that $J^{PC}=0^{++}$~\cite{babar_jpc_x3915}.  BaBar (and the PDG) identify this state
as the $\chi_{c0}(2P)$, the first radial excitation of the $\chi_{c0}$ charmonium state. This assignment
has some serious problems: the mass is too high; the total width is too narrow; decays to $D\bar{D}$
final states, which should be the dominant decay mode for the $\chi_{c0}(2P)$, are not seen; and the
production rates in $B$ decays and $\gamma\gamma$ fusion are incompatible with a $\chi_{c0}(2P)$
assignment~\cite{guo,slo}. 

\item[The $\mathbf{X(3872)}$] was the first $XYZ$ meson to be discovered and is the most
well studied. I discuss its properties in some detail in the following section. 

\end{description}

\section{The $\mathbf{X(3872)}$}
\noindent
The $X(3872)$ was first seen by Belle as a narrow peak in the $\pipi\jp$ invariant mass distribution
in exclusive $B\rt K\pipi\jp$ decays~\cite{belle_x3872} (see Fig~\ref{fig:x3872}a). It is a well
established state that has been seen by (at least) six other
experiments~\cite{babar_x3872,cdf_x3872,d0_x3872,lhcb_x3872,cms_x3872,bes3_y4260_2_gx3872}.

\subsection{Properties of the $\mathbf{X(3872)}$}
\begin{figure}[htb]
\begin{minipage}[t]{48mm}
  \includegraphics[height=0.7\textwidth,width=0.9\textwidth]{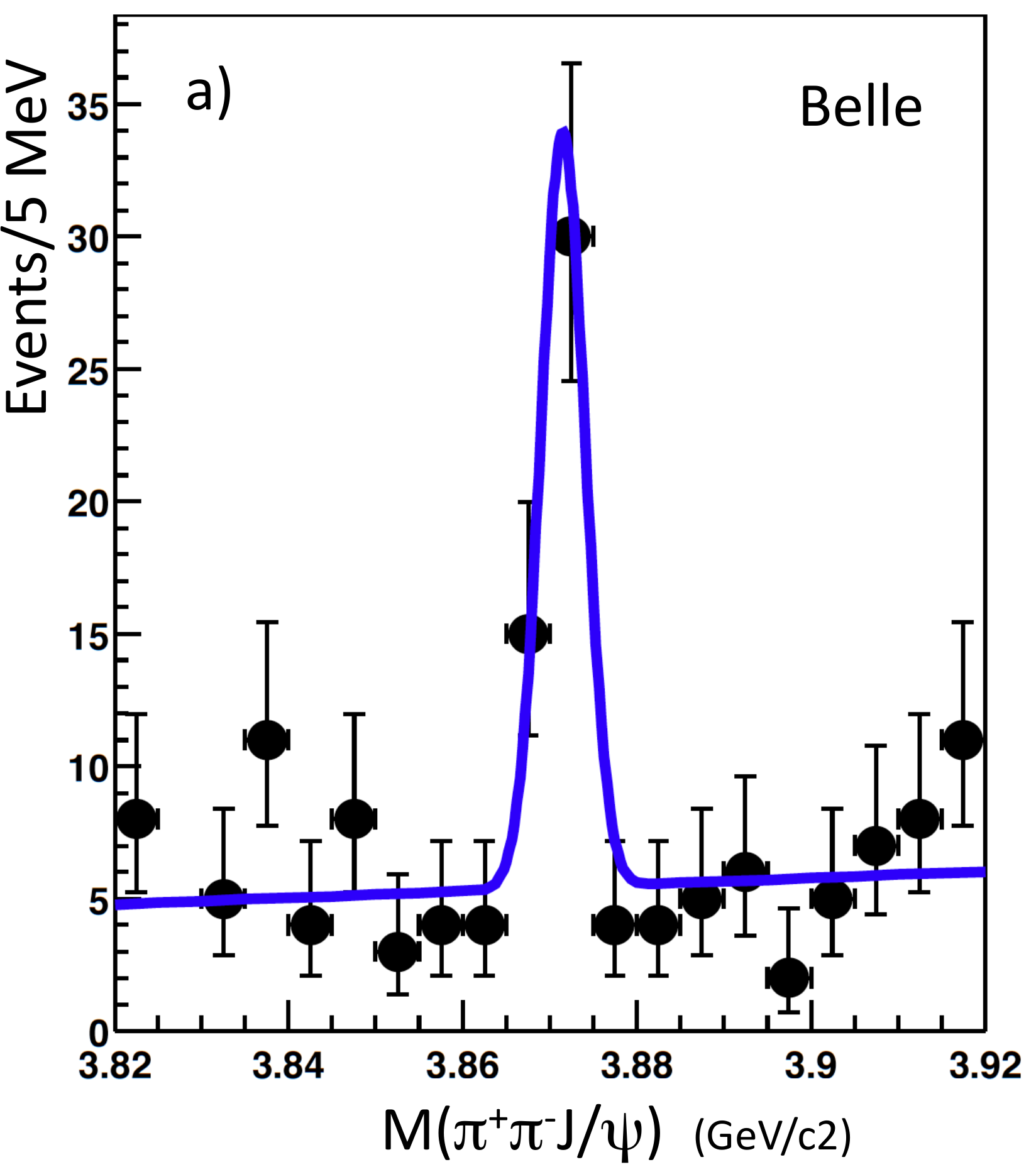}
\end{minipage}
\begin{minipage}[t]{48mm}
  \includegraphics[height=0.7\textwidth,width=0.9\textwidth]{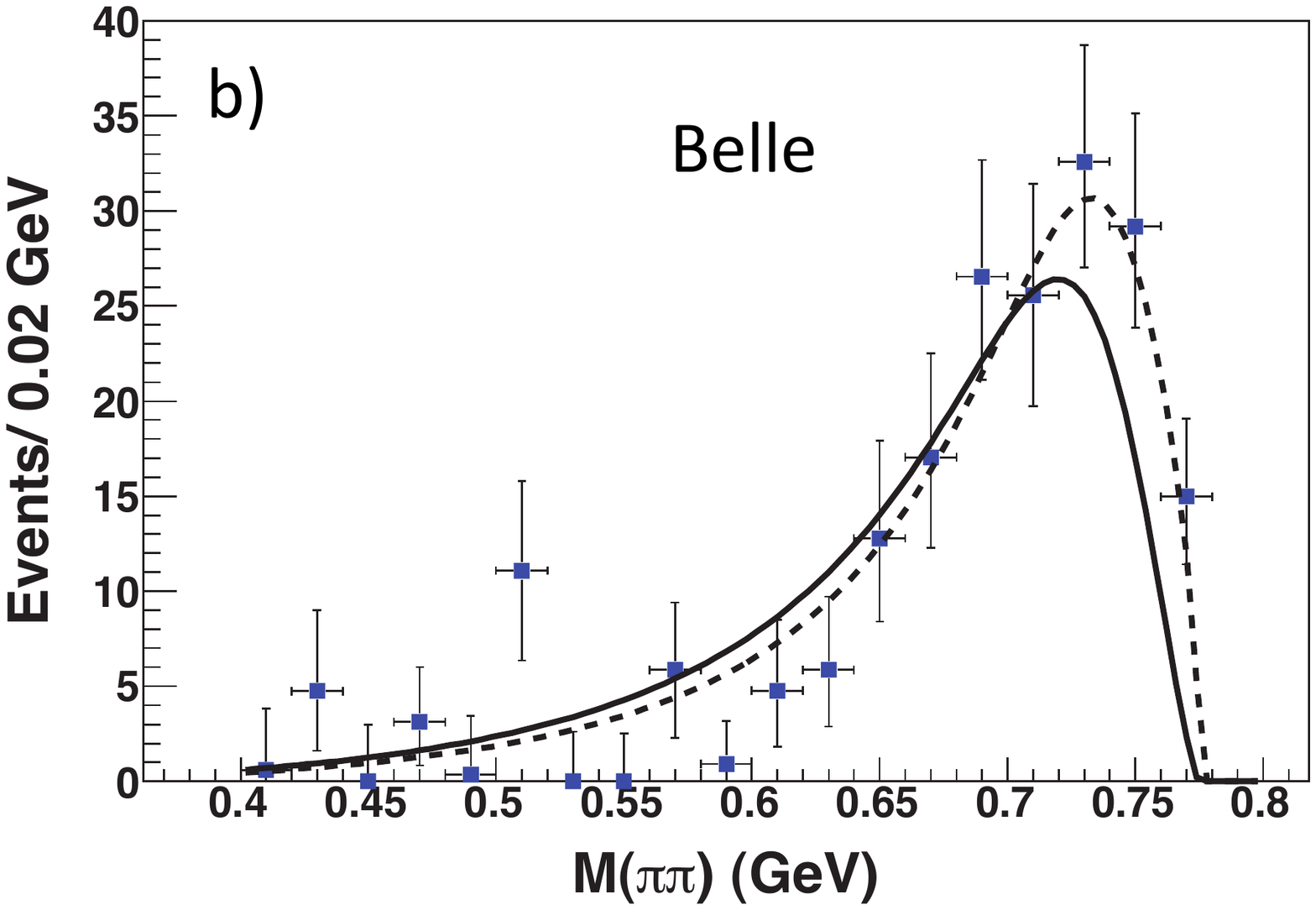}
\end{minipage}
\begin{minipage}[t]{48mm}
  \includegraphics[height=0.7\textwidth,width=0.9\textwidth]{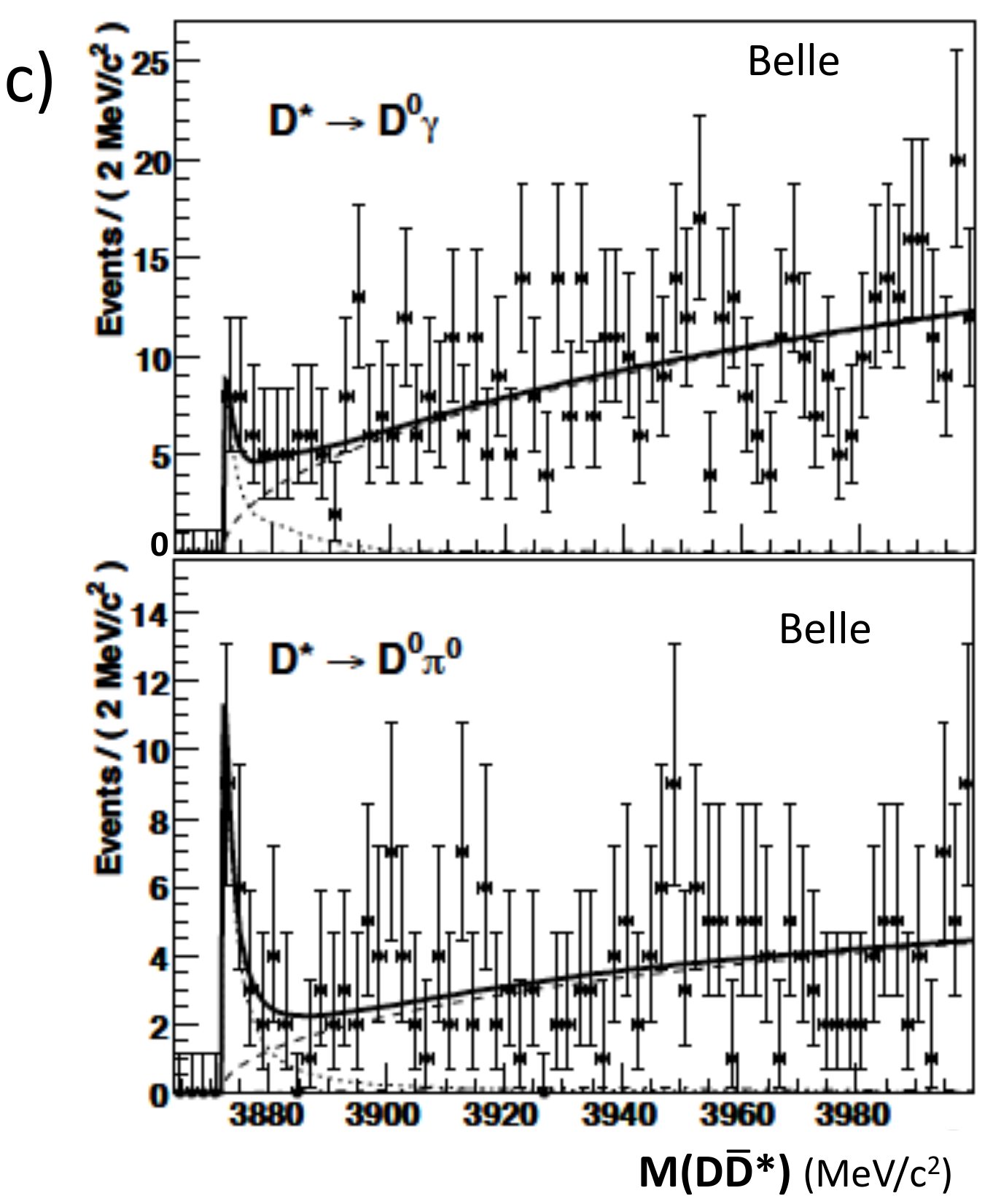}
\end{minipage}
\hspace{\fill}
\caption{\footnotesize {\bf a)} The $M(\pipi\jpsi)$ distribution for
$B\rt K\pipi\jpsi$ events from Belle's original $X(3872)$ paper~\cite{belle_x3872}. 
{\bf b)}
The $M(\pipi)$ distribution for $X(3872)\rt\pipi\jp$ events from Belle~\cite{belle_x3872-pipi}.
The curves shows results of fits to a $\rho\rt\pipi$ line shape including $\rho$-$\omega$
interference. The dashed (solid) curve is for even (odd) $X(3872)$ parity. 
{\bf c)}
$M(D^0\bar{D}^{*0})$ distributions for $B\rt K D^0\bar{D}^{*0}$ decays from Belle~\cite{belle_x3872-ddstr}.
The upper plot is for $\bar{D}^{*0}\rt \bar{D}^0\gamma$ decays, the lower plot is for
$\bar{D}^{*0}\rt \bar{D}^0\piz$ decays.  The peaks near threshold are attributed to 
$X(3872)\rt D^0\bar{D}^{*0}$ decays.
}
\label{fig:x3872}
\end{figure}

\noindent
The most striking feature of the $X(3872)$ is the virtual indistinguishability between its measured
mass, $M_{X(3872)}=3871.69\pm 0.17$~MeV~\cite{pdg}, and the sum  of the $D^0$ and $D^{*0}$ masses,
$m_{D^{0}} + m_{D^{*0}}=3871.69\pm0.09$~MeV~\cite{seth_mddstr}.  Also striking is its narrow total 
width, $\Gamma_{X(3872)}<1.2$~MeV (90\% CL)~\cite{belle_x3872-pipi}.  The $X(3872)\rt\gamma\jpsi$ 
decay mode has been seen by BaBar~\cite{babar_x3872-gjpsi}, Belle~\cite{belle_x3872-gjpsi},
and LHCb~\cite{lhcb_x3872-gjpsi} with a branching fraction that is $0.24\pm 0.05$ that for
$X(3872)\rt\pipi\jpsi$.  BaBar and LHCb group report signals for $X(3872)\rt\gamma\psip$ with
a branching fraction that is $2.7\pm 0.6$ times that for $X(3872)\rt\gamma\jpsi$. The $M(\pipi)$
distribution for $X(3872)\rt\pipi\jp$, shown in Fig.~\ref{fig:x3872}b, is consistent with
expectations for $\rho\rt\pipi$ decays~\cite{belle_x3872-pipi,cdf_x3872-pipi}.  The
$X(3872)\rt\omega\jpsi$ decay mode has been seen with a branching fraction that is similar to
that for $\rho\jpsi$~\cite{belle_x3872-wjpsi,babar_y3940} even though the decay phase space
only covers a small fraction of the $\omega$  resonance's low mass tail. These observations
clearly establish that the $C$-parity of the $X(3872)$ is $C=+1$ and that isospin is strongly
violated in its decays.  A CDF study of $X(3872)\rt\pipi\jpsi$ decays limited the $J^{PC}$ quantum
numbers to be either $1^{++}$ or $2^{-+}$~\cite{cdf_x3872-jpc}; a subsequent LHCb comparison of these
two possibilities using $X(3872)\rt\pipi\jpsi$ events produced via $B$ meson decay unambiguously
favored $J^{PC}=1^{++}$~\cite{lhcb_x3872-jpc}.  The $X(3872)$ has a significant coupling to
$D^0\bar{D}^{*0}$ that is seen by both Belle~\cite{belle_x3872-ddstr} and Babar~\cite{babar_x3872-ddstr}
as a pronounced threshold enhancement in the $M(D^0\bar{D}^{*0})$ distribution in $B\rt K D^0\bar{D}^{*0}$
decays; the Belle results are shown in Fig.~\ref{fig:x3872}c. The $X(3872)$ branching fraction to
$D^0\bar{D}^{*0}$ is $9.9\pm 3.2$ times larger than that for $X(3872)\rt\pipi\jpsi$.

\subsection{Prompt $\mathbf{X(3872)}$ production in high energy $\mathbf{\pbar (p)}$-$\mathbf{p}$ collisons}
\begin{figure}[htb]
\begin{minipage}[t]{70mm}
  \includegraphics[height=0.7\textwidth,width=0.8\textwidth]{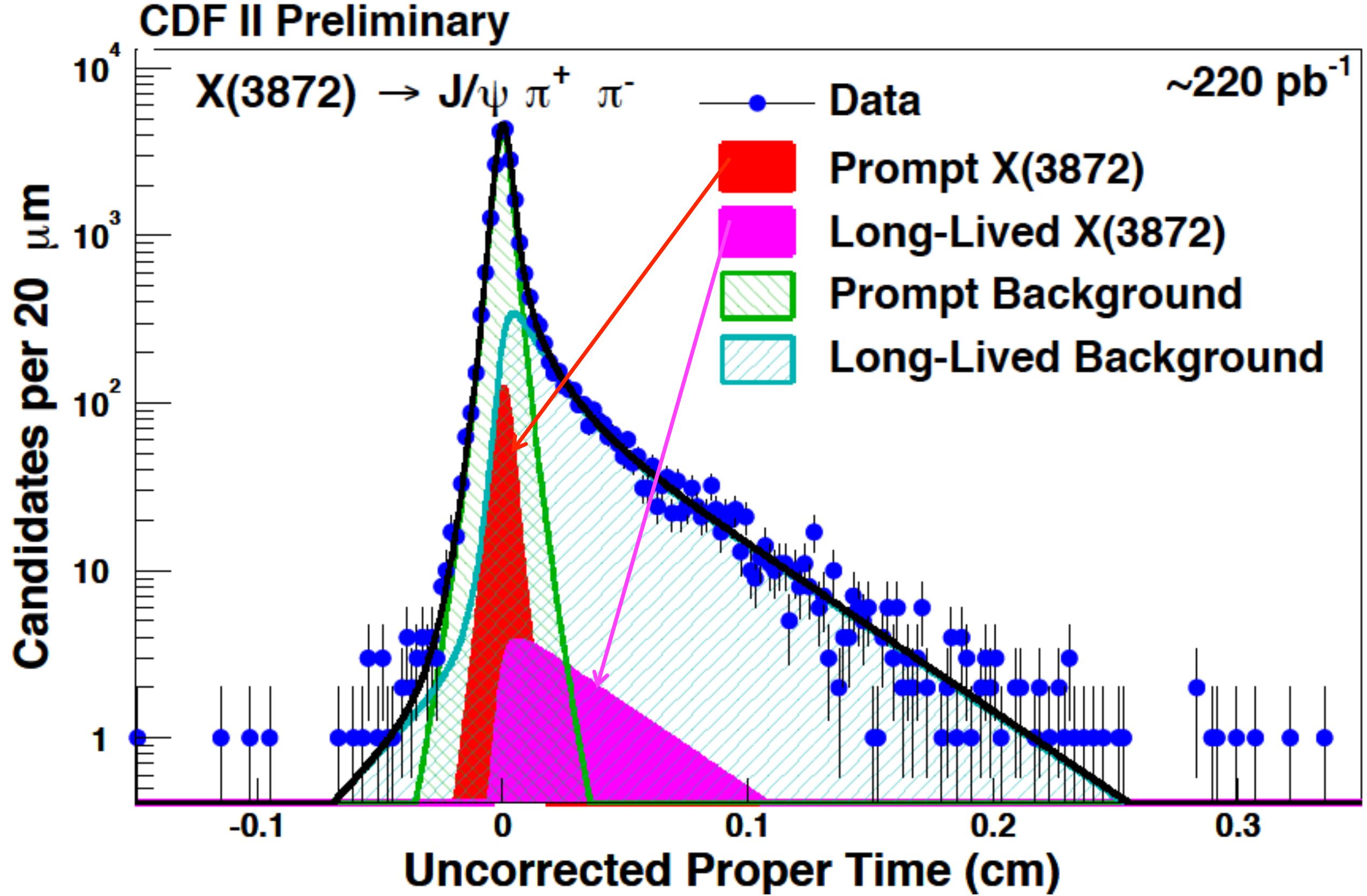}
\caption{\footnotesize{The uncorrected proper time distribution for $X(3872)$ production from
CDF~\cite{cdf_x3872-ctau}}}
\label{fig:cdf_x3872-prod}
\end{minipage}
\hspace{\fill}
\begin{minipage}[t]{70mm}
  \includegraphics[height=0.7\textwidth,width=0.8\textwidth]{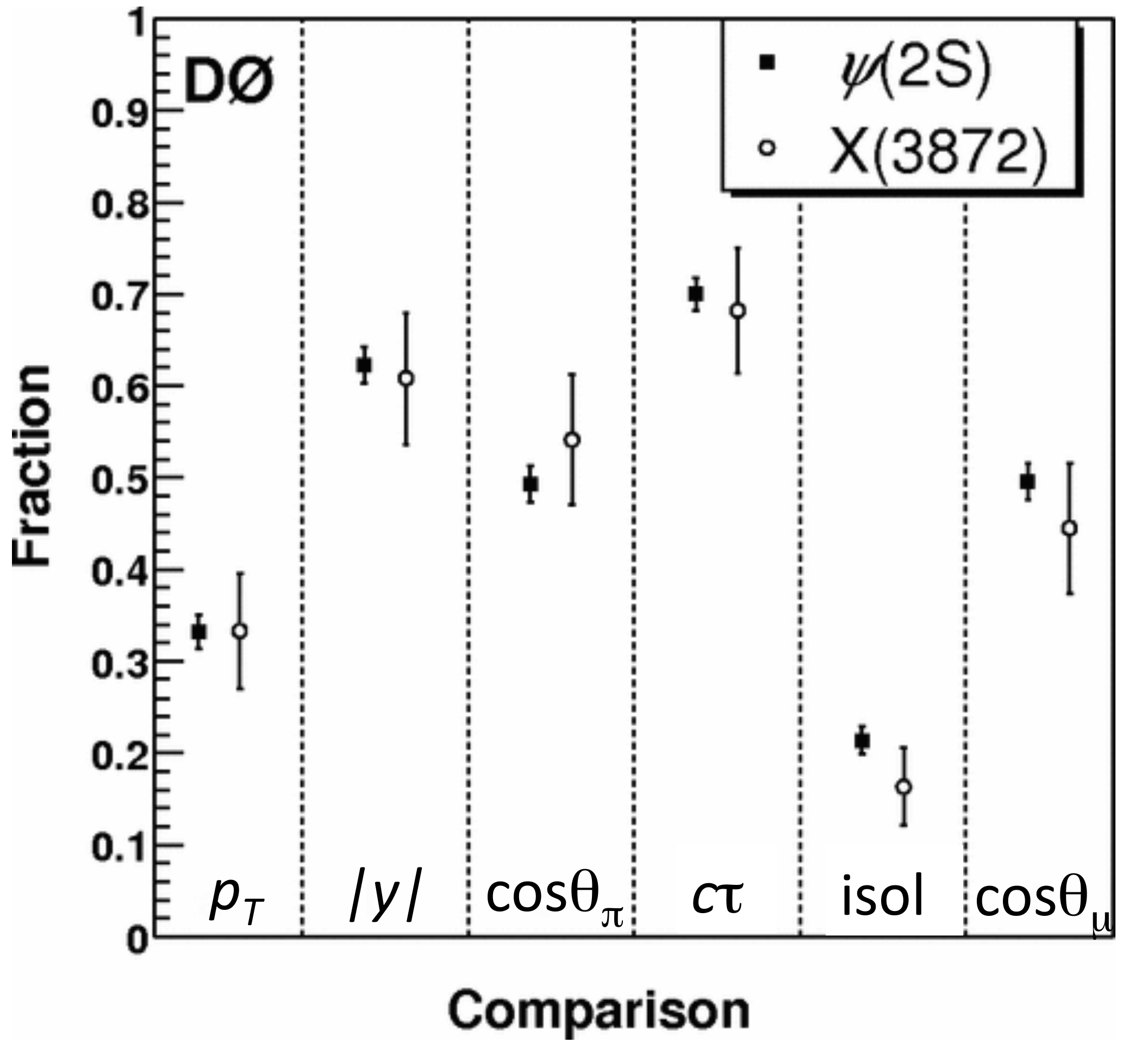}
\caption{\footnotesize{Comparison of $X(3872)$ and $\psip$ event-yield fractions for different
measured quantities~\cite{d0_x3872}. (See text for details.)}}
\label{fig:d0_x3872-prod}
\end{minipage}
\end{figure}

\noindent
The $X(3872)\rt\pipi\jpsi$ signals seen in 1.96~TeV $\ppbar$~\cite{cdf_x3872} and 7~TeV
$pp$~\cite{cms_x3872}) collisions are $7\sim 10\%$ those for $\psip\rt\pipi\jpsi$. 
Figure~\ref{fig:cdf_x3872-prod} shows CDF results for the proper-time dependence of $X(3872)$
production in inclusive $p\bar{p}\rt\pipi\jpsi + X$ annihilations at $E_{\rm cm}=1.96$~TeV~\cite{cdf_x3872-ctau}.
They find that only a small fraction of the $X(3872)$ signal, shown in magenta, has a displaced vertex
distribution that is characteristic of the $B$ meson lifetime;  ($84\pm 5$)\% of the $X(3872)$ signal,
shown in red, is produced promptly.  A similar study of the $\psip\rt\pipi\jpsi$ signal found ($72\pm 1)$\%
of the $\psip$ signal is produced promptly. Results from a $D0$ comparison of the properties of $X(3872)$ and
$\psip$ production at the same energy are shown Fig.~\ref{fig:d0_x3872-prod}b~\cite{d0_x3872}.  Here the
open circles show the fractions of the $X(3872)$ signal that have: transverse momentum $p_T > 10$~GeV/$c$;
pseudorapidity in the range $|y|<1$; pion (muon) helicity angle in the region $\cos\theta_{\pi (\mu)}<0.4$;
and isolation<1, where isolation is the ratio of $X(3872)$ momentum to the summed momenta of all
other charged tracks within $\Delta R = 0.5$ of the $X(3872)$ direction
($\Delta R\equiv \sqrt{\Delta y)^2+(\Delta \phi)^2}$).  These fractions agree quite well with
the corresponding quantities for $\psip$ production, which are shown in the figure as solid squares. 
 
\subsection{Charmonium assignment for the $\mathbf{X(3872)}$?}
\noindent
The only unassigned conventional $1^{++}$ charmonium state that is expected to have a mass that is 
anywhere near 3872~MeV is the $\chi_{c1}(2P)$ (commonly called the $\chi_{c1}^{\prime}$), {\it i.e.}, the first radial
excitation of the $\chi_{c1}$.  This assignment for the $X(3872)$ has some problems.  The
$\chi^{\prime}_{c2}$, the $J=2$ spin-multiplet partner of the the $\chi^{\prime}_{c1}$, is well established
with a measured mass of $3927.2\pm 2.6$~MeV~\cite{pdg}.   An $X(3872)=\chi^{\prime}_{c1}$ assignment
would imply a $\chi_{c2}(2P)$-$\chi_{c1}(2P)$ mass splitting of $\Delta M_{2-1}(2P)=55.5\pm 2.6$~MeV,
which is larger than the ``ground-state'' splitting $\Delta M_{2-1}(1P)=45.5 \pm 0.1$~MeV.  This
behavior is contrary to potential model expectations, where these splittings are due to tensor
and spin-orbit forces that decrease with increasing radius (and, therefore, radial quantum
number)~\cite{gi}.  For states above open-charmed threshold,  like the $\chi^{\prime}_{c2}$
and, depending on its mass, the $\chi^{\prime}_{c1}$, potential model predictions are modified by
couplings to on-mass-shell, open-charmed meson pair configurations.  However, three different
methods for computing these effects all find that they tend to suppress the $\chi^{\prime}_{c2}$ mass
while increasing that of the $\chi^{\prime}_{c1}$ and, thereby, reducing (not increasing) this splitting
to values that are below potential-model~based expectations~\cite{li-chao_2009,elq_2004,wang_2014}.
A second problem with the $X(3872)=\chi^{\prime}_{c1}$ assignment is that the measured upper limit on
its natural width ($\Gamma_{X(3872)}<1.2$~MeV) is only slightly above the natural width of the
``ground-state'' $\chi_{c1}$: $\Gamma_{\chi_{c1}} = 0.84\pm 0.04$~MeV.  Since the $\chi_{c1}^{\prime}$
could access any of the $\chi_{c1}$ decay channels with significantly increased phase-space and
have a number of additional decay channels, including decays to open-charmed mesons and hadronic \&
radiative transitions to $1P$ \& $1S$ charmonium states, it is expected that its natural width would
be substantially broader than that of the $\chi_{c1}$.  (All of the identifed $X(3872)$ decay channels,
which account for at least one third of its total decay width, are to final states that are kinematically
inaccessible to the $\chi_{c1}$, namely $D^0\bar{D}^{0}\pi^{0}(\gamma )$, $\rho\jp$, $\omega\jp$ and
$\gamma\psip$~\cite{pdg}.)  A third problem with the $X(3872)=\chi^{\prime}_{c1}$ assignment is that the
decay $\chi^{\prime}_{c1}\rt\rho\jp$ violates isospin and is, therefore, expected to be strongly suppressed
and unlikely to be a ``discovery channel'' for the $\chi_{c1}^{\prime}$.

\subsection{If not charmonium, then what?}
\noindent
For these reasons, the $X(3872)$ is expected to have a more complex sub-structure than the simple
$\ccbar$  configuration that is expected for the $\chi^{\prime}_{c1}$ in charmonium potential models.

The near coincidence of its mass with the $D^0\bar{D}^{*0}$ mass threshold has led to considerable
speculation that it is predominantly a molecule-like
$X(3872)=(D^{*0}\bar{D}^{0} + {D}^{0}\bar{D}^{*0})/\sqrt{2}$ configuration in which the (color-singlet)
$D$ and $\bar{D}^*$ mesons are loosely bound by Yukawa-like nuclear forces~\cite{molecule} (see
Fig.~\ref{fig:diquarks}d), an idea that has been around for some time~\cite{molecule_pre1994}. 

Other authors have interpreted the $X(3872)$ as a nearly point-like, tetraquark combination consisting
of an anticolored diquark and a colored diantiquark in an $S$-wave and tightly bound by the QCD color
force~\cite{maiani1,nielsen} (see Fig.~\ref{fig:diquarks}c).

Problems with the two above-mentioned pictures have inspired a number of charmonium-molecular hybrid
models,\footnote{This ``hybrid'' is not the same as the QCD hybrid shown in Fig.~\ref{fig:diquarks}d.}
in which the $X(3872)$ is a quantum mechanical mixture of $\ccbar$, $D^0\bar{D}^{*0}$ and $D^+ D^{*-}$
components~\cite{hybrids,takeuchi,rupp}, where the $\ccbar$ component is (mostly) the $\chi_{c1}^{\prime}$.  
Hadronic production and radiative transitions to the $\psip$ and $\jpsi$ are hypothesized to proceed 
ia the $\ccbar$ component and this could explain why $X(3872)$ production properties are similar to
those of the $\psip$ and its decay width to $\gamma\psip$ is larger than that for $\gamma\jpsi$
(because of the closer overlap of the $\chi^{\prime}_{c1}$ and $\psip$
radial wavefunctions).

Friedmann eschews potential model ideas for meson spectra entirely, including
the notion of radial excited states and manages to reproduce the entire spectrum of measured meson and
baryon states, including the XYZ meson candidates, with a uniform picture that is based only on quarks
and diquarks~\cite{friedmann}.

In the following I compare molecule, QCD-tetraquark and ``hybrid'' models
to measurements. (I have no comments on Friedmann's unified model because no phenomenological
consequences are currently available.)

\subsubsection{A molecule?}
\noindent
The spatial extent of a $D^0\bar{D}^{*0}$ ``molecule'' with the $X(3872)$ mass would be characterized by its
scattering length $a_{0}=\hbar/\sqrt{\mu \delta E_{0}}$~\cite{braaten_swave}, where $\mu = 970$~MeV is the
$D\bar{D}^{*}$ reduced mass and $\delta E_{0}\equiv |M_{X(3872)}-(m_{D^0}+m_{D^{*0}})|=0.003\pm 0.192$~MeV~\cite{seth_mddstr}.
The close proximity of the $X(3872)$ mass to the $m_{D^0}+m_{D^{*0}}$ threshold implies a characteristic
size of the $D^0\bar{D}^{*0}$ system of $a_{0}>10$~fm, {\it i.e.}, more than ten times the rms radius of the
$\psip$,  $\langle r_{\psip}\rangle \sim 0.8$~fm~\cite{li-chao_2009,eiglsperger}.  In contrast, 
$\delta E_{\pm} \equiv |M_{X(3872)}-(m_{D^+}+m_{D^{*-}})| = 8.2\pm 0.2$~MeV~\cite{pdg}, and $a_{\pm}\sim 2$~fm.  This
$D^0\bar{D}^{*0}$-$D^+ D^{*-}$ difference easily accounts for the strong isospin violations in $X(3872)$
decays.  On the other hand, it is hard to imagine that such a large, weakly bound system would be
produced in ultra-high energy $\pbar p$ collisions with a cross section and production properties
that so closely match those of the compact and tightly bound $\psip$ charmonium state.  In fact,
a detailed examination~\cite{bignamini} confirms this intuitive expectation and shows that a
$D\bar{D}^{*}$ molecule-like structure could not be promptly produced in high energy hadron collisons
with characteristics that are in any way similar to those for the $\psip$.

Another problem with a purely molecular picture for the $X(3872)$ is the above-mentioned result 
${\mathcal B}(X(3872)\rt\gamma\psip)=(2.5\pm 0.7)\times{\mathcal B}(X(3872)\rt\gamma\jpsi)$.
In specific molecule models, $\gamma\psip$ decays are suppressed relative to $\gamma\jpsi$ by more
than two orders of magnitude~\cite{gpsip_molecule}.

\begin{figure}[htb]
\begin{minipage}[t]{47mm}
  \includegraphics[height=0.9\textwidth,width=1.0\textwidth]{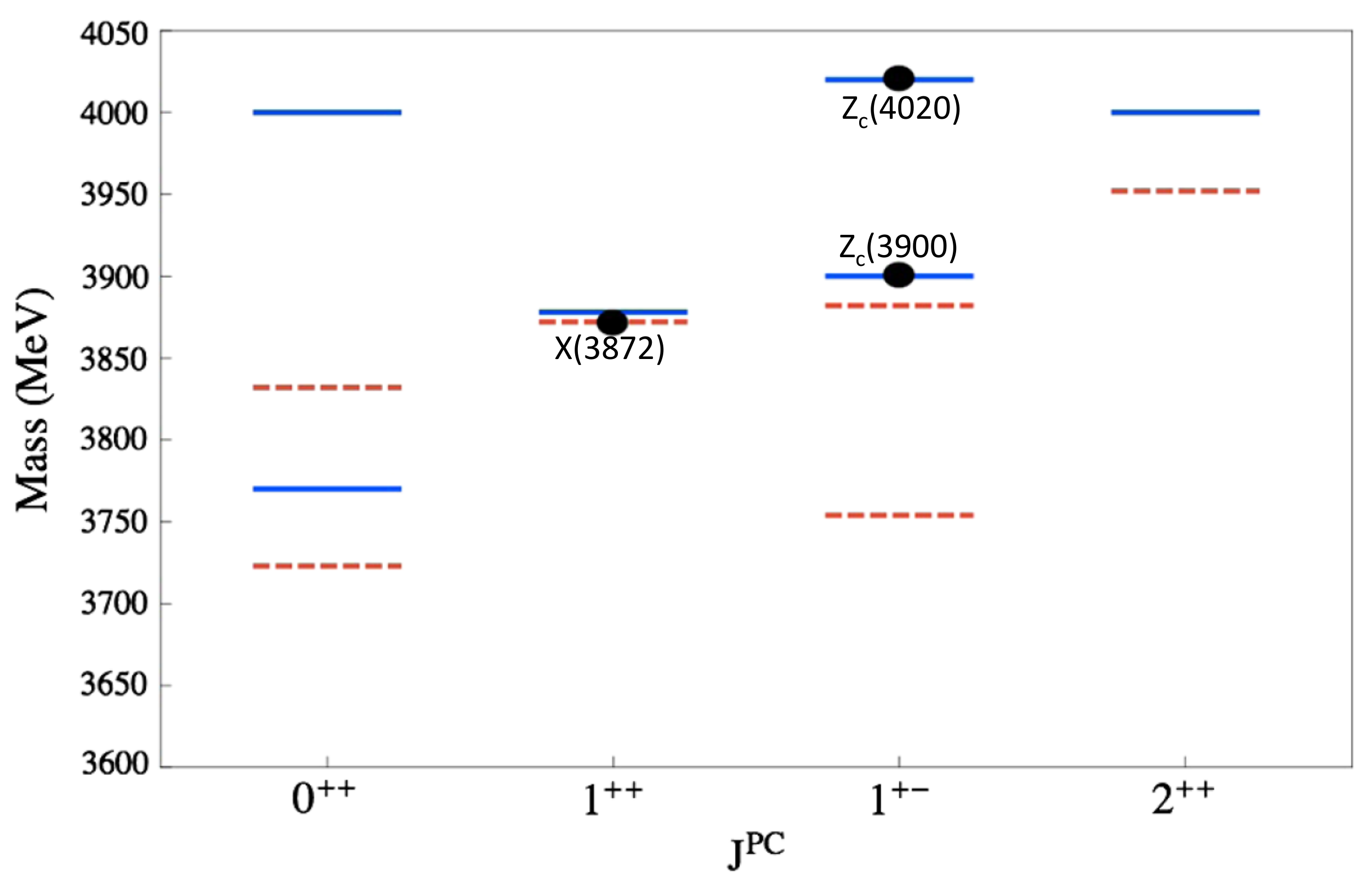}
\caption{\footnotesize{Predicted diquark-diantiquark $S$-wave states 
from Ref.~\cite{maiani3} (blue lines), with black dots indicating the levels assigned to the
$X(3872)$, $Z_c(3900)$  and $Z_c(4020)$.  Red dashes show an earlier version of the model~\cite{maiani1}.  
}}
\label{fig:QCD-4quark-spectrum}
\end{minipage}
\hfill
\begin{minipage}[t]{94mm}
  \includegraphics[height=0.5\textwidth,width=1.1\textwidth]{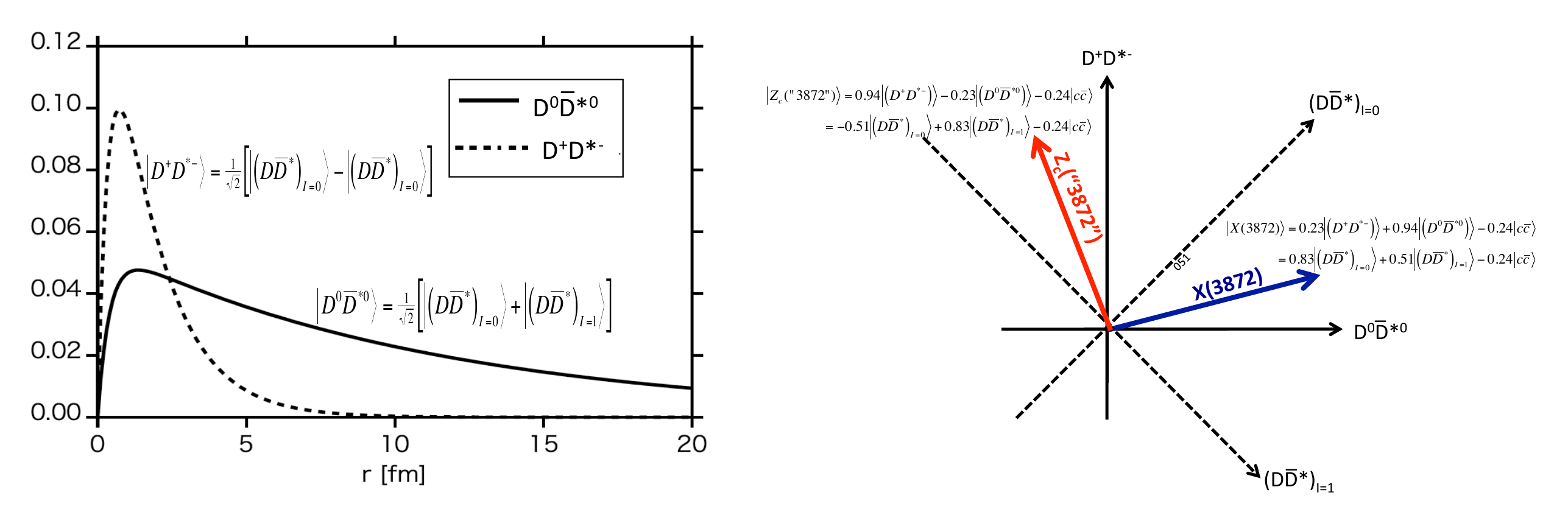}
\caption{\footnotesize{{\bf Left} 
The radial wave functions for the $D^0\bar{D}^{*0}$ (solid curve) and $D^+ D^{*-}$ (dashed curve)
components of the $X(3872)$ in the charmonium-molecule hybrid model of Ref.~\cite{takeuchi}.
{\bf Right} The blue arrow indicates the isospin comonents of the $X(3872)$ state vector given
in Ref.~\cite{takeuchi}. The red arrow is the orthogonal (mostly isovector) counterpart of the
$X(3872)$ (assuming equal $\ccbar$ components).  
}}
\label{fig:x3872-ispin}
\end{minipage}
\end{figure}

\subsubsection{A QCD tetraquark?}
\noindent
In the diquark-diantiquark picture, a charmoniumlike tetraquark has a $cq_i\cbar\qbar_j$ configuration,
where $q_{1(2)}=u(d)$. For the $X(3872)$, $i=j$ and two configurations are expected:
either $cu\cbar\ubar$ and $cd\cbar\dbar$ or linear combinations of the two~\cite{maiani1,maiani3}. 
In addition to two neutral states, two charged states, where $i\neq j$ are also expected. Searches for 
nearby neutral~\cite{cdf_x3872-mass} and charged~\cite{belle_x3872-pipi,babar_x3872-pipi0} partners
of the $X(3872)$ have come up empty.  This model predicts the existence of $S$-wave diquark-diantiquark 
states with $J^{PC} = 0^{++}, 1^{+-}$~and~$2^{++}$, as indicated in Fig.~\ref{fig:QCD-4quark-spectrum}.  The
recently discovered $Z_c(3900)$ and $Z_c(4020)$ are identified as the expected $1^{+-}$ states, although
the initial version of the model, shown as dashed red lines, specifically predicted that the $Z_c(4020)$
mass would be lower, and not higher, than $M_{Z_c(3900)}$~\cite{maiani2}.   Many predicted states remain
unseen, including a $0^{++}$ state with a mass that is close to the $D\bar{D}$ open-charmed threshold,
which suggests that it might be narrow and relatively easy to see.  Note that all of  the indicated levels
correspond to isospin triplets, so, if this model is correct, lots of additional states remain to be found.

\subsubsection{A $\mathbf{\ccbar}$-$\mathbf{ D\bar{D}}$  ``hybrid?''}
\label{sect:hybrid}
\noindent
The $\ccbar$-$D\bar{D}^*$ hybrid model accommodates the measured properties of the $X(3872)$, including
large isospin violations, production properties in high energy $\ppbar$ collisions, and the relatively
large $\gamma\psip$ decay width. In a specific version of this model, the authors of Ref.~\cite{takeuchi}
introduce both a mutual interaction between the $D$ and $\bar{D}^*$ and a coupling between the $\ccbar$
``core'' and $D\bar{D}^*$ systems. This results in a $X(3872)$ state vector of the form:
\begin{equation}
|X(3872)\rangle = \alpha_{0}|D^0\bar{D}^{*0}\rangle + \alpha_{\pm}|D^+D^{*-}\rangle + \alpha_{\rm core}|\ccbar\rangle.
\label{eqn:x3872-wavefcn}
\end{equation}
Because of the disparate mass differences between the $X(3872)$ and the $D^0\bar{D}^{*0}$ and $D^+D^{*-}$
thresholds, the amplitudes and wave functions of these two components are quite different, as shown for a
specific example in the left-hand panel of Fig.~\ref{fig:x3872-ispin}. This means $\alpha_{0}\neq\alpha_{\pm}$
and, thus, an $X(3872)$ state with mixed isospin:
\begin{equation}
|X(3872)\rangle = \frac{\alpha_{0}+\alpha_{\pm}}{\sqrt{2}}|(D\bar{D}^{*})_{I=0}\rangle
+ \frac{\alpha_{0}-\alpha_{\pm}}{\sqrt{2}}|(D\bar{D}^{*})_{I=1}\rangle| + \alpha_{\rm core}\ccbar\rangle .
\label{eqn:x3872-wavefcn-ispin}
\end{equation}
In the Ref.~\cite{takeuchi} calculation, $\alpha_{0}=0.94$, $\alpha_{\pm}=0.23$ and $\alpha_{\rm core}= -0.24$,
corresponding to probabilities of 68\% for $I=0$, 25\% for $I=1$ and 6\% for the $\ccbar$ core.  An
interesting feature of the Ref.~\cite{takeuchi} calculation is that the bulk of the attraction between
the $D$ and $\bar{D}^*$ mesons in the $X(3872)$ comes from the $\ccbar$-$D\bar{D}^*$ coupling.  The
mutual $D$-$\bar{D}^*$ attraction, which is the dominant term in pure molecular models, only plays a minor
role.  Similar conclusions are reported in Ref.~\cite{rupp}.    

The state vector given in Eq. (\ref{eqn:x3872-wavefcn-ispin}) has two related orthogonal counterparts.
Reference~\cite{takeuchi} discusses one that is mostly $\ccbar$ and probably should be
considered to be the physical manifestation of the $\chi_{c1}^{\prime}$ charmonium state. This is found
to have a mass that is well above both the $D^0\bar{D}^{*0}$ and $D^+D^{*-}$ thresholds and wide.  As a
result, it may not be experimentally easy to identify.   The third state would be predominantly
a $D\bar{D}^*$ isovector with a large $D^+D^{*-}$ component, as illustrated in the right-hand panel of 
Fig.~\ref{fig:x3872-ispin}.\footnote{This state would be distinct from the $Z_c(3900)$ because of its
opposite $C$-/$G$-parity.}  If this is a physical particle, it might be accessible in $B\rt KD\bar{D}^*$
decays.  The near-threshold $M(D^0\bar{D}^{*0})$ distributions for $B\rt KD^0\bar{D}^{*0}$ published
by Belle~\cite{belle_x3872-ddstr} and Babar~\cite{babar_x3872-ddstr} have limited statistics and are
inconclusive (see, {\it e.g.}, Fig.~\ref{fig:x3872}c).  To date, no results for charged
$(D\bar{D}^*)^+$ combinations have been published.

\section{The charged $\mathbf{Z}$ mesons}

\subsection{The $\mathbf{Z(4430)}$}
\begin{figure}[htb]
\begin{minipage}[t]{48mm}
  \includegraphics[height=0.8\textwidth,width=0.95\textwidth]{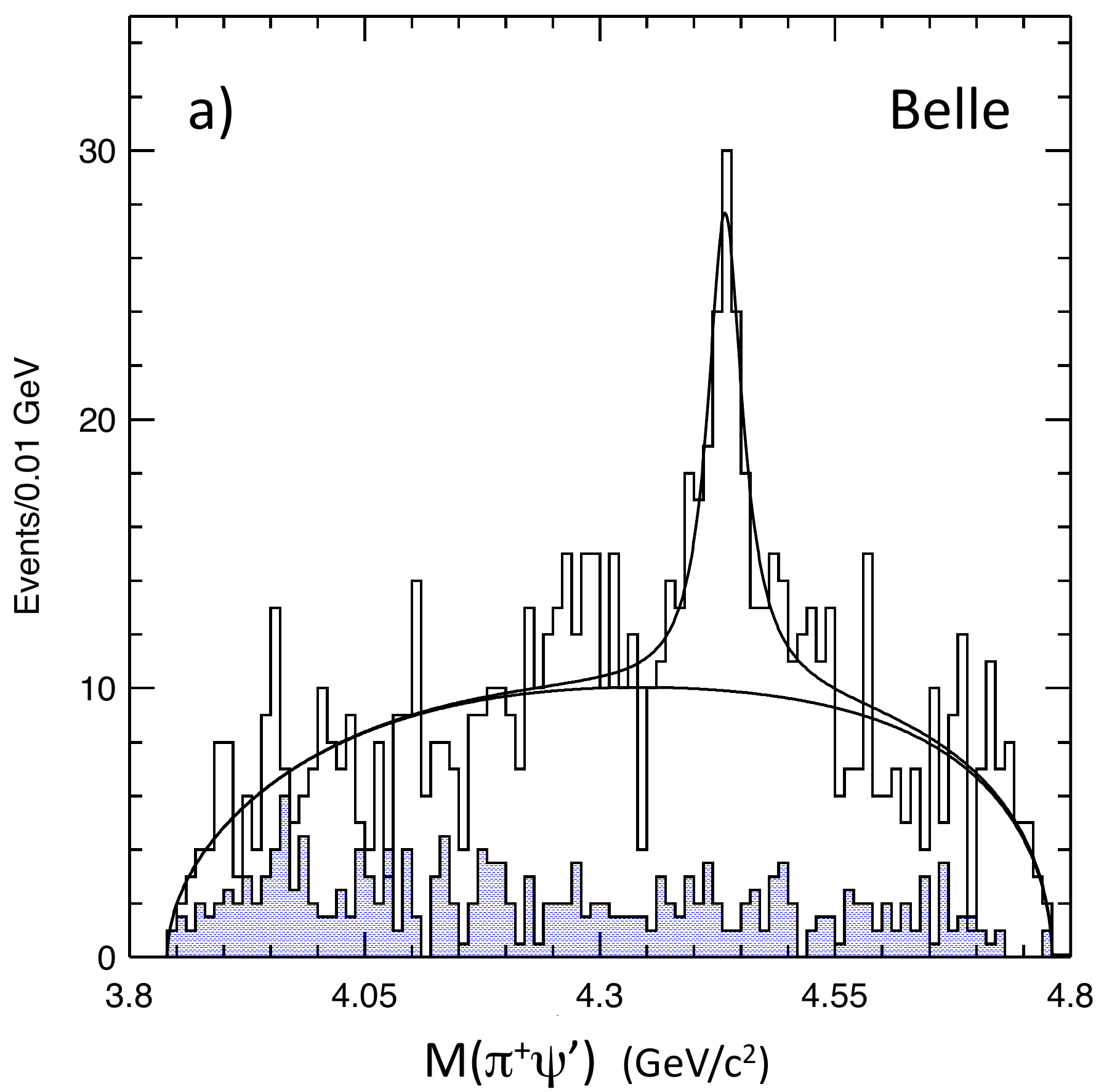}
\end{minipage}
\begin{minipage}[t]{48mm}
  \includegraphics[height=0.8\textwidth,width=0.95\textwidth]{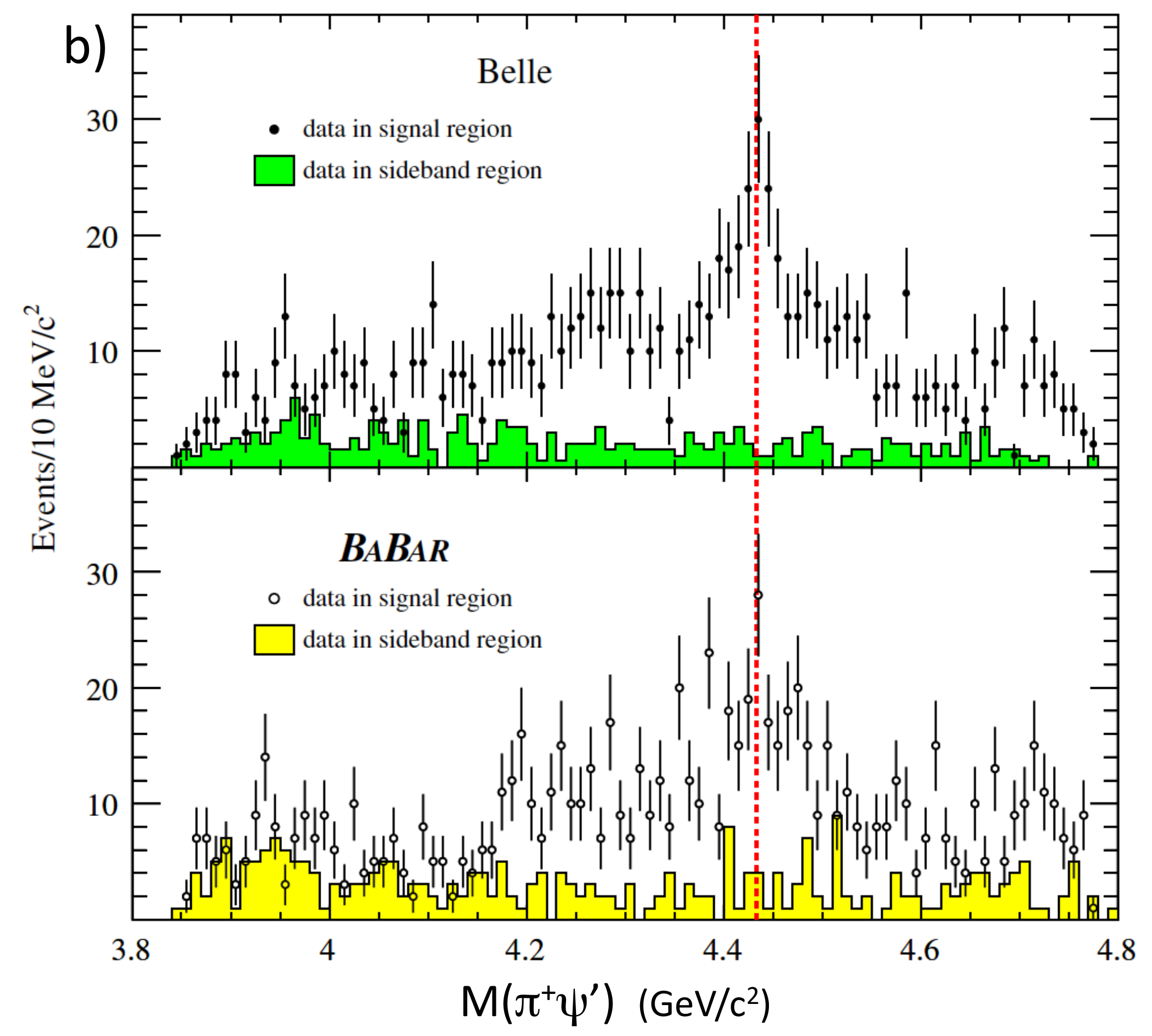}
\end{minipage}
\begin{minipage}[t]{48mm}
  \includegraphics[height=0.83\textwidth,width=0.98\textwidth]{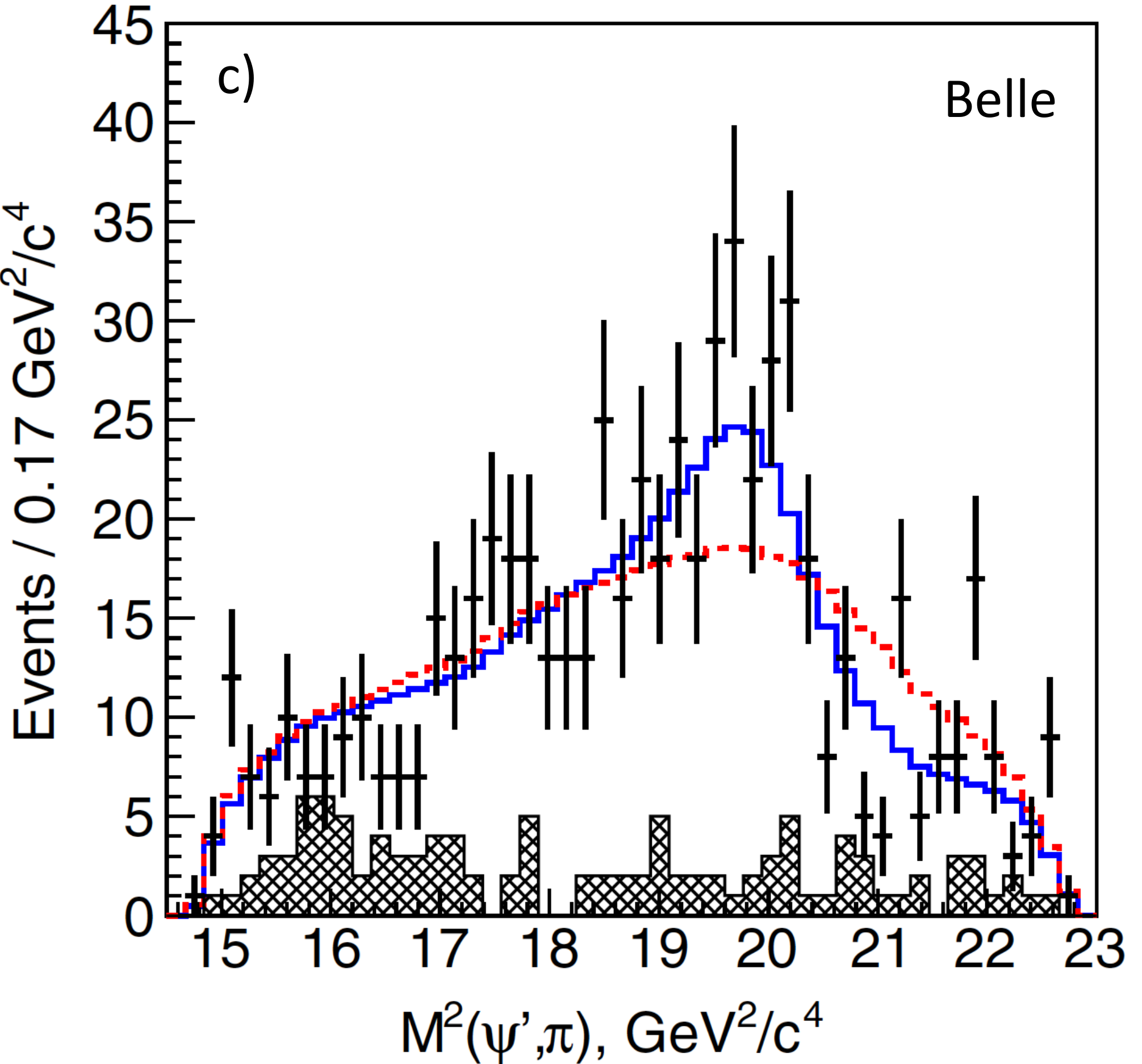}
\end{minipage}
\hspace{\fill}
\caption{\footnotesize
{\bf a)} 
The $\pip\psip$ invariant mass distribution from $B\rt K\pip\psip$ decays from
Belle~\cite{belle_z4430} for events with the $K^*$ veto requirement applied is shown as
the open histogram.  The shaded histogram is non-$\psip$ background, estimated from the 
$\psip$ mass sidebands.  The curves represent results of a fit that returned the mass and
width values quoted in the text.  
{\bf b)} A comparison of Belle {\it (upper)} and BaBar {\it (lower)} data~\cite{babar_z4430}
with the  $K^*$ and $K^*_2$ vetoed.
{\bf c)}  
The data points show the Belle $M^2(\pip\psip)$ distribution with the $K^*$ veto applied.
The solid blue histogram shows a projection of the Belle 4D fit results
with a $Z^+\rt\pip\psip$ resonance included~\cite{belle_z_dalitz2}.  The dashed red
curve shows fit results with no resonance in the $\pip\psip$ channel.  
}
\label{fig:z4430-1}
\end{figure}

\noindent
Figure~\ref{fig:z4430-1}a shows the $M(\pip\psip)$ distribution for $B\rt K\pip\psi'$ decays reported
by Belle in 2007~\cite{belle_z4430}.  Here, to reduce the influence of the dominant $B\rt K^*(890)\psip$
and $K^*_2(1430)\psip$ decay channels, events with $K\pi$ invariant masses within $\pm  100$~MeV of the
$K^*(890)$ or $K^*_2(1430)$ peaks have been excluded (the ``$K^*$ veto'').  The distinct peak is
fitted with a Breit-Wigner (BW) resonance on an incoherent background.  The BW signal from the
fit has a statistical significance of $\sim 8\sigma$, with a mass and width of $M=4433\pm 5$~MeV and
$\Gamma = 45^{+35}_{-18}$~MeV.  Since this peak structure has a non-zero electrical charge, if it is due to
a meson resonance, that meson must necessarily have a minimal $\ccbar u\bar{d}$ four-quark substructure.

A BaBar study of the $B\rt K\pip\psip$ decay channel did not confirm the Belle result~\cite{babar_z4430}.
A (nearly) direct comparison of the Belle and BaBar results for $M(\pip\psip)$ from $B\rt K\pip\psip$ 
with a $K^*$ veto is shown in Fig.~\ref{fig:z4430-1}b.  Although the BaBar plot shows an excess of events
in the same $M(\pip\psip)$ region as the Belle signal, their fit using Belle's mass and width values
yielded a statisically marginal ($\sim 2\sigma$)~$Z(4430)\rt\pip\psip$ signal.  Belle responded
to concerns about the possibility of $M(\pip\psip)$ reflection peaks due to interference between
different partial waves in the $K\pi$ resonance channels by doing two different coherent amplitude analyses
of the $B\rt K\pip\psip$ decay process. The first one used coherent amplitudes
that depended on two kinematic variables ($M(K\pip)$ and $M(\pip\psip)$)~\cite{belle_z_dalitz1} and the
second one used kinematically complete four-dimensional (4D) amplitudes that incorporated possible
dependence on the $\psip\rt\leplep$ decay helicity angle and the  angle between the $K\pip$ and
$\psip\rt\leplep$ decay planes~\cite{belle_z_dalitz2}.  Both reanalyses, which included all known $K\pi$
resonances and allowed for contributions from possible additional ones,
confirmed the existence of a resonance in the $\pip \psip$ channel with greater than $6\sigma$
significance, but with larger mass and width values than those from Belle's original
analysis~\cite{belle_z4430}: the results from the 4D analyses are $M=4485^{+36}_{-25}$~MeV and
$\Gamma=200^{+48}_{-58}$~MeV. 

The reason for the upward shifts in mass and width from Belle's originally published results can be
seen in Fig.~\ref{fig:z4430-1}c, which shows a comparison of projections of the 4D fit results with
the experimental $M^2(\pip\psip)$ distribution with the $K^*$ veto applied.  The dashed red histogram
shows the best fit results with no resonance in the $\pip\psip$ channel.  The solid blue histogram
shows results with the inclusion of a single $\pip\psip$ resonance, where strong interference effects 
that are constructive below, and destructive above, the resonance mass, are evident.  The original Belle
analysis neglected interference effects and only fitted the lower lobe of this double-lobed interference
pattern, and this resulted in a lower mass and narrower width.  

\subsubsection{LHCb confirmation of the $\mathbf{ Z(4430)}$}
\begin{figure}[htb]

\begin{minipage}[t]{48mm}
  \includegraphics[height=0.75\textwidth,width=0.95\textwidth]{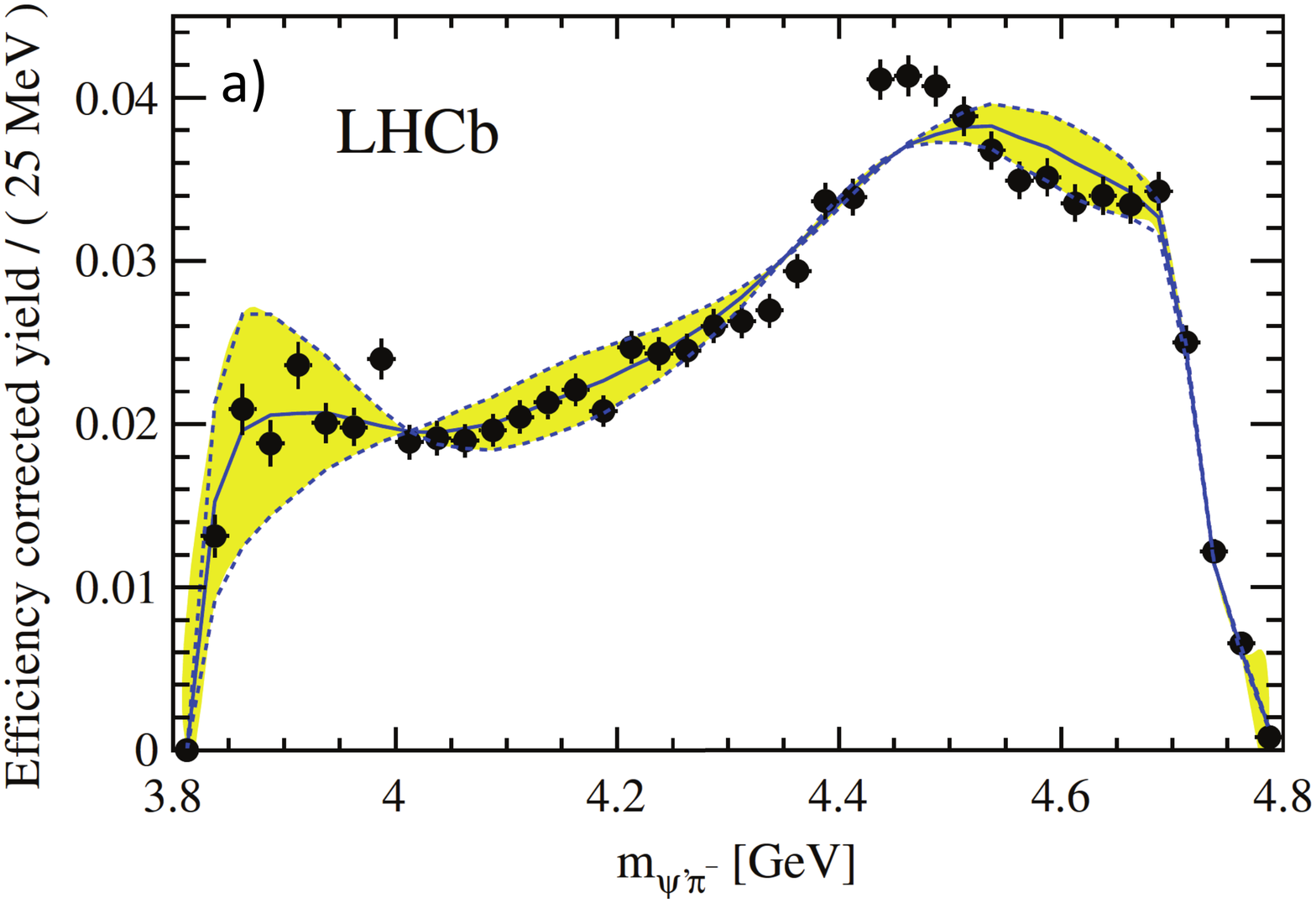}
\end{minipage}
\begin{minipage}[t]{48mm}
  \includegraphics[height=0.75\textwidth,width=0.95\textwidth]{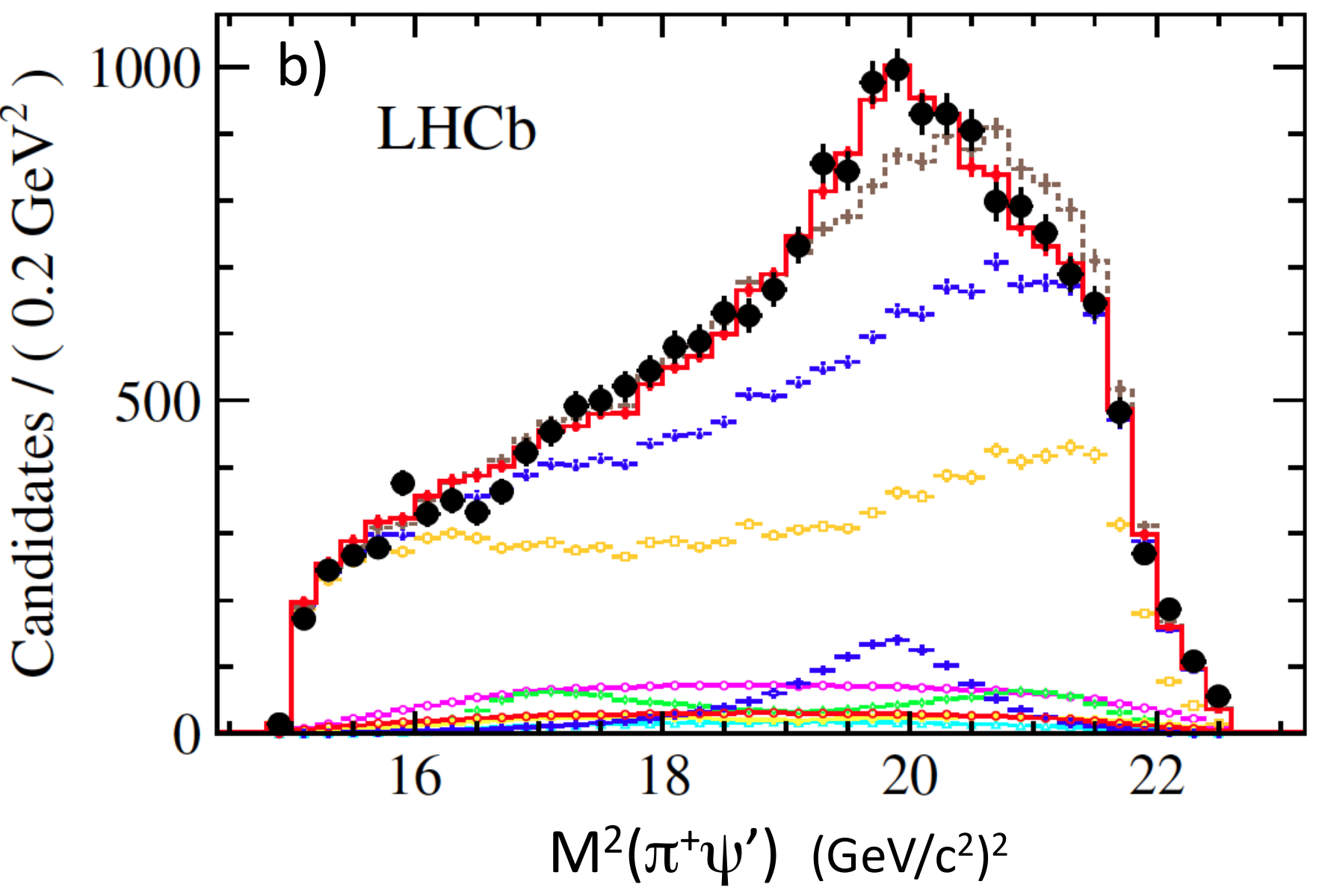}
\end{minipage}
\begin{minipage}[t]{48mm}
  \includegraphics[height=0.75\textwidth,width=0.95\textwidth]{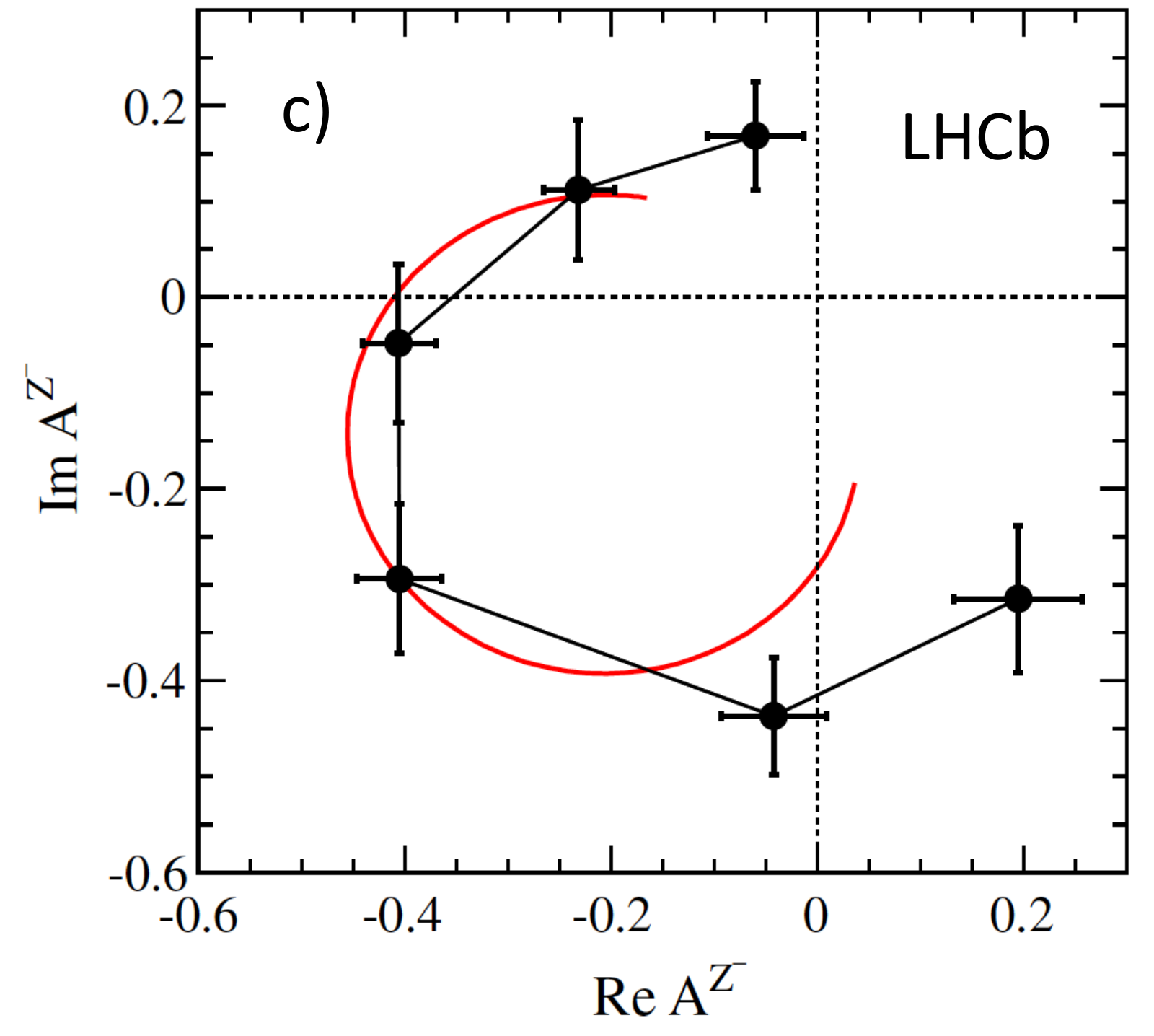}
\end{minipage}
\hspace{\fill}
\caption{\footnotesize
{\bf a)} The data points show LHCb's background-subtracted and efficiency corrected $M(\pip\psip)$
distribution.  The solid blue curve shows the result of the fit using model-independent reflections
from $\cos\theta_{K^*}$ moments up to fourth-order.  The shaded band indicates the range of errors
associated with the fit.
{\bf b)} The LHCb $M^2(\pip\psip)$ distribution for all events (no $K^*$ veto), together with
projections from the four-dimensional fits. The solid red histogram shows the fit that includes a 
$Z^+\rt\pip\psip$ resonance term; the dashed brown histogram shows the fit with no resonance in the
$\pip\psip$ channel.
{\bf c)} The Real (horizontal) and Imaginary (vertical) parts of the ($1^+$) $Z^+\rt\pip\psip$ amplitude
for six mass bins spanning, counter-clockwise, the $4430$~MeV mass region (from LHCb~\cite{LHCb_z4430}). The
red curve shows expectations for a BW resonance amplitude.
}
\label{fig:z4430-2}
\end{figure}

\noindent
The big news in 2014 was the confirmation of the Belle $Z(4430)\rt\pip\psip$ claims by the LHCb 
experiment~\cite{LHCb_z4430} based on a data sample containing $\sim$25K $B^0\rt K^-\pip\psip$
events, an order of magnitude larger than the event samples used by either Belle or BaBar. They find
that their $M(\pip\psip)$ mass distribution cannot be reproduced by reflections from the $K\pi$
channel either with a model-dependent assortment of $K\pi$ resonances up to J=3, or by a
model-independent approach that determines Legendre polynomial moments up to fourth order
($J_{K^*}\le 2$) in $\cos\theta_{K^*}$ in bins of $K\pi$ mass, where $\theta_{K^*}$ is the $K\pi$
helicity angle, and reflects them into the $\pip\psip$ channel.   Figure~\ref{fig:z4430-2}a
shows a comparison of the $M(\pip\psip)$ data with a fot that only uses reflections from the
$\cos\theta_{K^*}$ moments,
where a clear discrepancy  shows up in the $Z(4430)$ mass region. The application of the Belle
4D amplitude analysis procedure that includes a BW resonance amplitude in the $\pip\psip$ channel
results in a $Z(4430)$ signal with a huge, $\sim 14\sigma$,~statistical significance and mass \&
width values ($M=4475^{+17}_{-26}$~MeV \&
$\Gamma=172^{+39}_{-36}$~MeV) that are in close agreement with the Belle 4D analysis results.
A comparison of the LHCb fit results with the data is shown in Fig.~\ref{fig:z4430-2}b, where
strong interference effects, similar to those seen by Belle (Fig.~\ref{fig:z4430-1}c), are evident.

The LHCb group's large data sample enabled them to relax the assumption of a BW form for the
$Z^+\rt\pip\psip$ amplitude and directly measure the real and imaginary parts of the $1^+$ $\pip\psip$
amplitude in bins of $\pip\psip$ mass.  The results are shown as data points in the Argand plot in
Fig.~\ref{fig:z4430-2}c. There, the phase motion near the resonance peak agrees well with expectations
for a BW amplitude as indicated by the circular red curve superimposed on the plot.  This rapid phase
motion near amplitude-maximum is characteristic of a BW-like resonance.  (The orientation of the red
circle relects the phase angle between the $B\rt KZ$ and $B\rt K^*(890)\psip$ decay amplitudes.)  

The weighted averages of the LHCb and Belle mass and width measurements are $M_{Z(4430)}=4477\pm 20$~MeV
and $\Gamma_{Z(4430)}=181\pm 31$~MeV.  This mass is near $(m_{D}+m_{D(2600)})=4479\pm 6$~MeV, where
the $D(2600)$ is a candidate for the $D^*(2S)$, the first radial excitation of the $D^*$, that
was reported by BaBar in 2010~\cite{babar_d2600}.

\subsubsection{The recently discovered $\mathbf{Z_c(4200)}$ and observation of $\mathbf{Z(4430)\rt\pi^+J/\psi}$}

\begin{figure}[htb]

\begin{minipage}[t]{48mm}
  \includegraphics[height=0.48\textwidth,width=0.75\textwidth]{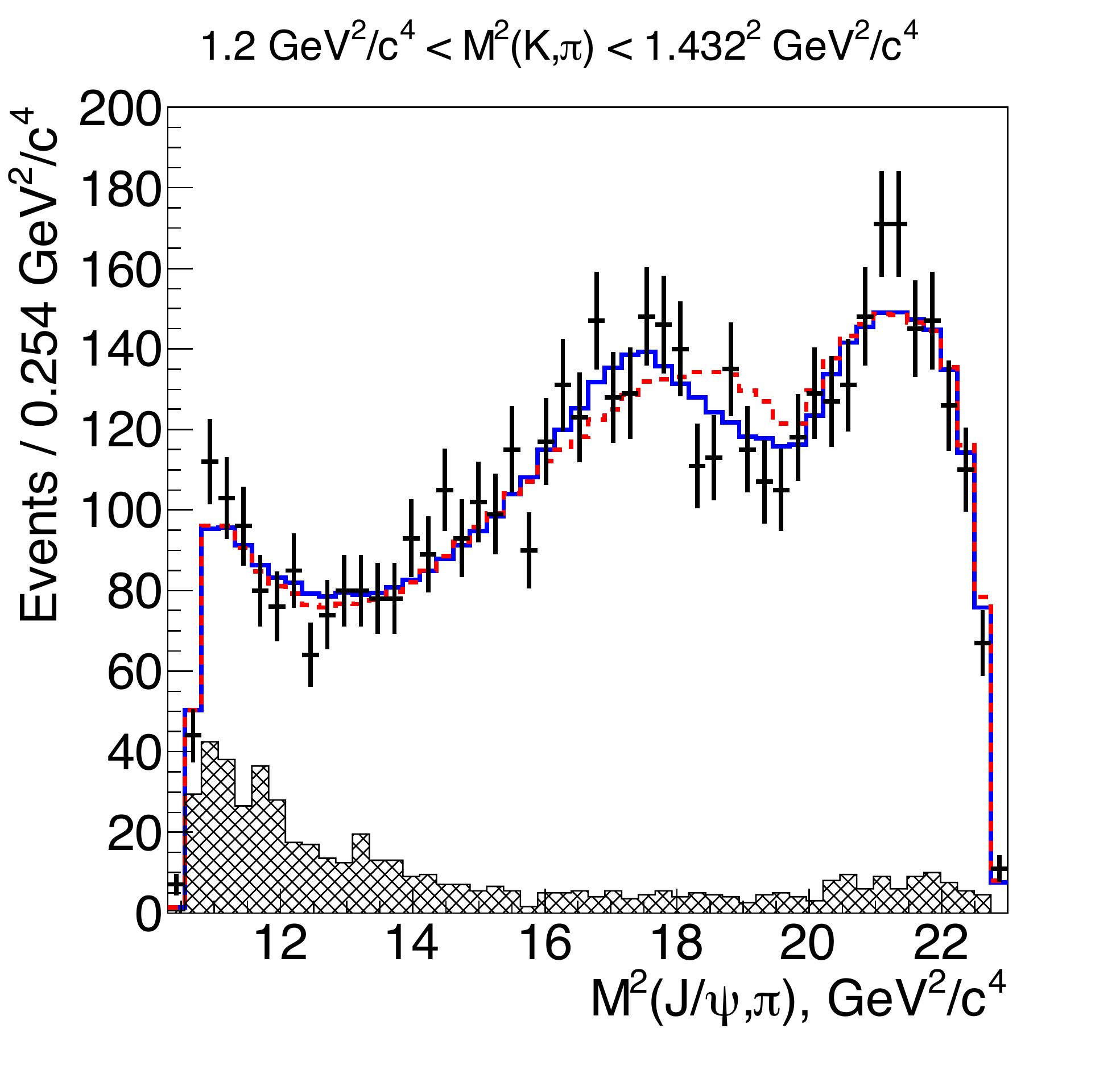}
\end{minipage}
\begin{minipage}[t]{48mm}
  \includegraphics[height=0.48\textwidth,width=0.9\textwidth]{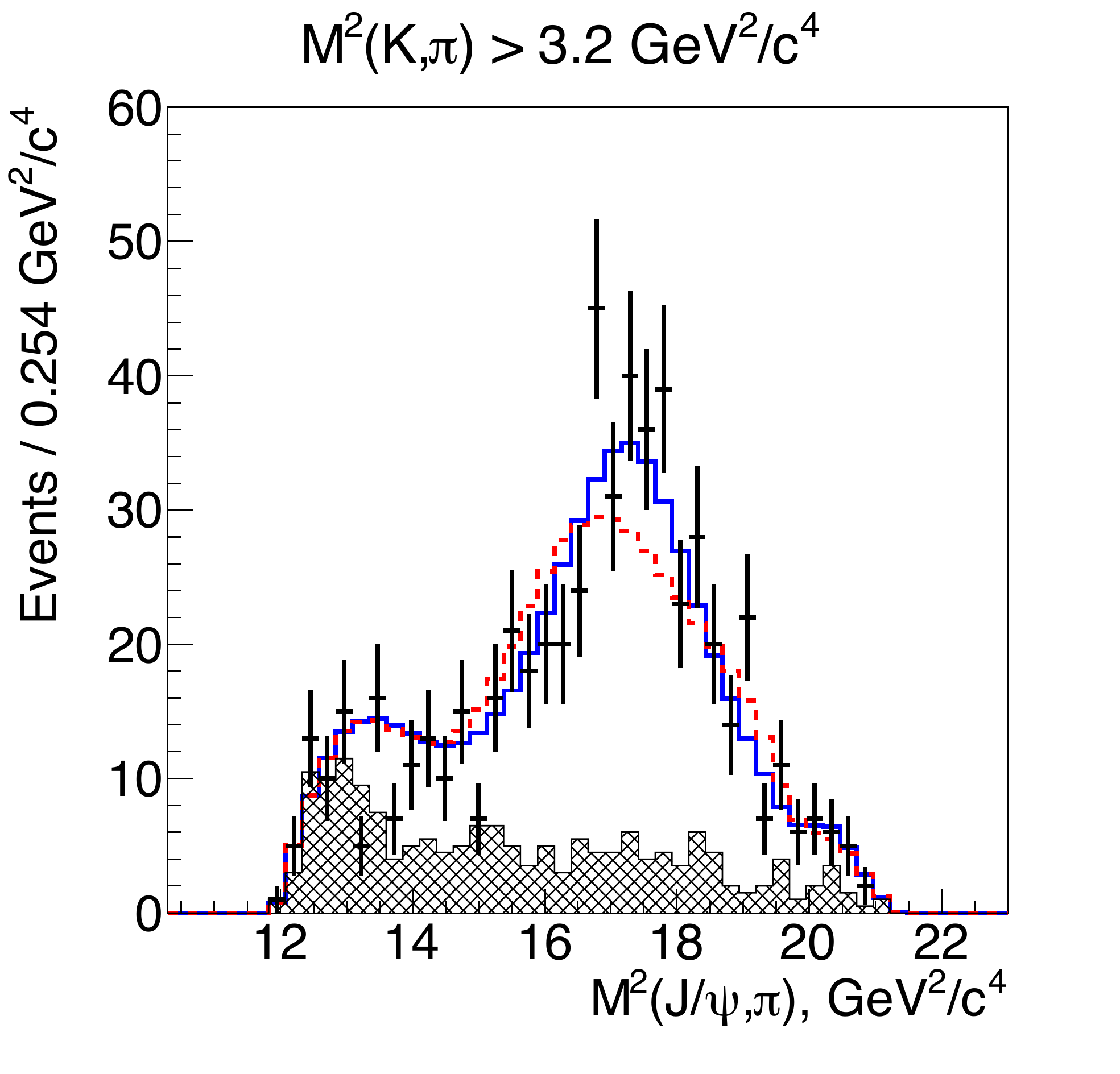}
\end{minipage}
\begin{minipage}[t]{48mm}
  \includegraphics[height=0.48\textwidth,width=0.75\textwidth]{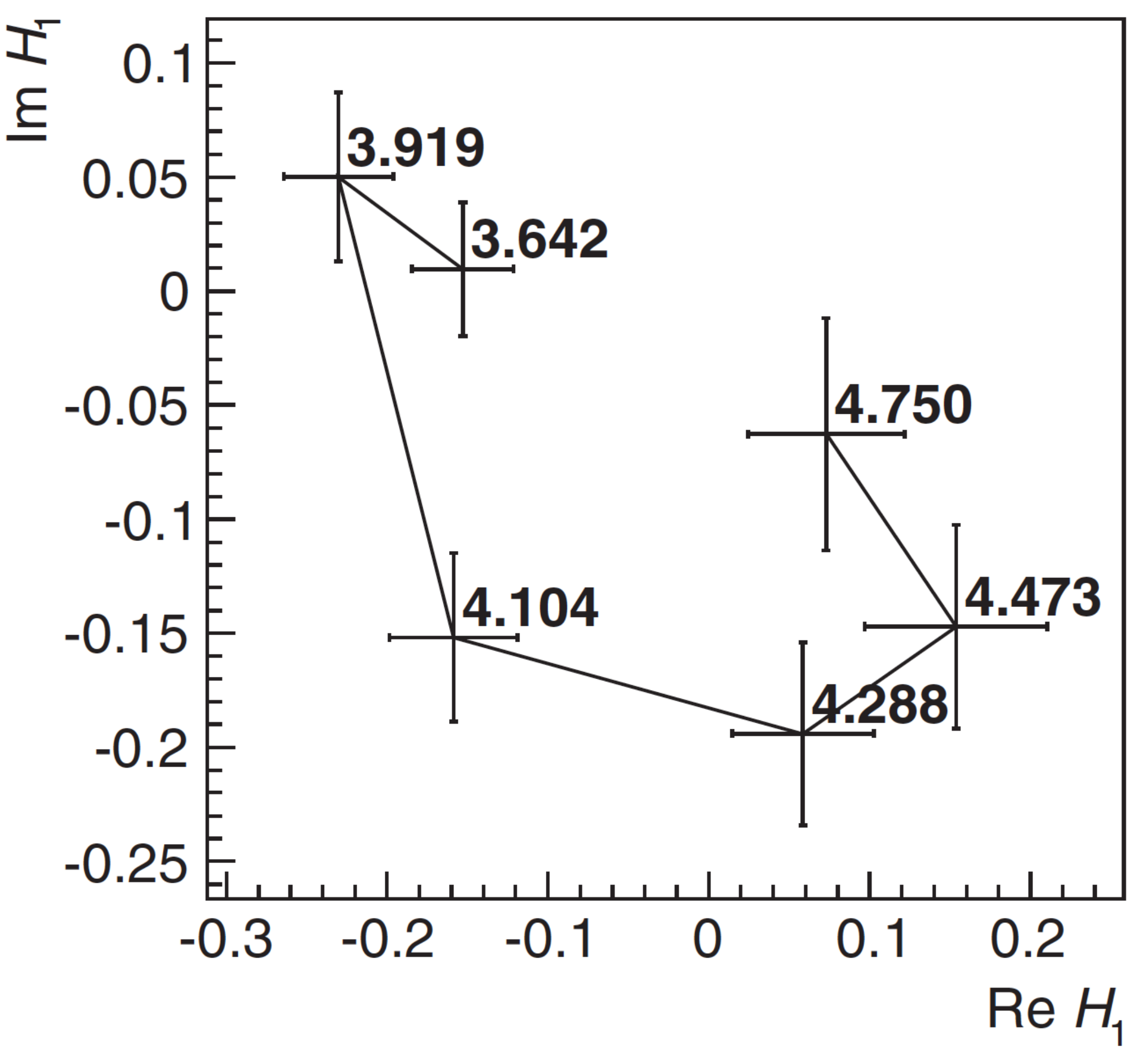}
\end{minipage}
\hspace{\fill}
\caption{\footnotesize
{\bf a)} The data points show Belle's $M^2(\pip\jpsi)$ distribution for $B^0\rt K^-\pip\jpsi$ events
with $M(K\pi)$ between the $K^*(890)$ and $K_2^*(1432)$ resonance regions.  The dashed red histogram
shows the projection of the results of a fit with no resonances in the $\pi\jpsi$ channel and
solid blue histogram is the projection of the fit that includes $Z_c(4200)$ and $Z(4430)$ BW
amplitudes. {\bf b)} The $M^2(\pip\jpsi)$ distribution for events with $M(K\pi)$ above the
$K_2^*(1432)$ resonance region. The histograms are fit projections with and without $Z$ resonance
amplitudes.
{\bf c)} The Real (horizontal) and Imaginary (vertical) parts of the $1^+$ zero-helicity
$Z^+\rt\pip\jpsi$  amplitude
for six mass bins spanning the $4200$~MeV mass region (from Belle~\cite{belle_z4200}).
}
\label{fig:z4200}
\end{figure}

New at this meeting are results from a Belle 4D amplitude analysis of $B^0\rt K^-\pip\jpsi$
decays~\cite{belle_z4200}, based on a nearly background-free data sample containing $\simeq 30K$ events.
The main result from this analysis is a $6.2\sigma$ signal for a broad $\pip\jpsi$ resonance, dubbed
the $Z_c(4200)$, with mass and width $M=4196^{+31~+17}_{-29~-13}$~MeV and $\Gamma= 370^{+70~+70}_{-70~-132}$~MeV,
and a preferred quantum number assignment of $J^{P}=1^{+}$.  Figure~\ref{fig:z4200}a shows Belle's
$M^2(\pip\jpsi)$ distribution for events with $K\pi$ masses that lie between the $K^*(890)$ and
$K_2^*(1432)$ resonance regions, with a projection from the best fit for a model in which the
$Z(4430)$ (with mass and width set at the Ref.~\cite{belle_z_dalitz2} values) is the only resonance
in the $\pip\jpsi$ channel (dashed red histogram) and results from a fit that includes an
additional $\pip\jpsi$ resonance (solid blue histogram).  Figure~\ref{fig:z4200}b shows similar results
for events with $K\pi$ masses above the $K_2^*(1432)$ resonance region.  Figure~\ref{fig:z4200}c shows an
Argand plot for the (dominant) Helicity=1, $J^P=1^+$ $\pip\jpsi$ amplitude in the 4200~MeV mass region, 
where rapid phase motion near 4200~MeV is evident. The LHCb group~\cite{LHCb_z4430} reported evidence for
a broad $\pip\psip$ resonance in this mass region with $J^P=0^-$ or $1^+$, which may be an indication of
a $\pip\psip$ decay mode of the $Z_c(4200)$.  There are no open-charmed meson-antimeson combinations that
could form a $1^+$ $S$-wave resonance with a mass threshold that are within $\sim\pm 100$~MeV of the
$Z_c(4200)$, which speaks against a molecule-like interpretation for this peak. 
 
The Belle $B^0\rt K^-\pip\jpsi$ analysis also found a $4\sigma$ signal for $B^0\rt K^- Z(4430)^+$;
$Z(4430)^+\rt \pip\jpsi$ with a product branching fraction
\begin{equation}
{\mathcal B}(B^0\rt K^- Z(4430)^+)\times {\mathcal B}(Z(4430)^+\rt\pip\jpsi)=5.4^{+4.0~+1.1}_{-1.0~-0.9}\times 10^{-6},
\end{equation} 
which is an order of magnitude smaller (albeit with large errors) than the corresponding value
for  $B^0\rt K^- Z(4430)^+$; $Z(4430)^+\rt \pip\psip$ decays: $6.0^{+1.7~+2.5}_{-2.9 -4.9} \times 10^{-5}$.
A search for $B^0\rt K^-Z_c(3900)^+)$; $Z_c(3900)^+\pip\jpsi$ found no signal; a product branching
fraction upper limit of $<9\times 10^{-7}$ (90\% CL) was established.

\subsection{The $\mathbf{Z_c(3900)}$ and $\mathbf{Z_c(4020)}$}
\noindent
The discovery and early measurements of the $Y(4260)$ were based on measurements of
the initial state radiation process, $\ee\rt \gamma_{\rm isr} Y(4260)$ at $E_{\rm cm}\simeq 10.6$~GeV.  
This reaction requires that either the incident $e^-$ or $e^+$ radiates a $\sim 4.5$~GeV photon
prior to annihilating, which results in a strong reduction in event rate.  However, since the
PEPII and KEKB $B$-factories ran with such high luminosities 
(${\mathcal L} > 10^{34}\ {\rm cm}^{-2}{\rm s}^{-1}$), the measurements were feasible.  A more
efficient way to produce $Y(4260)$ mesons would be to operate a high luminosity $\ee$ collider
as a ``$Y(4260)$ factory,'' {\it i.e.}, at a cm energy of 4260~MeV, corresponding to the peak
mass of the $Y(4260)$.  This was done at the two-ring Beijing electon-positron
collider (BEPCII)~\cite{bepcii} in 2013, and large numbers of $Y(4260)$ decays were detected
in the BESIII spectrometer~\cite{besiii}.  This resulted in the discoveries of two additional
charged charmoniumlike states: the $Z_c(3900)$ and $Z_c(4020)$.

\subsubsection{The $\mathbf{Z_c(3900)}$}
\begin{figure}[htb]
\begin{minipage}[t]{48mm}
  \includegraphics[height=1.0\textwidth,width=1.0\textwidth]{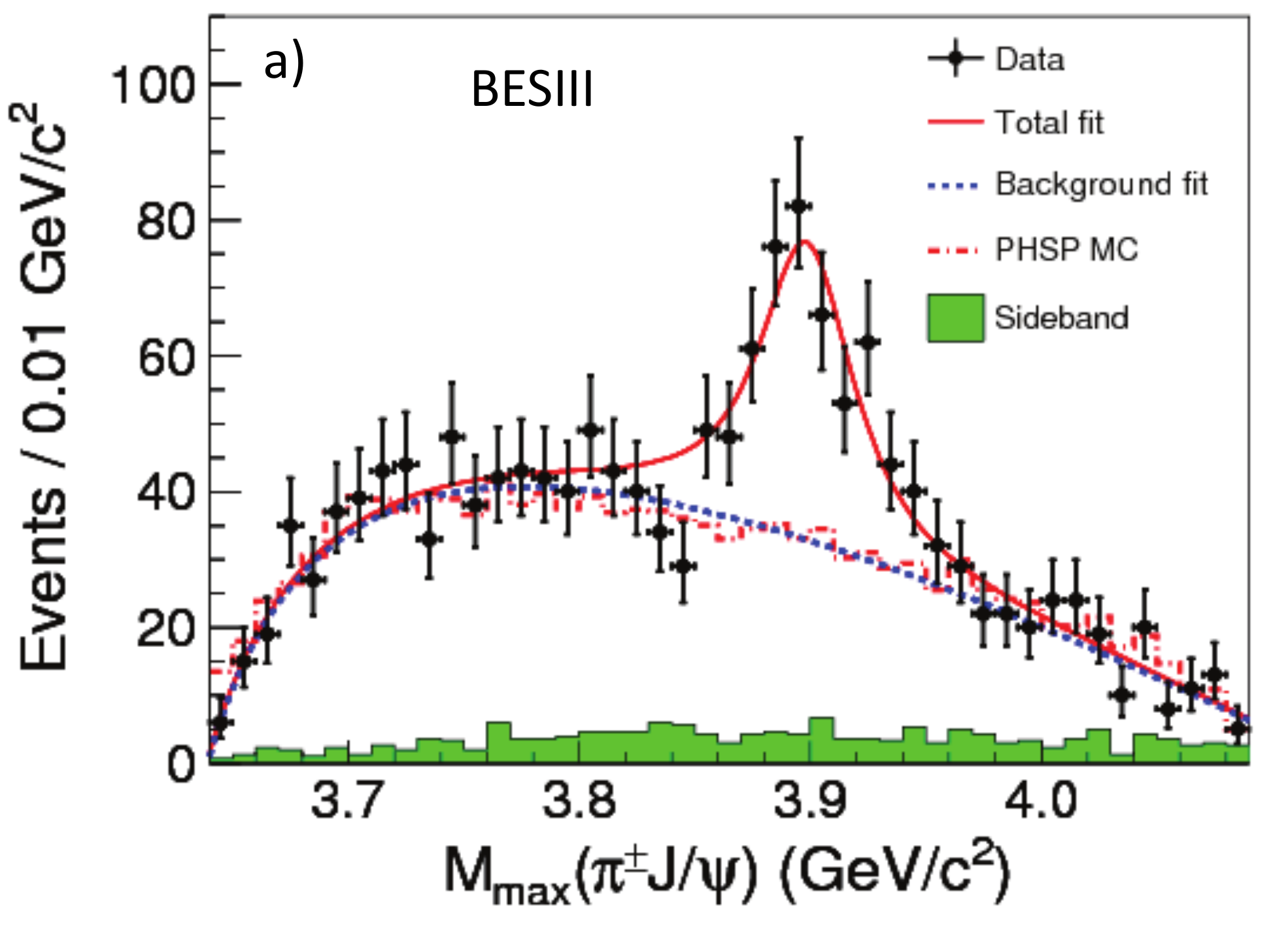}
\end{minipage}
\begin{minipage}[t]{48mm}
  \includegraphics[height=1.0\textwidth,width=1.0\textwidth]{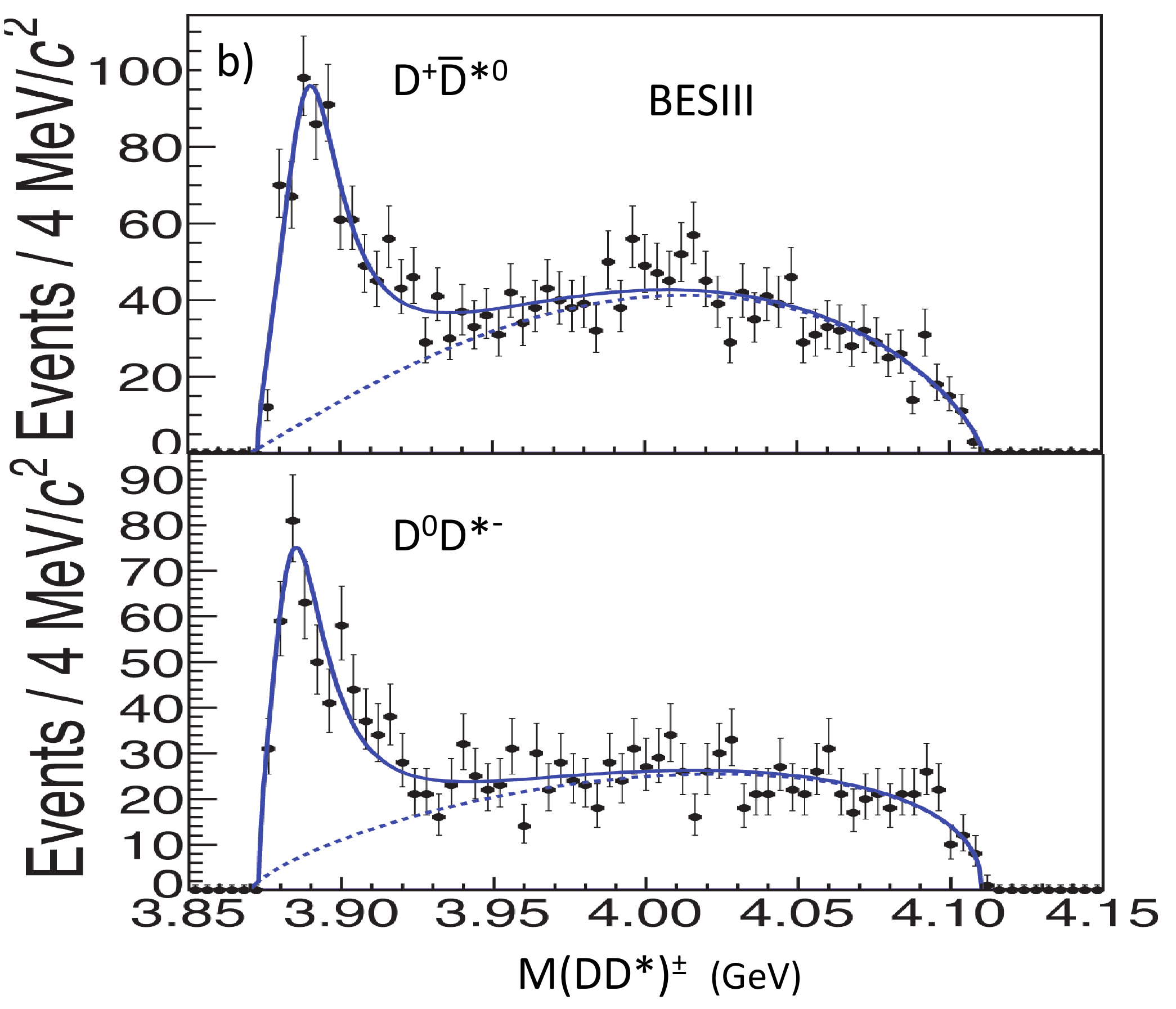}
\end{minipage}
\begin{minipage}[t]{48mm}
  \includegraphics[height=1.0\textwidth,width=1.0\textwidth]{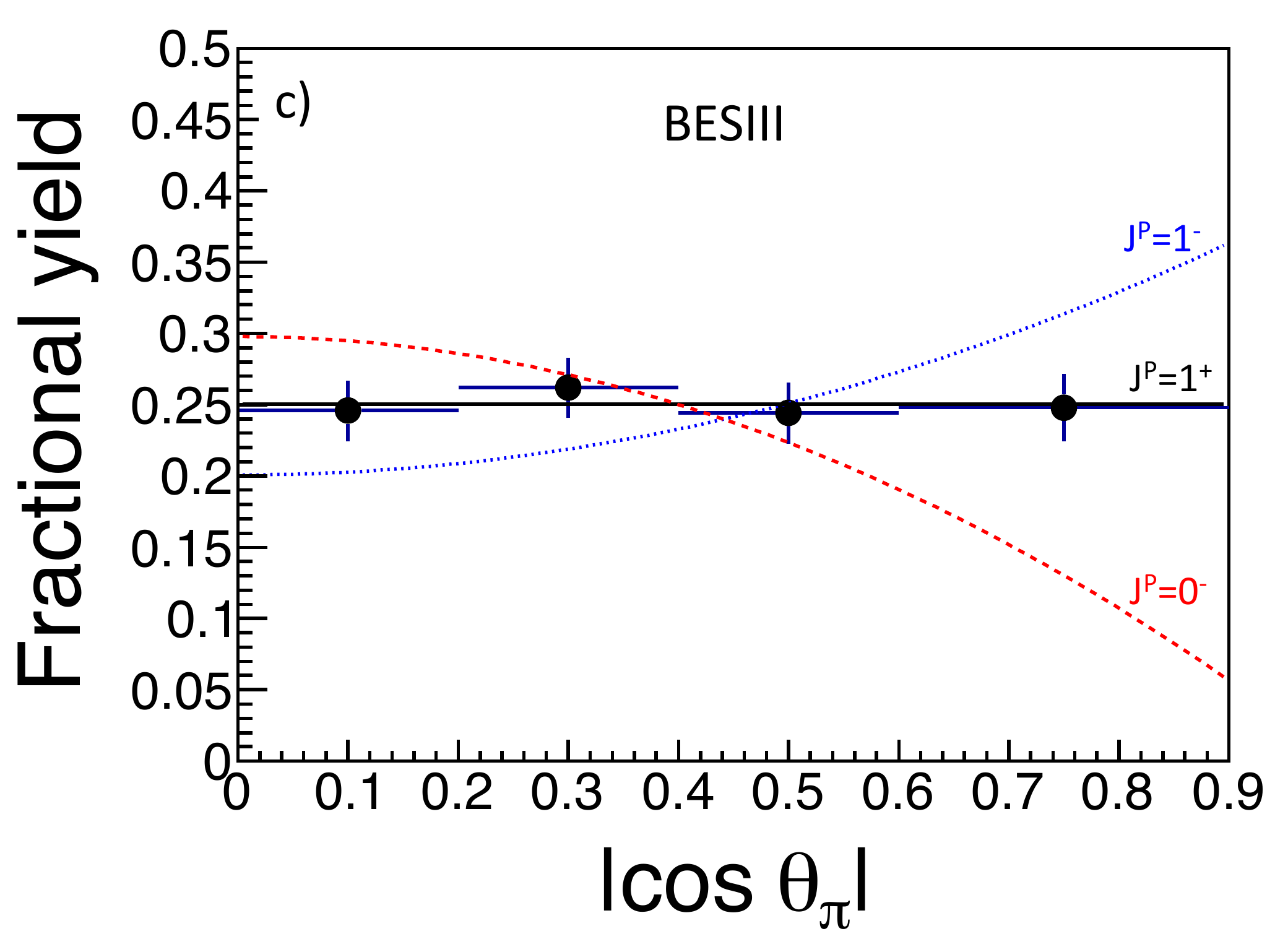}
\end{minipage}
\hspace{\fill}
\caption{\footnotesize {\bf a)} Invariant mass distributions for $\pip\jp$ from $\ee\rt\pipi\jp$ events
from Ref.~\cite{bes_z3900};
{\bf b)} $M(D^+\bar{D}^{*0}$ (top) and $M(D^0 D^{*-})$ (bottom) for $\ee\rt (D\bar{D}^*)^{\pm}\pi^{\mp}$ events from
Ref.~\cite{bes_z3885}; 
{\bf c)} the efficiency corrected production angle distribution compared with predictions for
$J^P = 0^-$ (dashed-red), $J^P=1^-$ (dotted blue) and $J^P=1^+$ (solid black) quantum number assignments.
}
\label{fig:zc3900}
\end{figure}

\noindent
The first channel to be studied with the $E_{\rm cm}=4260$~MeV data was $\ee\rt\pipi\jp$, where
a distinct peak, called the $Z_c(3900)$, was seen near $3900$~MeV in the distribution of the
larger of the two $\pi^{\pm}\jp$ invariant mass combinations in each event ($M_{\rm max}(\pi\jp)$),
as can be seen shown in Fig.~\ref{fig:zc3900}a~\cite{bes_z3900}.  A fit using a mass-independent-width
BW function to represent the $\pi^{\pm}\jp$ mass peak yielded a mass and width of 
$M_{Z_c(3900)}=3899.0\pm 6.1$~MeV and $\Gamma_{Z_c(3900)}= 46 \pm 22$ ~MeV, which is $\sim 24$~MeV
above the $m_{D^{*+}} + m_{\bar{D}^{0}}$ (or $m_{D^+} + m_{\bar{D}^{*0}}$) threshold. The $Z_c(3900)$ was
observed by Belle in isr data at the same time~\cite{belle_z3900}.

A subsequent BESIII study of the $(\DDbar^*)^+$ systems produced in $(\DDbar^*)^{\pm}\pi^{\mp}$
final states in the same data sample, found very strong near-threshold peaks in  both the
$D^0D^{*-}$ and $D^+\bar{D}^{*0}$ invariant mass distributions~\cite{bes_z3885}, as shown in
Fig.~\ref{fig:zc3900}b. The curves show results of
fits to the data with threshold-modified BW line shapes to represent the peaks. The average values
of the mass and widths from these fits are used to determine the resonance pole position 
($M_{\rm pole} + i\Gamma_{\rm pole}$) with real and imaginary values of $M_{\rm pole}= 3883.9\pm 4.5$~MeV
and $\Gamma_{\rm pole}=24.8\pm 12$~MeV.

Since the pole mass position is $\simeq 2\sigma$ lower than the $Z_c(3900)$ mass reported in
Ref.~\cite{bes_z3900}, BESIII cautiously
named this $D\bar{D}^*$ state the $Z_c(3885)$.  In the mass determinations of both the $Z_c(3885)$
and $Z_c(3900)$, effects of possible interference with a coherent component of the background are
ignored, which can bias the measurements by amounts comparable to the resonance widths, and this
might account for the different mass values. In any case, we consider it highly likely that the
$Z_c(3885)$ is the $Z_c(3900)$ in a different decay channel. If this the case, the partial width
for $Z_c(3900)\rt D\bar{D}^*$ decays is $6.2\pm 2.9$ times larger than that for $\jp\pip$, which
is small compared to open-charm {\it vs.} hidden-charm decay-width ratios for established charmonium
states above the open-charm threshold, such as the $\psi(3770)$ and $\psi(4040)$, where corresponding
ratios are measured to be more than an order-of-magnitude larger~\cite{pdg}.  

Since the $Z_c(3885)\rt D\bar{D}^*$ signals are so strong, the $J^P$ quantum numbers could be determined
from the dependence of its production on $\theta_{\pi}$, the polar angle of the bachelor-pion track relative
to the beam direction in the $\ee$ cm system. For $J^P=0^-$, $dN/d|\cos\theta_{\pi} |$ should go as
$\sin^2\theta_{\pi}$; for $1^-$ it should follow $1+\cos^2\theta_{\pi}$ and for $1^+$ it should be flat ($0^+$ is
forbidden by Parity). Figure~\ref{fig:zc3900}c shows the efficiency-corrected $Z_c(3885)$ signal yield
as a function of $|\cos\theta_{\pi}|$, together with expectations for $J^P=0^+$ (dashed red), $1^-$
(dotted blue) and $J^P=1^+$.  The $J^P=1^+$ assignment is clearly preferred and the $0^-$ and $1^-$
assignments are ruled out with high confidence. 

\subsubsection{The $\mathbf{Z_c(4020)}$}
\label{sect:zc4020}
\begin{figure}[htb]
\begin{minipage}[t]{36mm}
  \includegraphics[height=1.0\textwidth,width=1.0\textwidth]{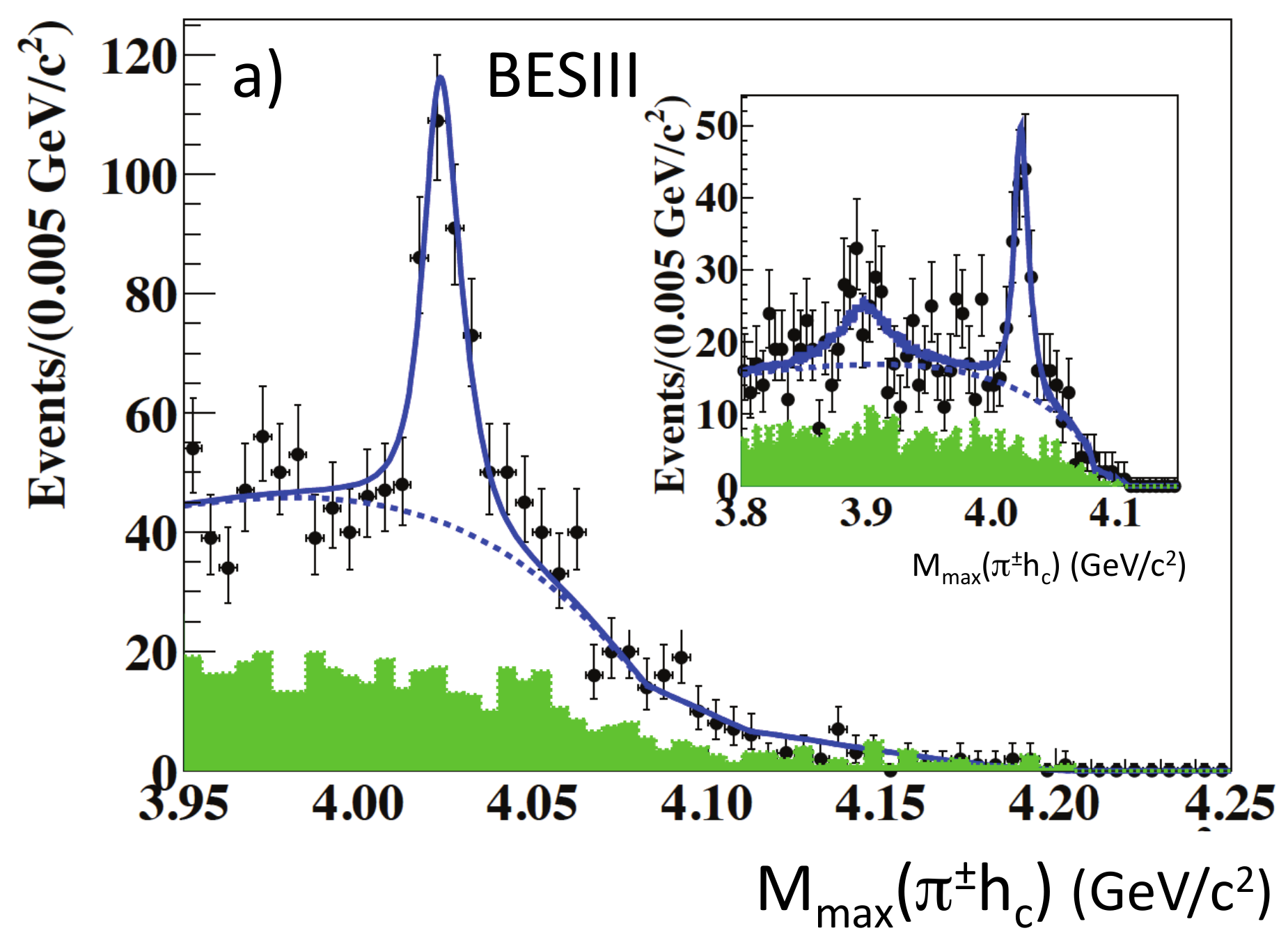}
\end{minipage}
\begin{minipage}[t]{36mm}
  \includegraphics[height=1.0\textwidth,width=1.0\textwidth]{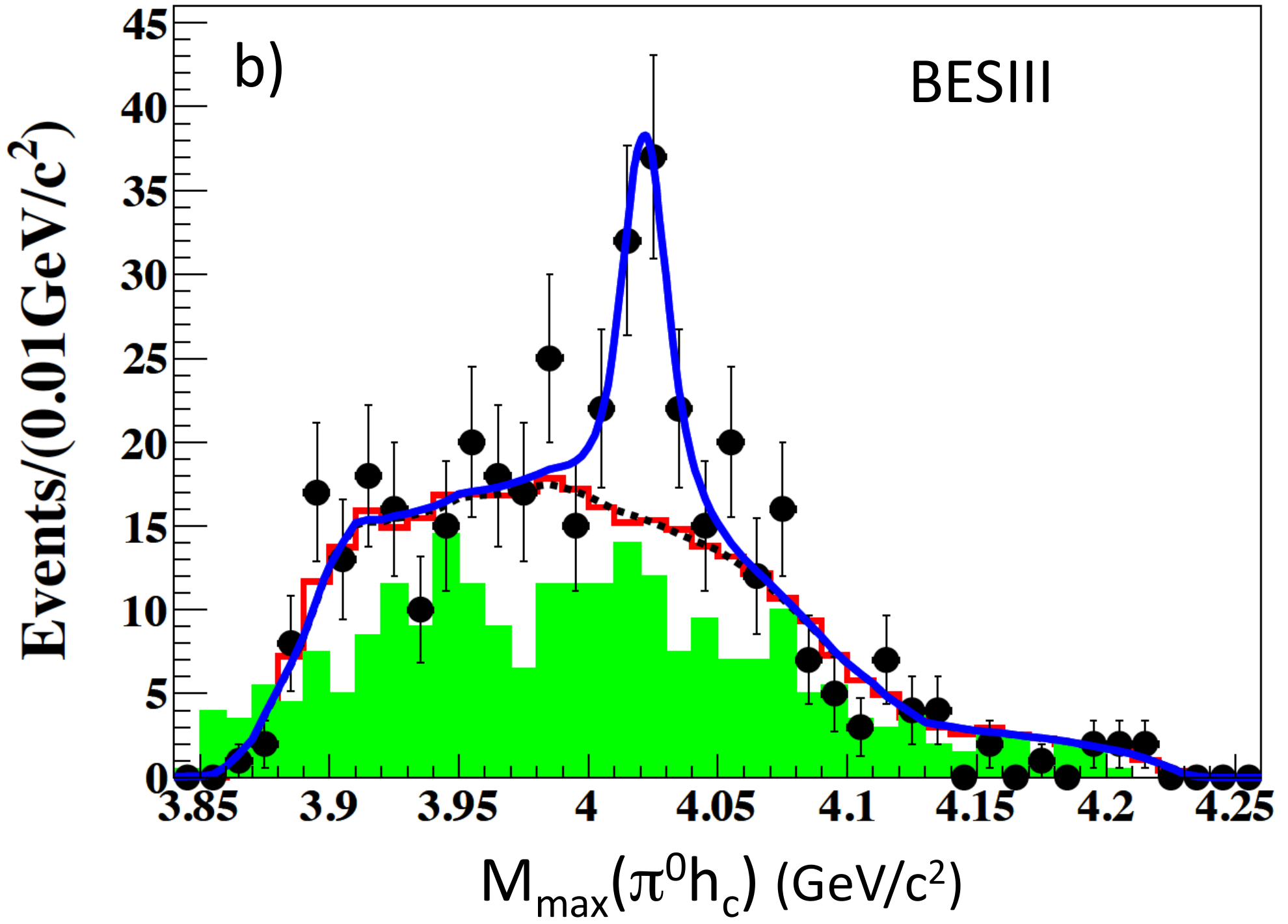}
\end{minipage}
\begin{minipage}[t]{36mm}
  \includegraphics[height=1.0\textwidth,width=1.0\textwidth]{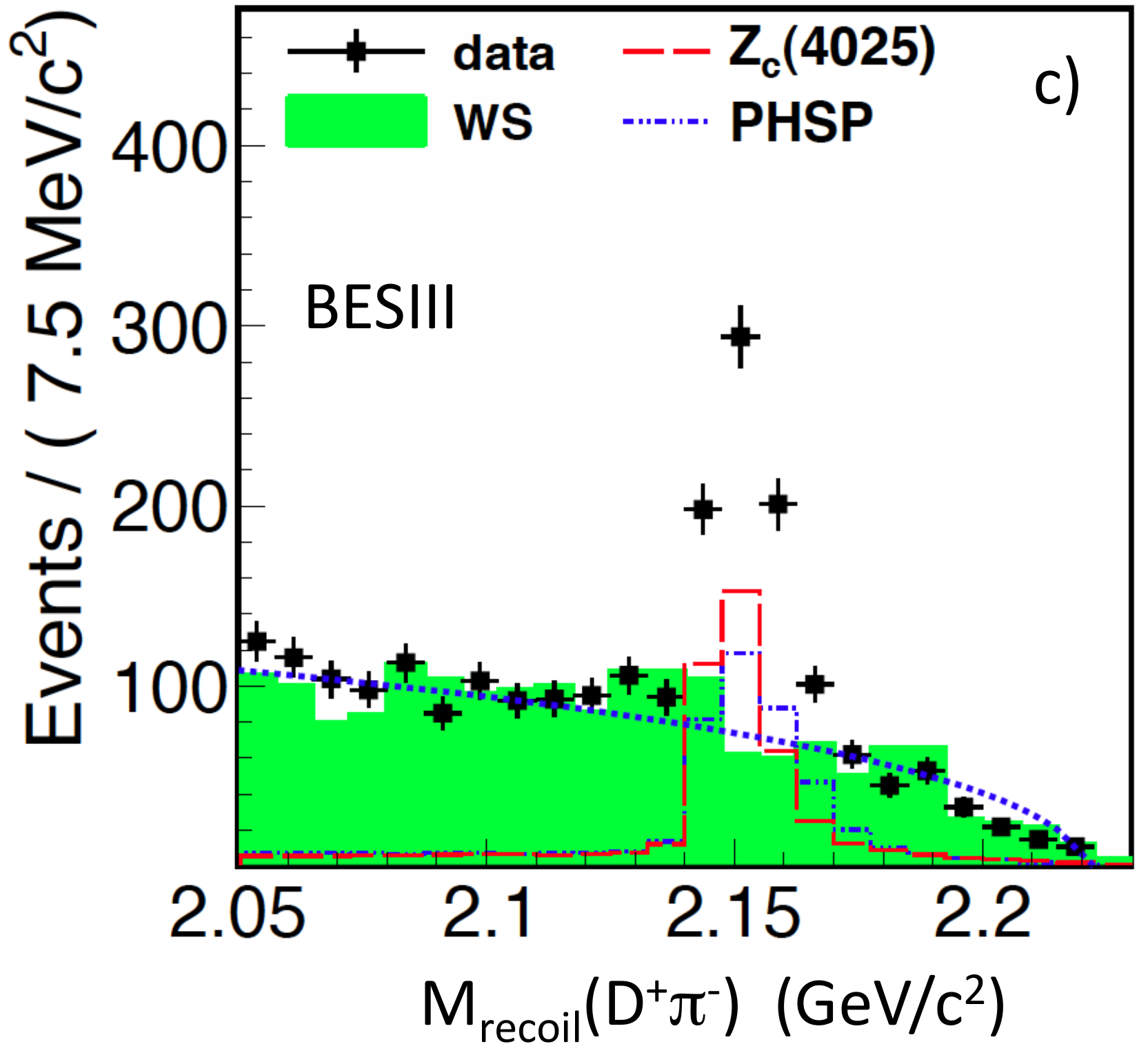}
\end{minipage}
\begin{minipage}[t]{36mm}
  \includegraphics[height=1.0\textwidth,width=1.0\textwidth]{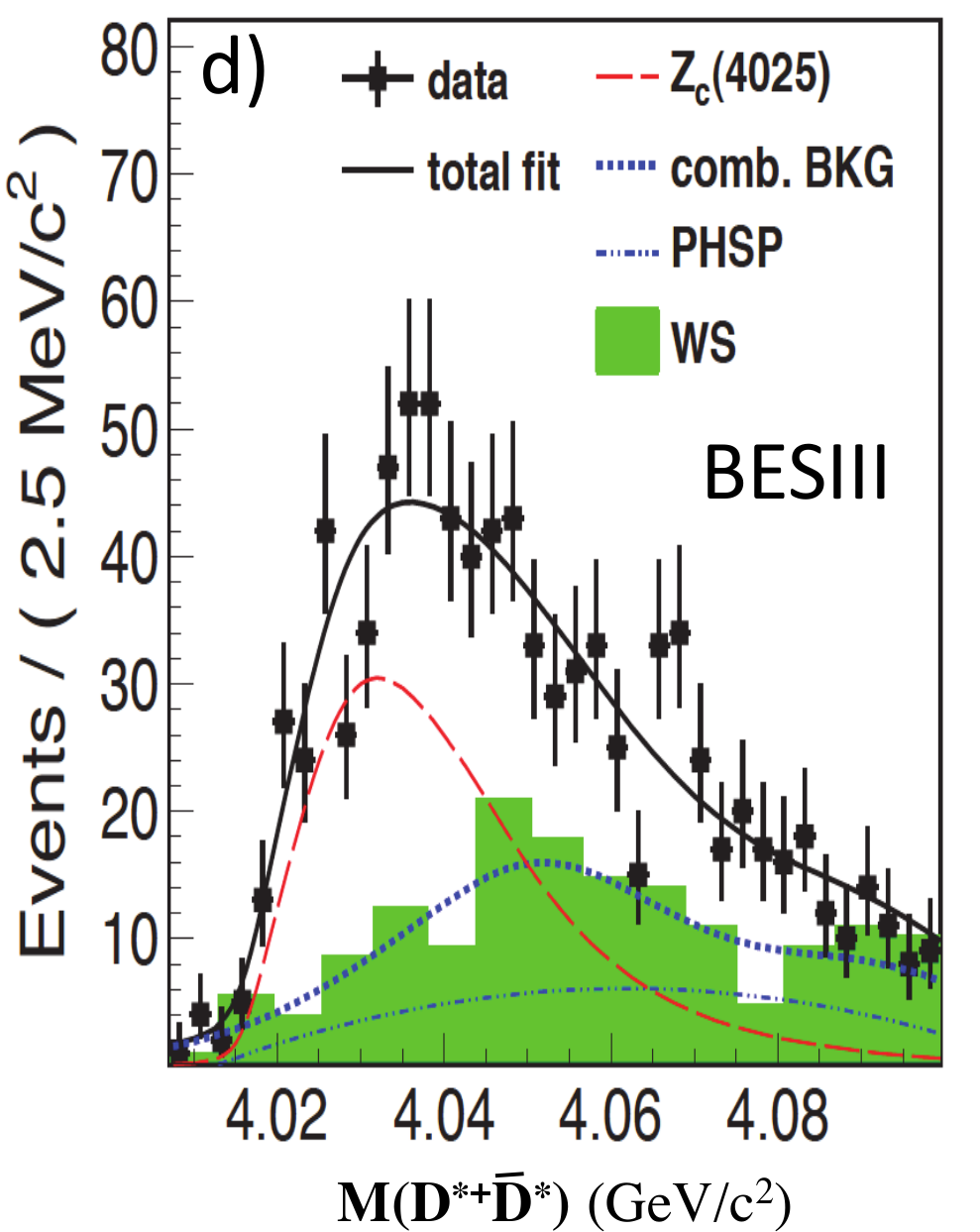}
\end{minipage}
\hspace{\fill}
\caption{\footnotesize {\bf a)} The $M_{\rm max}(\pip h_c)$ distribution for
$\ee\rt\pipi h_c$ events from BESIII.  The shaded histogram is background estimated from the $h_c$
mass sidebands.  The curves are results of fits described in the text.
{\bf b)} The corresponding $M_{\rm max}(\piz h_c)$ distribution for
$\ee\rt \pi^0\pi^0 h_c$ events from BESIII.
{\bf c)} The distribution of masses recoiling from a detected $D^+$ and
$\pim$ for $\ee\rt D^+\pim \piz X$ events at $\sqrt{s}=4260$~MeV.  The peak near 2.15~MeV corresponds
to $\ee\rt \pim D^{*+}\bar{D}^{*0}$ events.  The red dashed histogram shows the expected 
recoil mass distribution for $\ee\rt \pim Z_c$, with $M_{Z_c}=4025$~GeV;  the open, dash-dot histogram
shows results for MC $\pim D^{*+}\bar{D}^{*0}$ three-body phase-space events.  
The shaded histogram is combinatoric background from wrong-sign combinations in the data.
{\bf d)} $M(D^*\bar{D}^*)$ for $\ee\rt (D^*\bar{D}^*)^+\pim$ events, {\it i.e.}, events in the 
2.25~MeV peak in panel {\bf c}. 
The curves are described in the text. 
}
\label{fig:zc4020}
\end{figure}

\noindent
With data accumulated at the peaks of the $Y(4260)$, $Y(4360)$ and nearby energies,
BESIII made a study of $\pipi h_c(1P)$ final states. The exclusive $h_c(1P)$ decays were detected
via the $h_c\rt\gamma\eta_c$ transition, where the $\eta_c$ was reconstructed in 16 exclusive
hadronic decay modes.  With these data, BESIII observed a distinct peak near 4020~MeV in the
$M_{\rm max}(\pi^{\pm}h_c)$ distribution that is shown in Fig.~\ref{fig:zc4020}a.   A fit to this
peak, which the BESIII group called the $Z_c(4020)^+$, with a signal BW function (assuming $J^p=1^+$)
plus a smooth background, returns a $\sim 9\sigma$ significance signal with a fitted mass of
$M_{Z_c(4020)}=4022.9\pm 2.8$~MeV, about $ 5$~MeV above $m_{D^{*+}}+m_{\bar{D}^{*0}}$, and a width
of $\Gamma_{Z_c(4020)}= 7.9\pm 3.7$ ~MeV~\cite{bes_z4020}.  The product
$\sigma(\ee\rt\pim Z_c(4020)^+)\times {\mathcal B}(Z_c(4020)^+\rt\pip h_c) $ is measured to be
$7.4\pm 2.7 \pm 1.2$~pb at $E_{\rm cm}=4260$~MeV, where the second error reflects the uncertainty
of ${\mathcal B}(h_c\rt \gamma \eta_c$). 

The inset in Fig.~\ref{fig:zc4020}a shows the result of including a $Z_c(3900)^+\rt \pip h_c$ term
in the fit. In this case, a marginal $\sim 2\sigma$ signal for $Z_c(3900)^+\rt \pip h_c$ is
seen to the left of the $Z_c(4020)$ peak.  This translates into  an upper limit on the product 
$\sigma(\ee\rt\pim Z_c(3900)^+)\times {\mathcal B}(Z_c(3900)^+\rt\pip h_c) $ of 11~pb.  Since the
product $\sigma(\ee\rt\pim Z_c(3900)^+)\times {\mathcal B}(Z_c(3900)^+\rt\pip\jpsi)$ is measured to
be $62.9\pm 4.2 $ pb~\cite{bes_z3900}, this limit implies that the $Z_c(3900)^+\rt\pip h_c$ decay channel
is suppressed relative to that for $\pip\jpsi$ by at least a factor of five.

BESIII recently reported observation of the neutral member of the $Z_c(4020)$ isospin triplet~\cite{bes_z4020-0}.
The $M_{\rm max}(\piz h_c )$ distribution for $\ee\rt\pi^0\pi^0 h_c$ events in the same data set, shown in
Fig.~\ref{fig:zc4020}b, looks qualitatively like the $M_{\rm max}(\pip h_c )$ distribution
with a distinct peak near $4020$~MeV.  A fit to the data that includes a BW term with a width fixed at
the value measured for the $Z_c(4020)^+$ and floating mass returns a mass of $4023.9\pm 4.4$~MeV; this
and the signal yield are in good agreement with expectations based on isospin symmetry.  

BESIII studied $\ee\rt D^{*+}\bar{D}^{*0}\pim$ events in the $E_{\rm cm}=4.26$~GeV data sample using
a partial reconstruction technique that only required the detection of the bachelor $\pim$, the
$D^+$ from the $D^{*+}\rt \piz D^+$ decay and one $\piz$, either from the $D^{*+}$ or the
$\bar{D}^{*0}$ decay, to isolate the process and measure the $D^{*+}\bar{D}^{*0}$ invariant
mass~\cite{bes_z4025}. The signal for real $D^{*+}\bar{D}^{*0}\pim$ final states is the distinct peak
near 2.15~MeV in the $D^+\pim$ recoil mass spectrum shown on Fig.~\ref{fig:zc4020}c. The measured
$D^*\bar{D}^*$ invariant mass distribution for events in the 2.15~MeV peak, shown as data points
in Fig.~\ref{fig:zc4020}d, shows a strong near-threshold peaking behavior with a shape that cannot
be described by a phase-space-like
distribution, shown as a dash-dot blue curve, or by combinatoric background, which is
determined from wrong-sign (WS) events in the data ({\it i.e.}, events where the bachelor pion and
charged $D$ meson have the same sign) that are shown as the shaded histogram. The solid black curve
shows the results of a fit to the data points that includes an efficiency weighted $S$-wave BW
function, the WS background shape scaled to measured non-$D^{*+}\bar{D}^{*0}\pim$ background level
under the signal peak in Fig.~\ref{fig:zc4020}d, and a phase-space term. The fit returns
a $13\sigma$ signal with mass and width $M=4026.3\pm 4.5 $~MeV and $\Gamma=24.8\pm 9.5$~MeV,
values that are close to those measured for the $Z_c(4020)^+\rt\pip h_c $ channel. Although
BESIII cautiously calls this $(D^*\bar{D}^*)^+$ signal the $Z_c(4025)$, in the following we assume
that this is another decay mode of the $Z_c(4020)$.

From numbers provided in Ref.~\cite{bes_z4025}, we determine 
$\sigma(\ee\rt\pim Z_c(4020))\times {\mathcal B}(Z_c(4020)\rt D^*\bar{D}^*)=89\pm 19$~pb.
This implies that the partial width
for $Z_C(4020)\rt D^*\bar{D}^*$ is larger than that for $Z_c(4020)\rt\pi h_c$, but only by a factor
of $12\pm 5$, not by the large factors that are characteristic of open charm decays of conventional
charmonium.

The $J^P$ values of the $Z_c(4020)^+$ have not been determined. As mentioned in Lange's talk at this meeting,
charged bottomoniumlike states have been seen in the $b$-quark sector just above the $B\bar{B}^*$ and
$B^*\bar{B}^*$ open-bottom thresholds~\cite{belle_zb}, the $Z_b(10610)$ and $Z_b(10650)$, respectively, and
both have been determined to have $J^P=1^+$~\cite{belle_zb-jpc}.  Thus $J^P=1^+$ is probably a reasonable guess
for the $Z_c(4020)$.

\subsection{Are there non-resonant sources for the near threshold $\mathbf{Z_c(3900)}$ and $\mathbf{Z_c(4020)}$ peaks?}
\noindent
The $Z_c(3900)^+$ and the $Z_c(4020)^+$ are just above the $D\bar{D}^*$ and $D^*\bar{D}^*$ thresholds, and the
decay modes $Z_c(3900)\rt D\bar{D}^*$ and $Z_c(4020)\rt D^*\bar{D}^*$ have been seen. For $Z_c$ quantum
numbers of $J^{P}=1^+, $  the $D^{(*)}\bar{D}^*$ system are in an $S$-wave. In this case, the coupled-channel
process illustrated in diagram~b of Fig.~\ref{fig:cusp-1}~(left), can produce a sharp peaking structure in the
$\pi\jp~(h_c)$ invariant mass distribution just above the $D^{(*)}\bar{D}^*$ threshold.\footnote{The left-most
panels of Figs.~\ref{fig:cusp-1}~and~\ref{fig:cusp-2} apply specifically to the $Z_c(3900)$.
Diagrams for the $Z_c(4020)\rt D^*\bar{D}^*$ (and the $Z_b$) processes are similar.}  It has been
suggested that the observed $Z_c$ (and $Z_b$) peaks are not due to genuine mesons, but are, instead,
artifacts of this coupled-channel process~\cite{bugg,matsuki,swanson}.

\subsubsection{Cusps?}
\begin{figure}[htb]


\begin{minipage}[t]{48mm}
  \includegraphics[height=0.75\textwidth,width=0.75\textwidth]{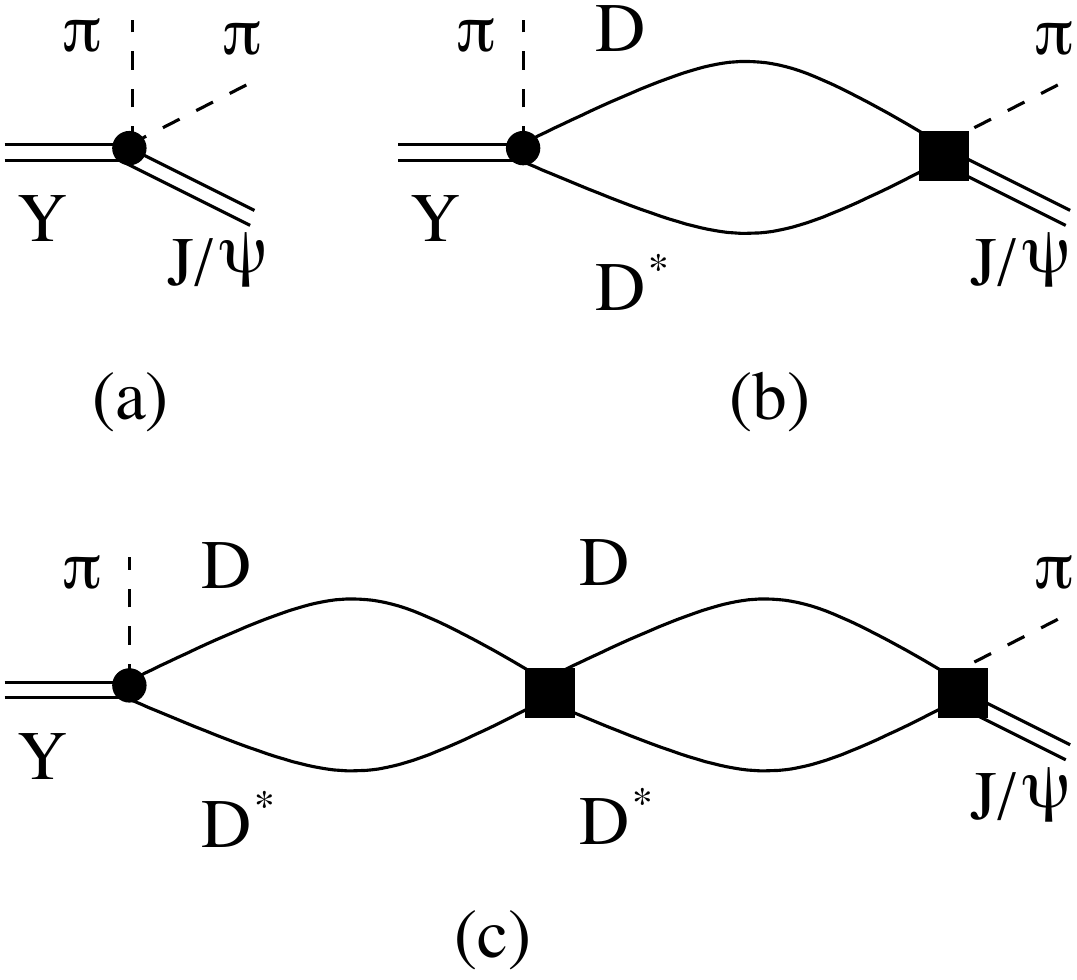}
\end{minipage}
\begin{minipage}[t]{48mm}
  \includegraphics[height=0.75\textwidth,width=0.9\textwidth]{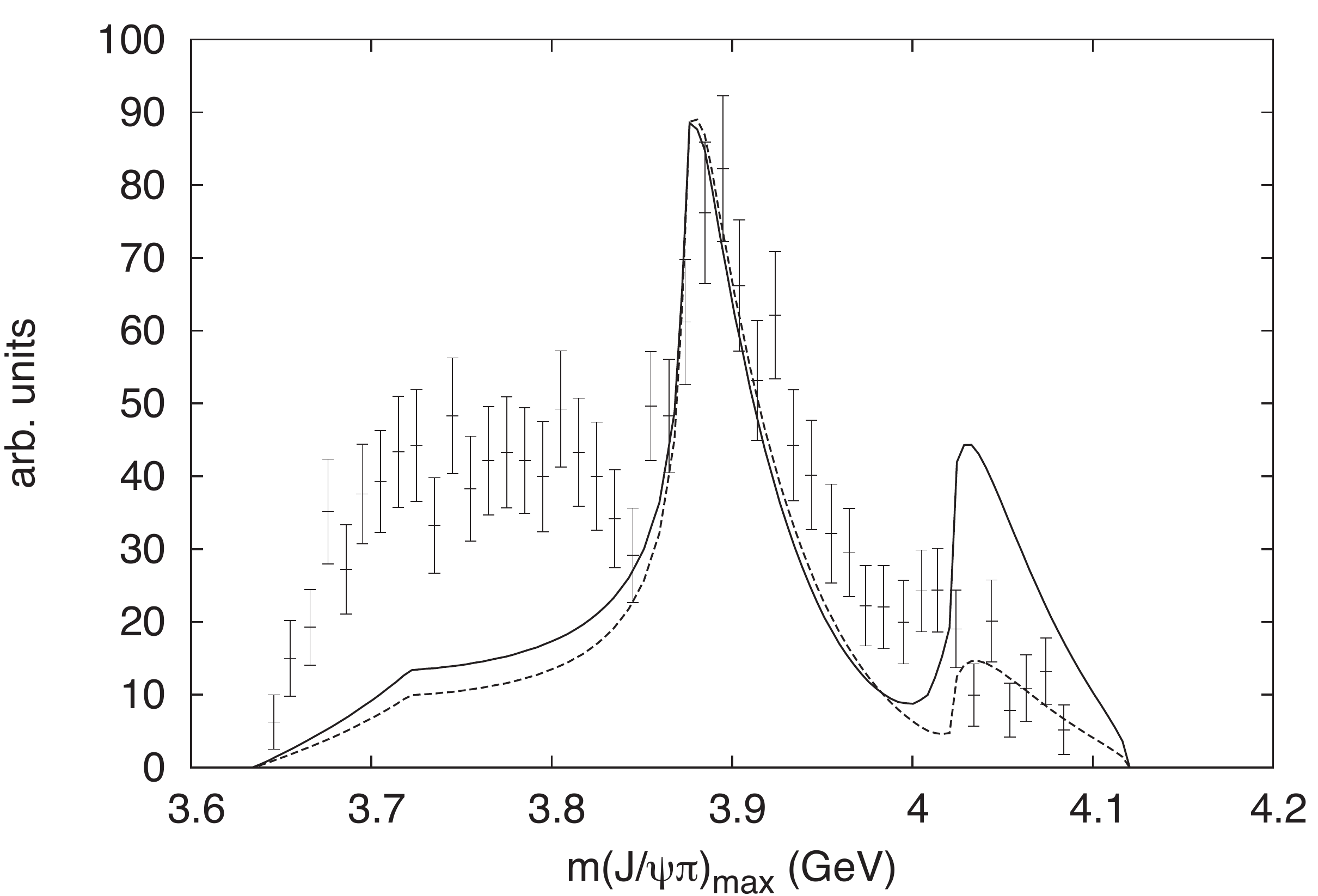}
\end{minipage} 
\begin{minipage}[t]{48mm}
  \includegraphics[height=0.75\textwidth,width=0.9\textwidth]{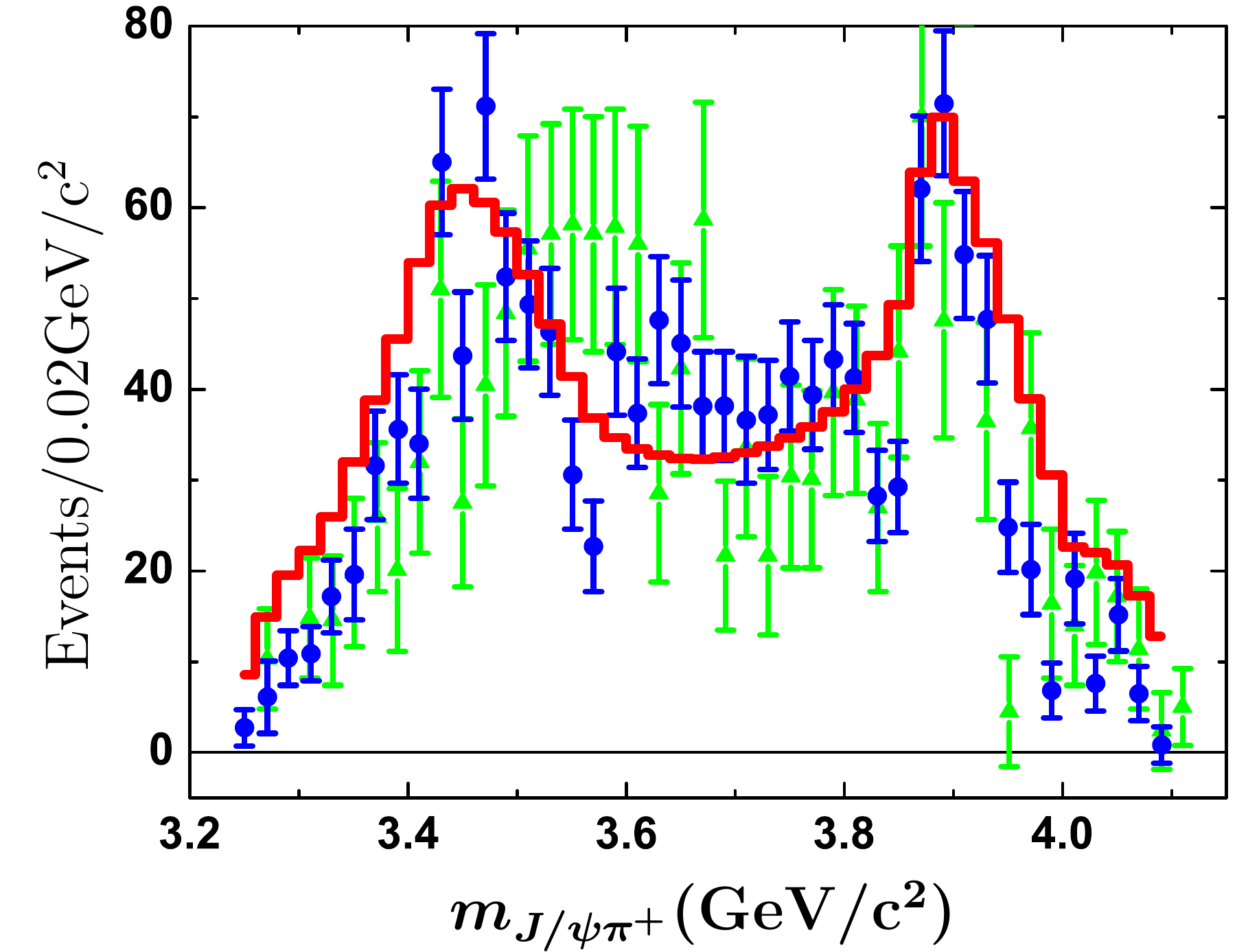}
\end{minipage}
\caption{\footnotesize {\bf Left:} (Figure~2 from Ref.~\cite{guo-hanhart}.) {\bf a)} Tree,
{\bf b)} one-loop and {\bf c)} two-loop diagrams for $Y(4260)\rt\pipi\jp$.
{\bf Center:} (Figure~6 from Ref.~\cite{swanson}.)  A comparison of BESIII $M(\pip\jpsi)$
data~\cite{bes_z3900} with the expectations for a cusp induced by a single $D\bar{D}^*$ loop that
is cut off with a Gaussian form factor.
{\bf Right:} (Figure~2 from Ref.~\cite{matsuki}.)  $M(\pi\jpsi)$ data with results from a fit to a
coupled-channel induced cusp produced by a single $D\bar{D}^*$ loop and cut off with a dipole form-factor,
plus a tree diagram with resonances in the $\pipi$ channel.  The round (blue) data points are from
BESIII~\cite{bes_z3900} and the triangular (green) data points are from Belle~\cite{belle_z3900}.}
\hspace{\fill}
\label{fig:cusp-1}
\end{figure}

\noindent
The $D\bar{D}^*$ loop in diagram~b of Fig.~\ref{fig:cusp-1}~(left) produces an imaginary
amplitude that rises rapidly starting at $M(\pi\jp)=m_{D}+m_{\bar{D}^*}$;  this rapid rise is
subsequently cutoff by a form-factor.   The net effect is a cusp-like peaking structure in the
$\pi\jpsi ~(h_c)$ ($\pi\Upsilon ~(h_b)$) invariant mass distibutions just above the $D\bar{D}^*$
($B\bar{B}^*$) threshold.  The authors of Refs.~\cite{matsuki} and~\cite{swanson} claim that these effects
can at least qualitatively  reproduce the general features of published $Z_c(3900)\rt\pi\jpsi$ data, as
shown in the center and right panels of Fig.~\ref{fig:cusp-1}.    

\subsubsection{Non-perturbative effects?}
\begin{figure}[htb]
\begin{minipage}[t]{48mm}
  \includegraphics[height=0.85\textwidth,width=0.85\textwidth]{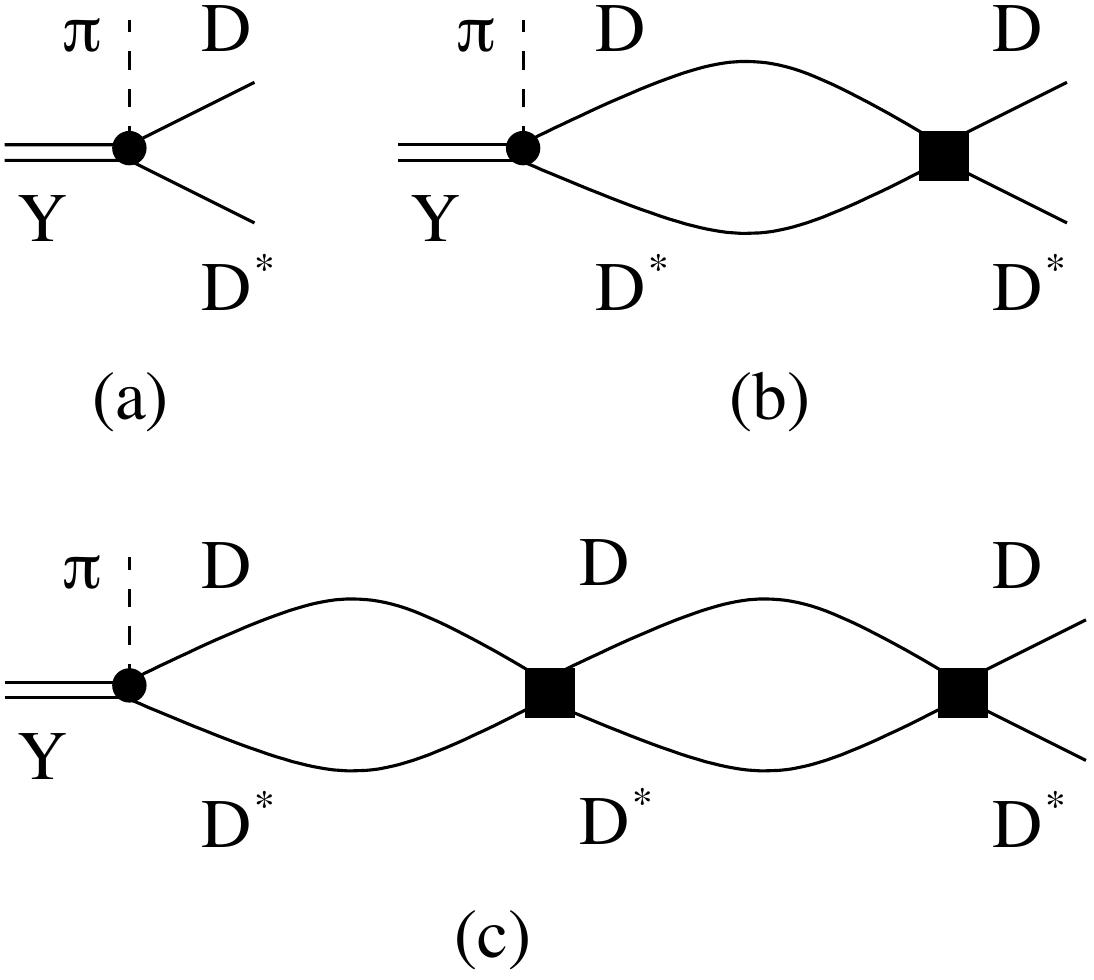}
\end{minipage}
\begin{minipage}[t]{48mm}
  \includegraphics[height=0.8\textwidth,width=0.95\textwidth]{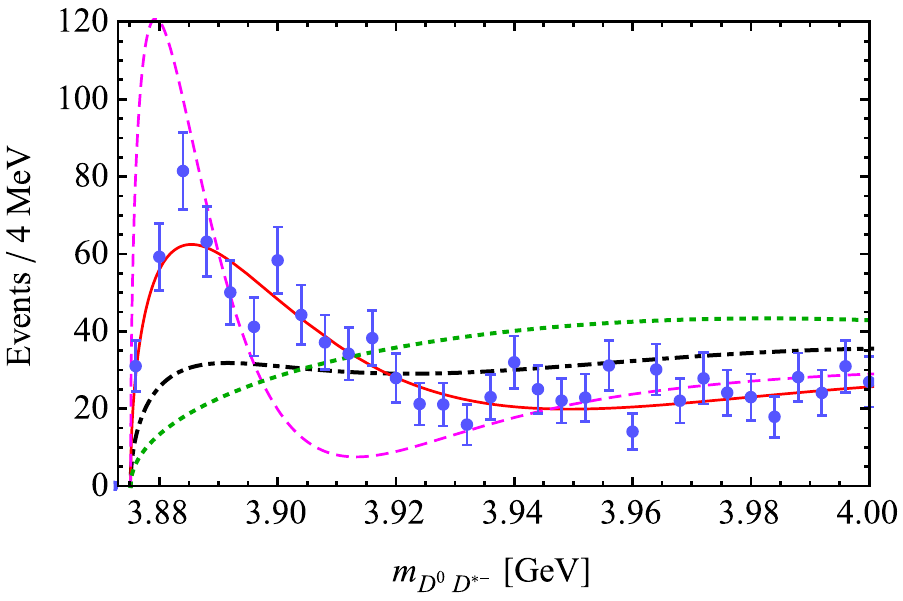}
\end{minipage} 
\begin{minipage}[t]{48mm}
  \includegraphics[height=0.8\textwidth,width=0.95\textwidth]{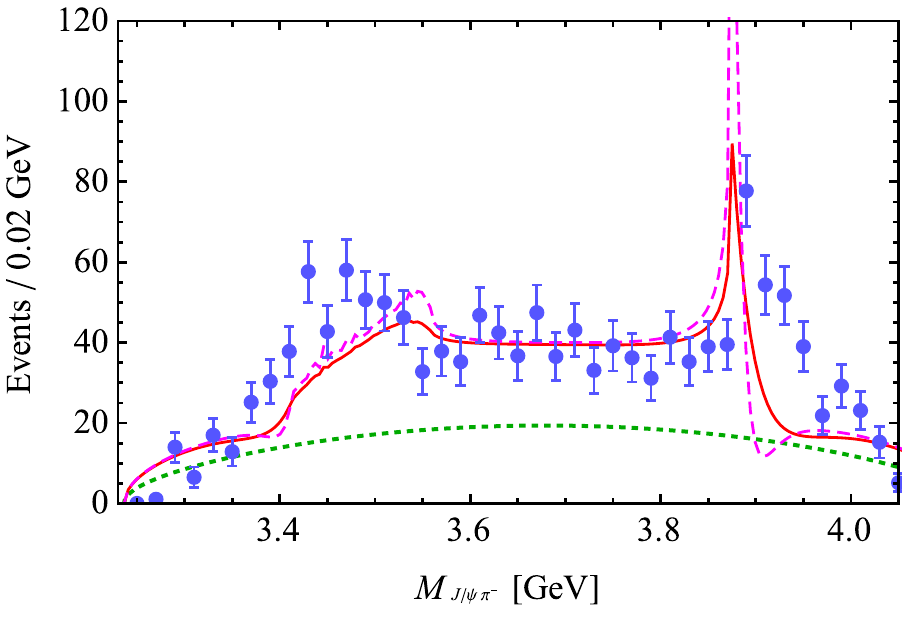}
\end{minipage}
\caption{\footnotesize {\bf Left:} {\bf a)} Tree, {\bf b)} one-loop
and {\bf c)} two-loop diagrams for $Y(4260)\rt \pi D\bar{D}^*$. (Figure~1 from Ref.~\cite{guo-hanhart}.) 
{\bf Center:}  The
solid (red) curve shows the Ref.~\cite{guo-hanhart} fit to the BESIII $M(D\bar{D}^*)$
distribution~\cite{bes_z3885} using the tree and single  $D\bar{D}^*$ loop (diagrams a~and~b
in the left panel of this figure) cut off with a Gaussian form factor. The dashed magenta curve includes
the two-loop term (diagram c) and the dot-dash black curve shows resuts when the perturbative expansion
is forced to converge.are described in the text.
(Figure~3 from Ref.~\cite{guo-hanhart}.)
{\bf Right:}   The  corresponding fit results applied to the BESIII $M(\pi\jp)$ data.
(Figure~4 from Ref.~\cite{guo-hanhart}.)}
\hspace{\fill}
\label{fig:cusp-2}
\end{figure}

\noindent
A more detailed study of this effect is discussed in Ref.~\cite{guo-hanhart}, where it is pointed
out that the closely related diagrams shown in Fig.~\ref{fig:cusp-2}~(left), with the same
form-factor and the same $Y$-$\pi D\bar{D}^*$ coupling, apply to the $Z_c\rt D\bar{D}^*$
channel, where they can produce threshold enhancements such as the $Z_c(3900)\rt D \bar{D}^*$ structure
reported by BESIII~\cite{bes_z3885}. The solid red curve in Fig.~\ref{fig:cusp-2}~(center) shows
results of a Ref.~\cite{guo-hanhart} fit to the BESIII $M(D\bar{D}^*)$ distribution that includes the
tree and single $D\bar{D}^*$ loop terms (diagrams a~and~b in Fig.~\ref{fig:cusp-2}(left)), and cut off
by a Gaussian form-factor; this fit shows that reasonable agreement with the data is possible. The
solid red curve in Fig.~\ref{fig:cusp-2}~(right) shows the results of a subsequent Ref.~\cite{guo-hanhart}
fit to the $M(\pi\jp)$ data from BESIII with the tree and single $D\bar{D}^*$ loop diagrams of
Fig.~\ref{fig:cusp-1} (left), for which the values of the $Y$-$\pi D\bar{D}^*$ coupling and the width of
the Gaussian form-factor are fixed at their $M(D\bar{D}^*)$-fit values.  Although the fit quality here
is poorer, the $Z_c(3900)\rt \pi\jp$ peak is at least qualitatively reproduced. (The dashed green lines
in the $M(D\bar{D}^*)$ and $M(\pi\jp)$ plots show results from fits that only use the tree diagram.)

It is emphasized in Ref.~\cite{guo-hanhart} that the comparisons with data shown in the center and
right panels of Figs.~\ref{fig:cusp-1}~and~\ref{fig:cusp-2} are  based on only the first two terms
of a perturbation series, {\it i.e.},  diagrams~a and~b in the left-hand panels of
Figs.~\ref{fig:cusp-1}~and~\ref{fig:cusp-2}, and neglect contributions from the double-loop terms
shown in diagrams c of the same figures as well as (not shown) three- and higher loop terms.   The
dashed magenta curves in the center and right panels of Fig.~\ref{fig:cusp-2} show the effects of
adding the double-loop amplitudes based on the parameters determined from the single-loop-only fits.
Here dramatic departures from the single-loop-only fit results for both the $M(D\bar{D}^*)$ and
$M(\pi\jp)$ distributions demonstrate that the neglect of the higher-order terms in the perturbation
series, which is implicit in the characterizations given in Refs.~\cite{bugg,matsuki,swanson}, is not
justified.  The dash-dot black curve in the $M(D\bar{D}^*)$ plot of Fig.~\ref{fig:cusp-2}~(center) shows
the result of an attempt to fit the $M(D\bar{D}^*)$ distribution with a full perturbation expansion that
is forced to converge; here the agreement with data is poor.  

Based on these results, the authors of Ref.~\cite{guo-hanhart} conclude that the near-threshold $Z_c$
(and $Z_b$ peaks) cannot be purely kinematic effects and must be due to the influence of a nearby pole
in the $S-$matrix, thereby qualifying them as legitimate meson states.

\subsubsection{Experimental tests}
\noindent
The question of whether the near-threshold $Z_{c(b)}$ peaks seen by BESIII and Belle are due
to genuine meson states or, instead, coupled-channel kinematic effects is too important
to be left to theorists\footnote{The theoretical situation remains unclear.  Three months after this
presentation, Swanson~\cite{swanson_2015} posted a rebuttal to the claims in Ref.~\cite{guo-hanhart}.
Shortly after that, the Ref.~\cite{guo-hanhart} authors responded wuth a rebuttal to Swanson's
rebuttal~\cite{cleven_2015}.} and should be settled, if possible, by experiment.
 To date, the BESIII group
has only done separate fits to the $M(\pi\jp)$ and $M(D\bar{D}^*)$ distributions for their
$Z_c(3900)\rt \pi\jp$ and $D\bar{D}^*$ data samples.  Simultaneous fits using amplitudes
suggested in refs.~\cite{bugg,matsuki,swanson,guo-hanhart} would probably be instructive. 
More critical would be phase measurements of the $Z_{c(b)}$ amplitudes.

\begin{figure}[htb]
\begin{minipage}[t]{74mm}
  \includegraphics[height=0.45\textwidth,width=1.0\textwidth]{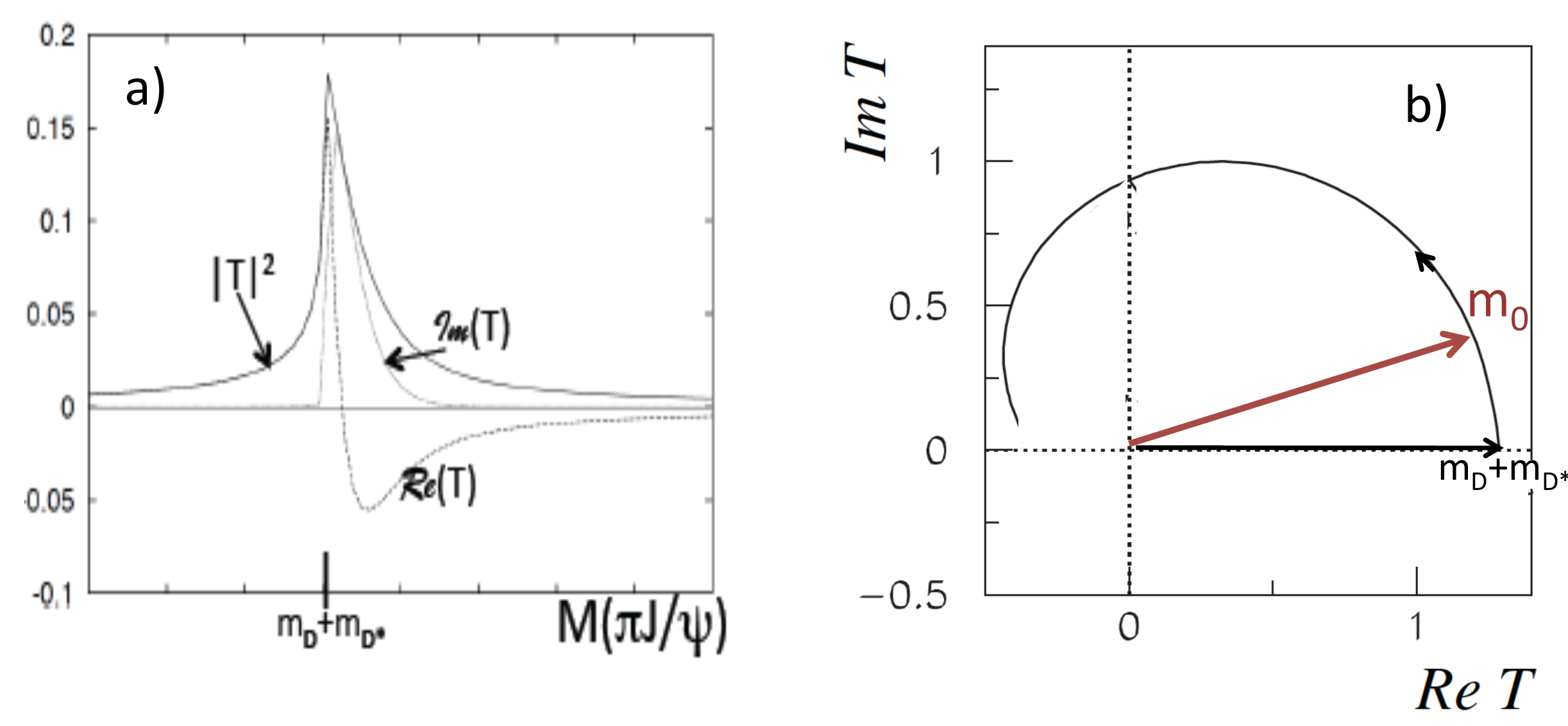}
\end{minipage}
\begin{minipage}[t]{72mm}
  \includegraphics[height=0.45\textwidth,width=1.0\textwidth]{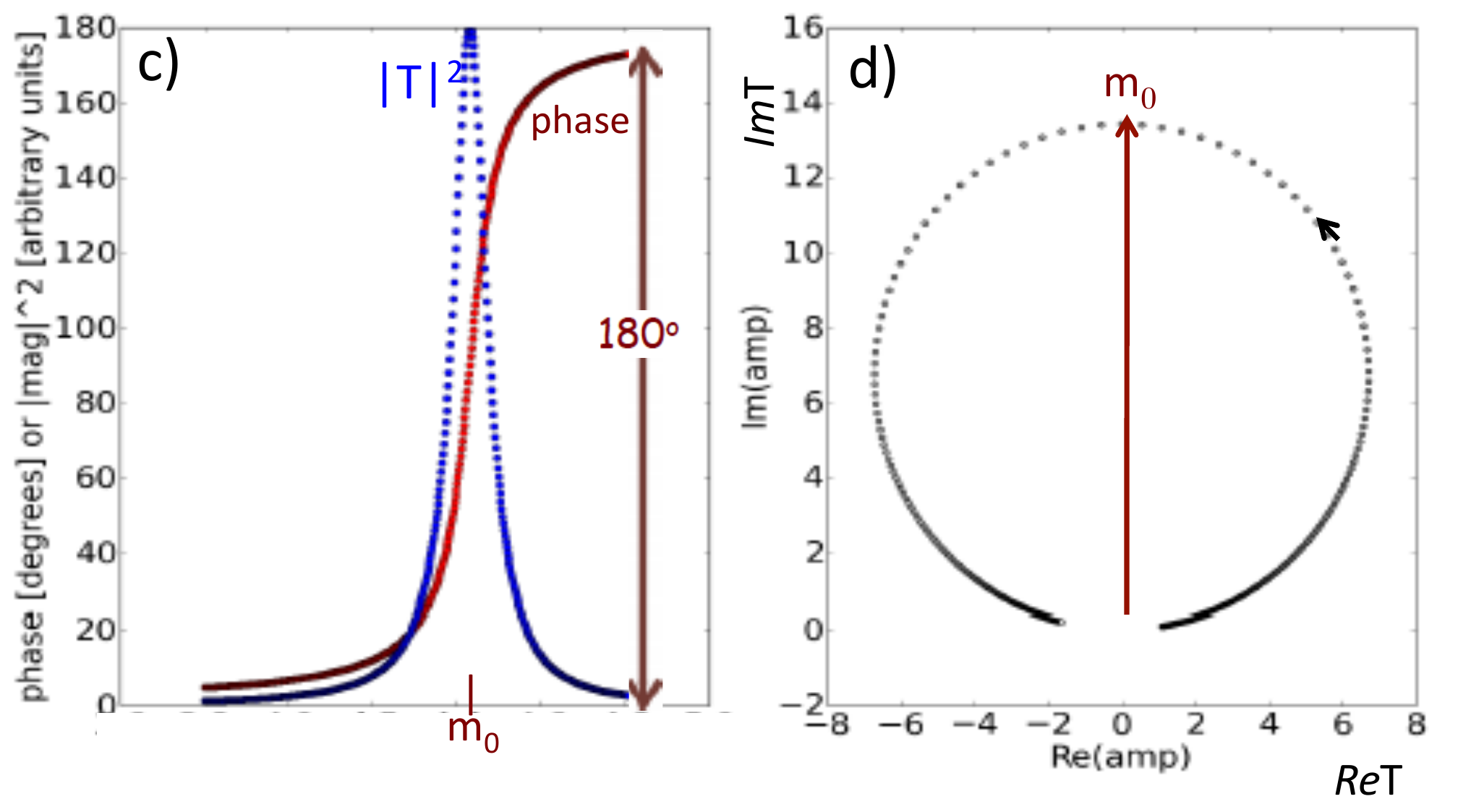}
\end{minipage}
\hspace{\fill}
\caption{\footnotesize{
{\bf a)} The real, imaginary and modulus squared for a $D\bar{D}^*$-loop-generated
peak in the $\pi\jpsi$ mass distribution (adapted from Ref.~\cite{swanson}).
{\bf b)} A sketch of the Argand diagram generated from the amplitudes shown in a), with the
peak location indicated by an arrow (adapted from Ref.~\cite{bugg}).
{\bf c} The modulus and phase of a Breit Wigner resonance.
{\bf d)} The Argand plot for a Breit Wigner resonance.
\label{fig:phase}}}
\end{figure}

The cusp models discussed above start with imaginary amplitudes generated by the loop diagrams shown
in the left panels of Figs.~\ref{fig:cusp-1}~and~\ref{fig:cusp-2};  the real parts of these
ampltudes can be determined by analyticity requirements.  The resulting phase motion, described
in some detail in Ref.~\cite{bugg}, is different than that of a BW amplitude. Figure~\ref{fig:phase}a
 shows the real and imaginary amplitudes and the modulus for a coupled-channel-generated peak
from Ref.~\cite{swanson}; Fig.~\ref{fig:phase}b shows a sketch of its associated Argand plot,
where the arrow indicates the location of the peak.  For comparison, Fig.~\ref{fig:phase}c shows
the modulus and phase of a BW resonance, and~\ref{fig:phase}d shows its associated Argand plot.
The latter two plots show that the BW amplitude has a rapid, $180^0$ phase change across the
resonance peak; this is not the case for the coupled-channel-generated peak, which has a
relatively small phase motion surrounding the peak. Thus, with sufficient statistics, amplitude
analyses of the $Z_c(3900)\rt\pi\jpsi$ and $D\bar{D}^*$ peaks should to able to distinguish
coupled-channel effects from a genuine resonance.   The BESIII group is currently doing a
Partial Wave Analysis of existing $Z_c(3900)\rt \pi\jp$ data that could address this question,
albeit with limited statistics~\cite{pingrg}.  There is also a proposal within the BESIII
collaboration to accumulate a much larger $Y(4260)\rt\pipi\jp$ data sample that could be
suitable for a definitive distinction between a resonance and coupled-channel-cusp
origin for the observed peaks~\cite{yuancz}.  

\section{Comments and speculations}
\noindent
\subsection{Comment on the partial widths for $\mathbf{Z_{(c)}\rightarrow  \pi (\ccbar)}$}
\begin{figure}[htb]
\begin{minipage}[t]{37mm}
  \includegraphics[height=0.9\textwidth,width=0.9\textwidth]{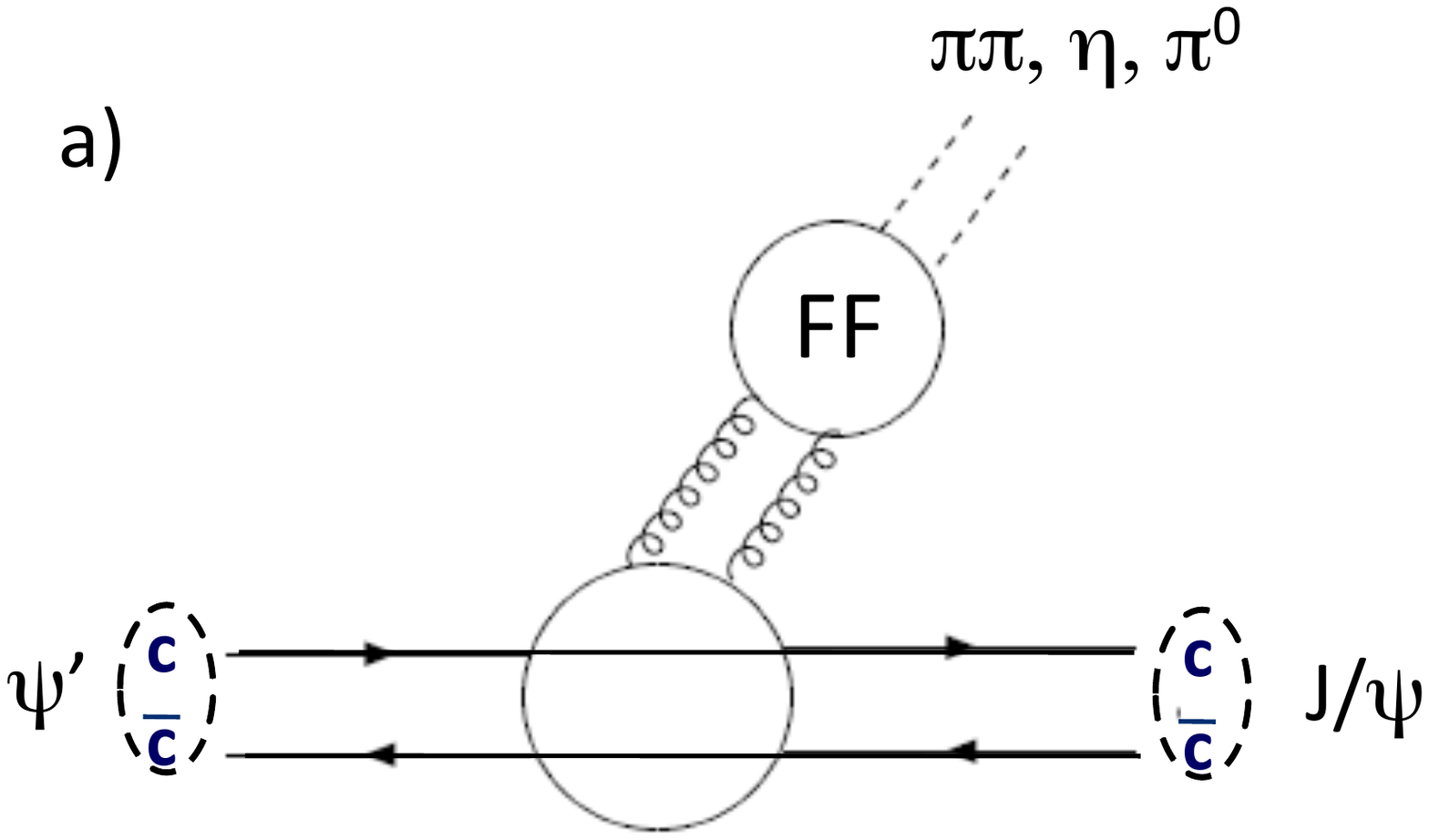}
\end{minipage}
\begin{minipage}[t]{37mm}
  \includegraphics[height=0.9\textwidth,width=0.9\textwidth]{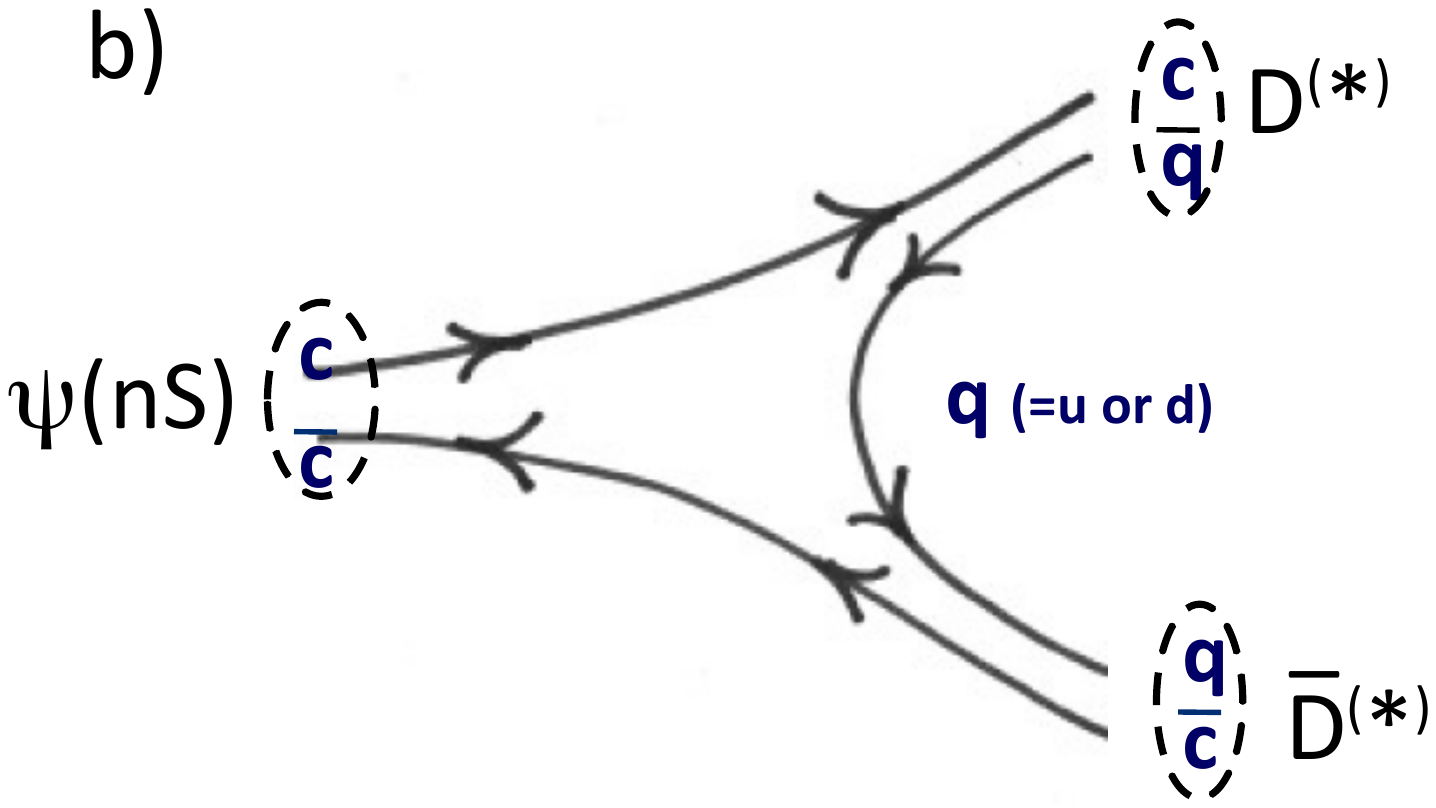}
\end{minipage}
\begin{minipage}[t]{37mm}
  \includegraphics[height=0.9\textwidth,width=0.9\textwidth]{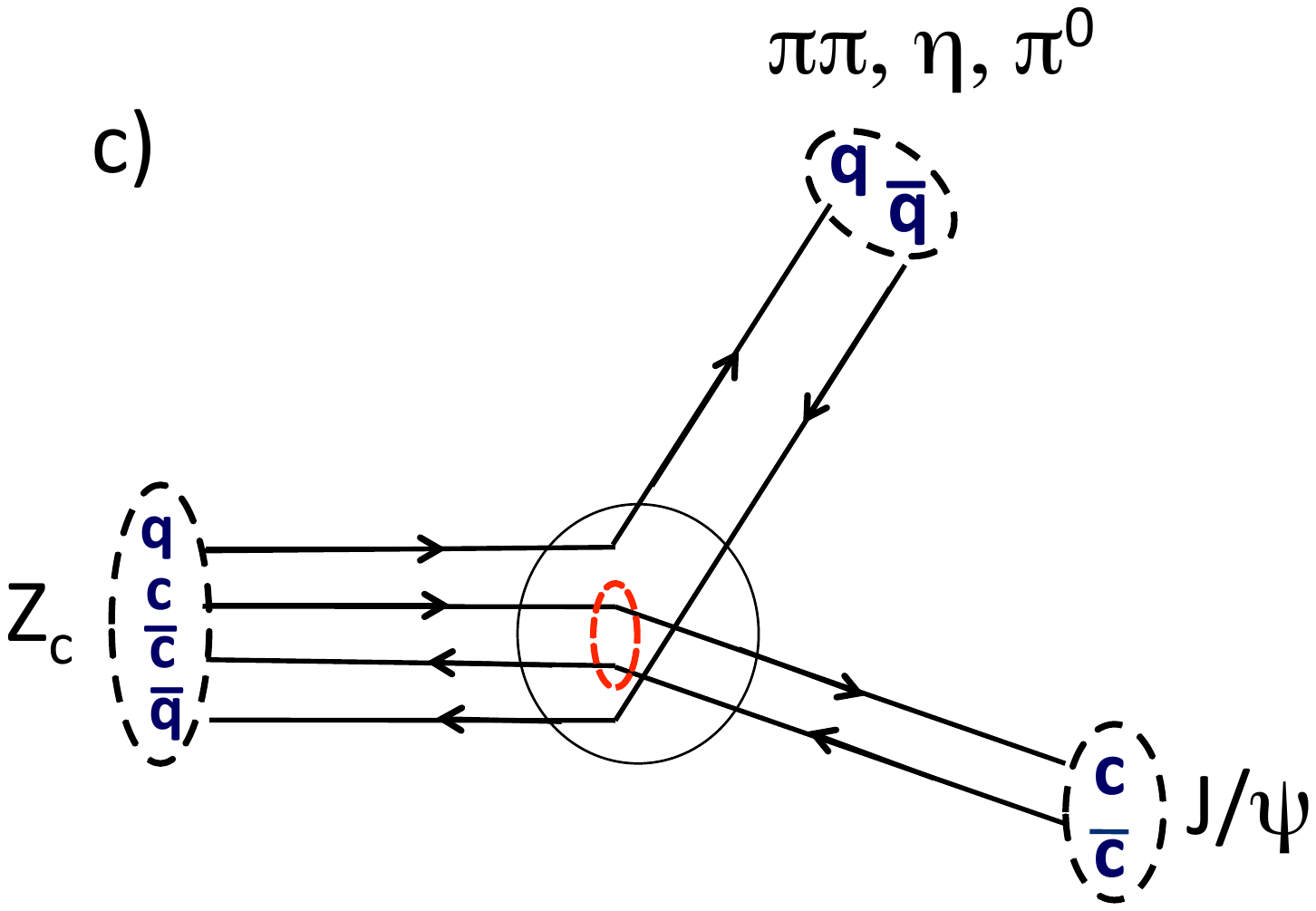}
\end{minipage}
\begin{minipage}[t]{37mm}
  \includegraphics[height=0.9\textwidth,width=0.9\textwidth]{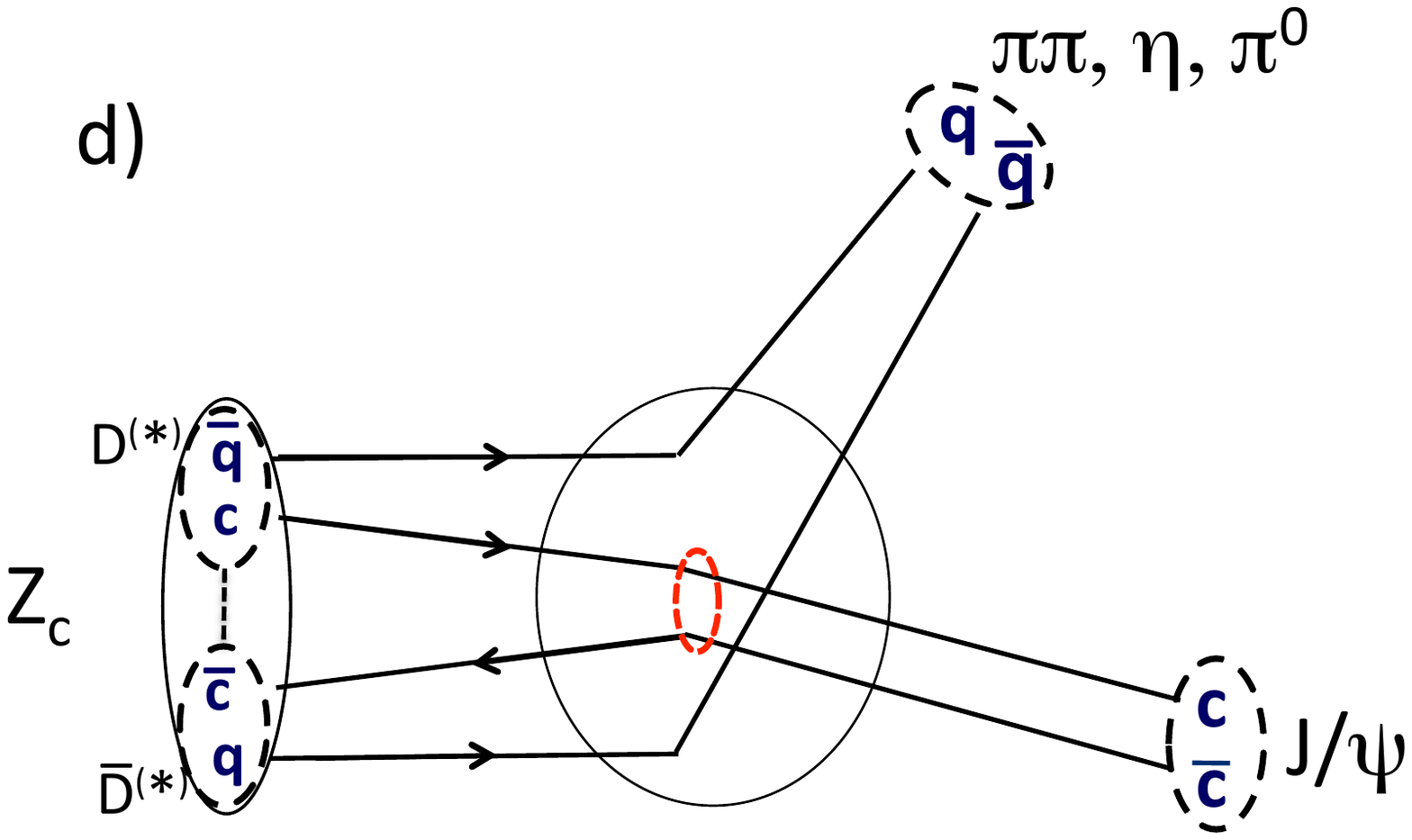}
\end{minipage}
\hspace{\fill}
\caption{\footnotesize{
{\bf a)} A diagram for $\psip\rt\pipi\jpsi$.  Since no quark lines connect the light
hadron system to the initial state, this process is expected to be OZI suppressed.
{\bf b)} The OZI-allowed ``fall-apart'' decays to open-charmed-meson pairs that are dominant for
$\ccbar$ states with masses that are above open-charmed threshold.
{\bf c)} OZI-allowed decays of a QCD-tetraquark to light hadrons plus a $\jpsi$.
{\bf d)} A cartoon of a $D^{(*)}\bar{D}^{(*)}$ ``molecule'' decaying to light hadrons plus a $\jpsi$.
\label{fig:feyn}}}
\end{figure}

\noindent
For standard $\ccbar$ mesons, the decay partial widths for hadronic transitions between
different charmonium states are typically of order of a few hundreds of keV or less.  The largest measured
one is $\Gamma(\psi(4040)\rt\eta\jpsi)=416\pm 76$~keV; others are smaller, {\it e.g.},
$\Gamma(\psip\rt\pipi\jpsi)=157\pm 5$~keV,  $\Gamma(\psi(3770)\rt\pipi\jpsi)=73\pm 11$~keV, and
$\Gamma(\chi_{c2}\rt\pipi\eta_c)<43$~keV.  This is generally understood to be a consequence of the OZI
rule~\cite{ozi}, which (in modern language) says that processes in which the Feynman diagram can be split
in two by only cutting internal gluon lines will be suppressed. This is the case for hadronic transitions
between charmonium states, as indicated in Fig.~\ref{fig:feyn}a.

For standard charmonium states that are above the open-charmed threshold, diagrams for decays to
$D^{(*)}\bar{D}^{(*)}$ final states, as shown in Fig.~\ref{fig:feyn}b, are not OZI suppressed and
partial widths for these ``fall-apart'' modes are substantially larger. For cases where they have been
measured, the OZI suppression factors are more than a hundred~\cite{pdg}:
\begin{equation}
\frac{\Gamma(\psi(3770)\rt D\bar{D})}{\Gamma(\psi(3770)\rt\pipi\jpsi)}\simeq 350;~~~~~
\frac{\Gamma(\psi(4040)\rt D^{(*)}\bar{D}^{(*)}}{\Gamma(\psi(4040)\rt\eta\jpsi)}\simeq 150.
\end{equation}

Since the $Z_c$ states are electrically charged, their minimal structure must be $cq_i\cbar\qbar_j$,
$q_{1(2)}=u(d)$, and OZI suppression is easily evaded.   Figure~\ref{fig:feyn}c shows a sketch of how
a QCD tetraquark could decay via an OZI-allowed hadronic transition to a $\jpsi$.  In a pure molecular
picture, one would expect the same transition to be suppressed relative to $D^{(*)}\bar{D}^{(*)}$ fall-apart
decays even though the OZI rule is not violated.  As illustrated in Fig.~\ref{fig:feyn}d, a molecular
configuration is an extended object in which the $c$- and  $\cbar$-quarks exist in
distinct, color-singlet $D^{(*)}$ and
$\bar{D}^{(*)}$ mesons, each with a spatial extent of order 1~fm. These $D^{(*)}$ and $\bar{D}^{(*)}$ mesons are
expected to be separated by a similar distance. Since they reside in distinct color-singlet systems, the colors
of the $c$- and $\cbar$-quarks are uncorrelated. In order to form a $\jpsi$, the $c$ and $\cbar$ colors must
match {\it and} they should have considerable overlap in a spatial region with volume of order $\langle r_{\jpsi}\rangle^3$,
where $\langle r_{\jpsi}\rangle\simeq 0.4$~fm is the mean $c$-$\cbar$ separation in the $\jpsi$~\cite{eiglsperger}. 
This is in contrast to a QCD tetraquark, in which the $c$ and $\cbar$ start out being both color correlated and
in close proximity.

The partial widths for $Z_c(3900)\rt\pi\jpsi$ and $Z_c(4020)\rt\pi h_c$ are smaller than those for $D\bar{D}^*$
and $D^*\bar{D}^{(*)}$, respectively, but not by very large factors~\cite{bes_z3885,bes_z4025}: 
\begin{equation}
\frac{\Gamma(Z_c(3900)\rt D\bar{D}^*)}{\Gamma(Z_c(3900)\rt\pi\jpsi)}=6.2\pm 3.0;~~~~~
\frac{\Gamma(Z_c(4020)\rt D^{*}\bar{D}^{*}}{\Gamma(Z_c(4020)\rt\pi h_c)}= 12\pm 5 .
\end{equation}
Although no data exist for either $Z_c(4200)$ or $Z(4430)$ decays to $D^{(*)}\bar{D}^{(*)}$ final states, if
one assumes that the branching fractions for $B\rt KZ_c(4200)$ and $B\rt KZ(4430)$ are no
larger than the PDG upper limit for ${\mathcal B}(B^+\rt K^+ X(3872))<3.2\times 10^{-4}$,
existing data~\cite{belle_z4200,belle_z_dalitz2} can be used to infer branching
fraction {\it lower} limits of ${\mathcal B}(Z_c(4200)\rt \pi\jpsi)>4\%$ and
${\mathcal B}(Z(4430)\rt \pi\psip)$>7\%, which imply large partial widths of order 10~MeV or larger
for hadronic transitions to standard charmonium states.\footnote{The lowest-order diagram for
$B\rt K X(3872)$ is ``factorizable.'' In contrast, the lowest-order $B\rt KZ_c(4200)$ and $B\rt KZ(4430)$
decay processes are non-factorizable. Non-factorizable processes are expected to be suppressed relative to
factorizable ones. For a discussion about factorization in $B$-meson decays, see Ref.~\cite{factorization}.}
This suggests that these states are not pure $D^{(*)}\bar{D}^{(*)}$ molecules but, instead, are hybrid-like
structures that contain a tightly bound diquark-diantiquark core. The mass spectrum of these states would
then reflect the underlying diquark-diantiquark dynamics, modified by the influence of nearby
$D^{(*)}\bar{D}^{(*)}$ thresholds.

\subsection{The observed spectrum of $\mathbf{J^P=1^+}$ states}
\begin{figure}[htb]
\begin{center}
  \includegraphics[height=0.7\textwidth,width=0.8\textwidth]{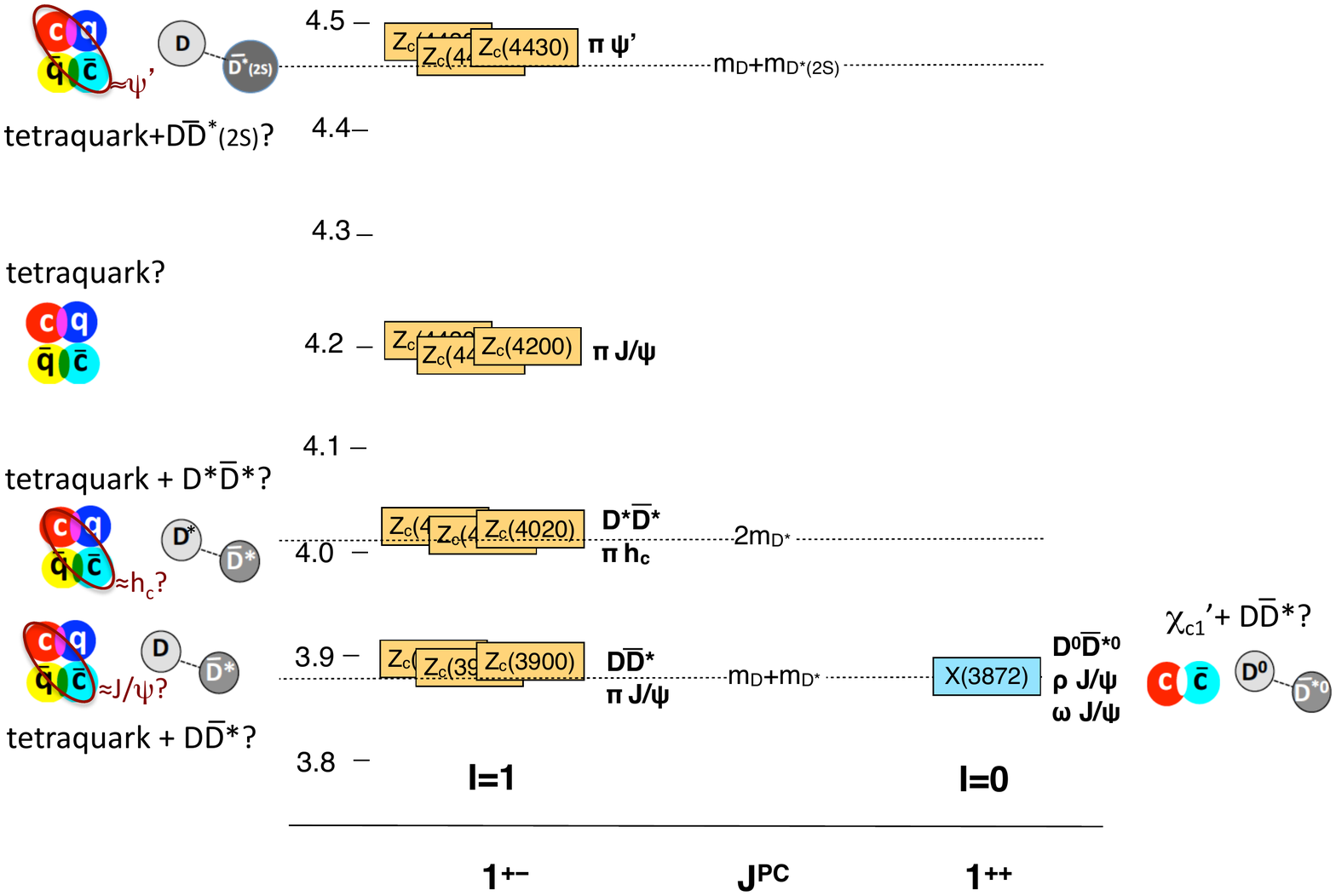}
\caption{\footnotesize{
A summary of the $J^P = 1^+$ $X$ and $Z$ charmoniumlike mesons that have been seen to date. All of
them are near open-charmed meson thresolds with the notable exception of the recently discovered
$Z_c(4200)$.  Possible core and meson components are indicated. Here the $C$-parity assignment of
the isovector states refers to that of the neutral member.
\label{fig:xyz_1+_seen}}}
\end{center}
\end{figure}

\noindent
Figure~\ref{fig:xyz_1+_seen} shows the spectrum of $J^P=1^+$ states discussed in the previous two
sections, along with their dominant decay modes .  The horizontal dashed lines indicate the
$m_D+m_{D^*}$, $2m_{D^*}$, and $m_D+m_{D^*(2S)}$ open-charmed thresholds.  All of the states lie near an 
open-charmed threshold with the notable exception of the recently discovered $Z(4200)$.  In accord with the hybrid
model for the $X(3872)$ discussed above in Section~\ref{sect:hybrid} and in the spirit of Gell-Mann's
Totalitarian Principle for Quantum Mechanics: ``{\it Everything not forbidden is compulsory}~\cite{gellmann},''
I attach cartoons next to each state suggesting a QCD core component that mixes with open charmed meson-antimeson
pairs ($D^{(*)}\bar{D}^{(*)}$) if their threshold is nearby in mass.  For the $X(3872)$, the simplest assumption
for the core component is the $\chi^{\prime}_{c1}$, although this could probably coexist with some admixture of
$cu\cbar\ubar$ and $cd\cbar\dbar$ tetraquarks.  For the various isovector $Z$ states, the simplest
core states would be $cq_i\cbar\qbar_j$, where $q_{1(2)}=u(d)$.

Since the tetraquark core components are not bound by the OZI rule, and have color-correlated $c$- and
$\cbar$-quarks in close proximity, these could account for the large hidden-charm decay partial widths
that are seen for the $Z$ states.  The effect of coupled-channel $D^{(*)}\bar{D}^{(*)}$ pairs might be
forcing some of the $Z$ states toward the open-charmed thresholds, similar to the way the $\ccbar$-$D\bar{D}^*$
couplings lower the mass of the $\ccbar$ core in the hybrid model for the $X(3872)$~\cite{takeuchi}.
Also, if, somehow, the $\ccbar$ pairs in the core states are somehow mostly configured in triplet $1S$, 
singlet $1P$, and triplet $2S$ configurations for the $Z_c(3900)$, $Z_c(4020)$ and $Z(4430)$, respectively,
that could cause the peculiar pattern where $\pi \jpsi$, $\pi h_c$ and $\pi \psip$ decays dominate for the
three different states. (Note that the $Z(4430)$-$Z_c(3900)$ mass splitting ($587\pm 20$~MeV) is close to
$m_{\psip}-m_{\jpsi}=589$~MeV.)

\subsection{Additional states?}
\begin{figure}[htb]
\begin{center}
  \includegraphics[height=0.6\textwidth,width=0.7\textwidth]{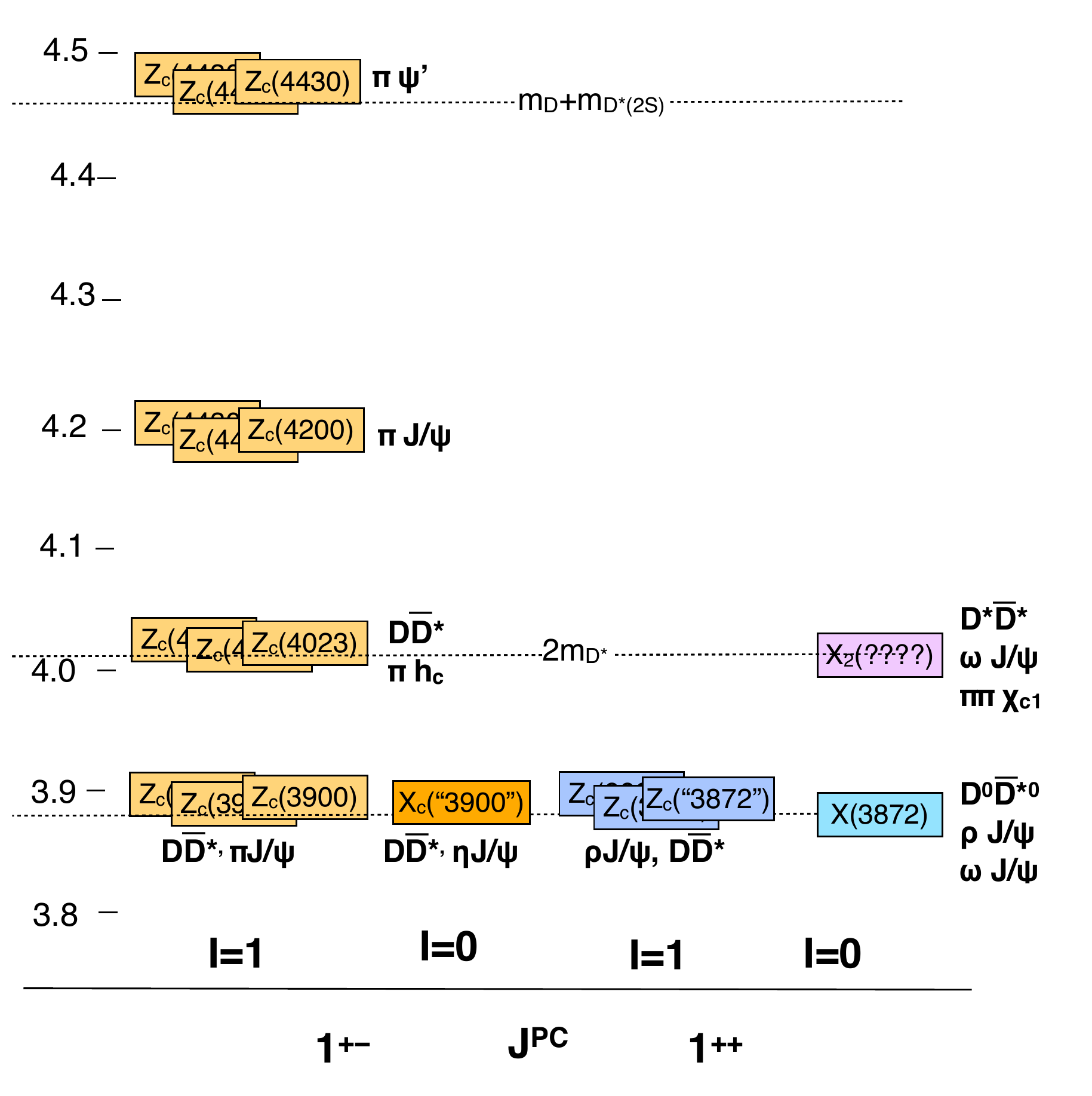}
\caption{\footnotesize{
Possibly additional low-lying $XYZ$ states discussed in the text are indicated. Here $Z_c(``3872")$ indicates
the possible $1^{++}$, mostly isovector partner of the $X(3872)$ discussed in Section~4.4.3, 
the $X_c(``3900")$ would be a $1^{--}$ mostly isoscalar partner of the $Z_c(3900)$ and the $X_2(``4020")$ would
be a counterpart of the $X(3872)$ near the $D^*\bar{D}^*$ threshold.
\label{fig:xyz-new}}}

\end{center}
\end{figure}

\noindent
The spectrum depicted in Fig.~\ref{fig:xyz_1+_seen} suggests the possibility of other related
states, some of which are indicated in Fig.~\ref{fig:xyz-new}.  These are labeled $Z_c(``3872")$,
the mostly isovector partner of the $X(3872)$ discussed above in Section~\ref{sect:hybrid}, 
$X_c(``3900")$, an isoscalar $1^{+-}$ partner of the $Z_c(3900)$, and $X_2(``4020")$ a version of the
$X(3872)$ located at the $D^*\bar{D}^*$ threshold.   Here I briefly discuss each of these.
\begin{description}
\item[The $\mathbf{Z_c(``3872")}$]  As discussed above, this might be a mostly isovector hybrid state
with a large $D^+ D^{*-}$ component.  The $\rho \jpsi$ decay mode would be isospin favored and, if it
were significant, this state would probably have been already found.  However, if the $Z_c(``3872")$
mass is near or above $m_{D^+}+m_{D^{*-}}=3879.9$~MeV, decays to $D\bar{D}^*$ final states might be strong,
resulting in a wide natural width and a small branching fraction for $\rho\jpsi$. The $Z_1(4050)\rt\pi\chi_{c1}$
peak reported by Belle~\cite{belle_z1z2} could have the correct $J^{PC}$ quantum numbers (to date, nothing
is known about its $J^{P}$ values), but its mass seems too high.
\item[The $\mathbf{X_c(``3900")}$] If this state exists and, in analogy to the $X(3872)$, has a hybrid
$\ccbar$-$D\bar{D}^*$ structure, the $h_c(2P)$ ($h_c^{\prime}$) would have the right mass and quantum
numbers to be its $\ccbar$ core state.  No evidence is seen for a structure in the $M(\eta\jpsi)$ distribution
for $B\rt K\eta\jpsi$ decays~\cite{belle_etajpsi}.  However, $1^{+-}$ quantum numbers do not seem to be
strongly produced in this $B$ decay process; $h_c(1P)$ production has not been been in $B$ meson decays
and a 90\% CL upper limit of ${\mathcal B}(B^+\rt K^+h_c) < 0.037\times{\mathcal B}(B^+\rt K^+\jpsi)$
has been established~\cite{pdg}.  One strategy might be to look for $\ee\rt\pipi X_c(``3900")$;
$X_c(``3900")\rt\eta\jpsi$ or $D\bar{D}^*$ in the $\sqrt{s}=4.36$~GeV BESIII data. This would be far enough
above threshold for a $\pipi X_c(``3900")$ to be detectable and a substantial, $\sim$50~pb~cross
section for the related $\ee\rt\pipi h_c$ process has been reported at this energy~\cite{bes_z4020}. 
\item[The $\mathbf{X_2(``4020")}$] There are no reports of a structure in the $\pipi\jpsi$ invariant mass
distribution in the vicinity of the $D^*\bar{D}^*$ mass threshold even though many experiments have studied
this sytstem.  However, the $Q$ value for transitions between a state with mass near $2m_{D^*}$
and the $\jpsi$ would be $\simeq 920$~MeV, well above the mass of the $\omega$ and, therefore,
the isospin-conserving $\omega\jpsi$ transition would likely be dominant.  The $\omega\jpsi$ invariant
mass distribution has not been well studied.  The BaBar experiment studied $\omega\jpsi$ systems produced
in $B\rt K\omega\jpsi$ decays using their full, 426~fb$^{-1}$ data sample.  They show an
intriguingly high data point in the $M(\omega\jpsi)$ distribution near 3990~MeV but with limited
statistical significance~\cite{babar_y3940}.  The only reported Belle study of the same channel
is based on a 253~fb$^{-1}$ data sample, which is only about one third of the full Belle data set.  
Another promising avenue for this search might be the $D^*\bar{D}^*$ system in $B\rt K D^*\bar{D}^*$
decays.  The only reported results for this channel are BaBar measurements of the branching fractions
${\mathcal B}(B\rt K D^{*}\bar{D}^{*})$~\cite{babar_kdstrdstr}, which are large: {\it e.g.},
${\mathcal B}(B^0\rt K^+ D^{*-}D^{*0}) = (1.06\pm 0.03\pm 0.09)\%$.  
\end{description}

\section{Summary}
\noindent
I think that now it is safe to conclude that four-quark states have been observed.  In fact, there are a
sufficient number of of established $J^P=1^+$ four-quark candidate states to reconstruct at least a partial
mass spectrum.  The initially proposed purely molecule-like and purely diquark-diantiquark explanations
for these states cannot reproduced their measured properties.  Some of the observed states are near
meson-antimeson thresholds, and have many properties that are similar to those expected for kinematically
induced threshold cusps.  However, these explanations fail to stand up well under close scutiny. 

The data seem to be telling us that rather than simple molecules of diquark-diantiquark substructures, the
observed states are hybrid configurations that consist of molecule-like meson-antimeson pairs coupled to a
tightly bound quark-antiquark or diquark-diantiquark core.   

This remains a data driven field where significant progress depends mainly on experimental observations of
additional states and better measurements  of the properties of existing states. To date, most initial
observations have involved final states containing a $\jpsi$ or a $\psip$ (or a narrow $\Upsilon$ state),
mostly because these are the simplest channels to access experimentally.  However, the BESIII experiment has
managed to isolate high-statistics, exclusive  $h_c$ signals with rather small backgrounds, and this resulted
in the discovery of the $Z_c(4020)$ More comprehensive studies of $D^{(*)}\bar{D}^{(*)}$ final states will be
difficult experimentally, but may be well worth the effort.  

There is a high interest in this subject and I expect it will continue to be a major emphasis of the
BESIII, CMS and LHCb research programs.  We can also look forward to future results from BelleII~\cite{belleii}
and PANDA~\cite{panda}. 

\section{Acknowledgements}
\noindent
I congratulate the organizers of Bormio-2015 for arranging another informative and stimulating meeting, and
thank them for giving me the opportunity to present this talk. This work was supported by the Institute for
Basic Science (Korea) under Project Code IBS-R016-D1.


\end{document}